%% file: Draft.tex
\documentclass[%
 reprint,
 amsmath,amssymb,
pra,
floatfix
]{revtex4-2}

\usepackage{graphicx}
\usepackage{dcolumn}
\usepackage{bm}
\usepackage{hyperref}
\hypersetup{
    colorlinks=true,       
    linkcolor=red,          
  citecolor=magenta,        
    filecolor=magenta,      
    urlcolor=cyan,           
    runcolor=cyan
}
\usepackage[export]{adjustbox} 

\usepackage{xcolor}
\usepackage{lipsum}
\usepackage[caption=false]{subfig}
\usepackage{comment}
\usepackage{multirow}
\usepackage{braket}
\usepackage[capitalise,compress]{cleveref}
\crefname{section}{Sec.}{Secs.}
\Crefname{section}{Section}{Sections}

\usepackage{subfiles}

\newcommand{\tr}{\mathrm{Tr}}

\begin{document}

\preprint{APS/123-QED}

\title{Provably Optimal Control for Multiplicative Amplitude Control Noise}

\author{Colin Trout}
\email{colin.trout@jhuapl.edu}
\author{Kevin Schultz}
\author{Paraj Titum}
\author{Leigh Norris}
\author{Gregory Quiroz}
\author{B. David Clader}
\thanks{Current Affiliation: Goldman, Sachs \& Co, New York, NY}%
\affiliation{%
 Johns Hopkins University Applied Physics Laboratory\\
 11100 Johns Hopkins Road, Laurel, MD, 20723, USA\\
 (Dated: \today)
}%

\begin{abstract}
\subfile{Sections/0-Abstract}
\end{abstract}

\maketitle

\section{\label{sec:intro} Introduction}
\subfile{Sections/1-Introduction}

\section{\label{sec:DCGderive} Gate-based Approach to Quantum Control}
\subfile{Sections/2-GateBasedControl}

\section{\label{sec:ErrorsControlNoiseDom} 
Error Dynamics for Control Noise Dominant Systems}
\subfile{Sections/3-ControlNoiseExpressions}

\section{\label{sec:controlham}Control Noise Only}
\subfile{Sections/4-ControlNoiseAnalysis}

\section{\label{sec:controldephasing} Noise Mitigation for Control Noise Dominant Systems}
\subfile{Sections/5-DephasingAddition}

\section{\label{sec:Discuss} Discussion and Future Work}
\subfile{Sections/6-Discussion}

\section{Acknowledgements}
\subfile{Sections/7-Acknowledgements}

\bibliography{Draft}

\onecolumngrid
\appendix

\newpage

\section{\label{sec:magnusexpansion} Magnus Expansion}
\subfile{Sections/A1-MagnusExpansion}

\newpage

\section{\label{sec:firstorderdephasing} Derivation of the First-Order Term for the Control Noise and Dephasing}
\subfile{Sections/A2-FirstOrderDerivation}

\newpage

\section{Control Noise Dominant Solution Examples}
\subfile{Sections/A3-ExampleSolutions}

\newpage

\section{\label{sec:highorderterms} Higher Order Terms in the Magnus Expansion}
\subfile{Sections/A4-HigherOrderTerms}

\end{document}

%% file: Sections/0-Abstract.tex
\noindent We provide a technique to obtain provably optimal control sequences for quantum systems under the influence of time-correlated multiplicative control noise. Utilizing the circuit-level noise model introduced in  [\href{https://journals.aps.org/prresearch/abstract/10.1103/PhysRevResearch.3.033229}{Phys. Rev. Research \bf{3}, 033229(2021)}], we show that we can map the problem of finding such a sequence to a convex optimization problem with guaranteed optimality that follows from the convexity. We also show that this technique is compatible with more general off-axis time-correlated dephasing noise. In spite of losing provable optimality, numerically optimized control sequences under this scenario can still achieve nearly optimal performance when the control noise is strong relative to the dephasing contribution. This approach will enable the development of optimal quantum logic gates in systems where noise due to amplitude drifts in the control is strong relative to dephasing such as in ion-trap based quantum computers or in the limit of fast control.

%% file: Sections/1-Introduction.tex
Precise and robust control of quantum systems is a requirement to take advantage of the great promise of quantum technologies; potentially surpassing theoretical bounds of the performance of current technologies for application in the areas of sensing \cite{2011GiovannettiMetrologyReview, 2017DegenSensingReview}, communication \cite{2021ChenQuantumComms}, and computing \cite{2021PreskillQCReview,2022WhitefiledQCcurrent}.  However, the sensitivity of quantum systems to undesired environmental influences and device imperfections has proven a major roadblock in the development of quantum technologies.  Because of this, approaches for making quantum device robust to these sources of noise have been developed at all levels of a quantum device software stack from hardware-level control protocols \cite{2010DongQuantumControlReview, 2021JamesOptControlReview} all the way up to software level techniques (quantum error correction \cite{2010GottesmanQEC, 2017CampbellFTreview}, for instance).

At the lowest level of the quantum software stack lies protocols for mitigating noise that are compiled using the language of the control hardware.
These approaches require engineering the applied control in a manner that generates a minimal system response to the sources of noise.
To engineer the proper control, first the features of the noisy environment must be measured and are typically represented by a spectrum of the noise which provides the frequency regimes where the noise is strong.  
Various approaches of estimating the noise spectrum through direct measurement of the quantum device \cite{2003ShoelkopfQuantumNoise,2004FaoroDynamicalSupp1fNoise,2012YoungQubitsSpectrometers} have been developed \cite{2016NorrisNonGaussianDephasing,2016SzannkowskiSpectroscopyofCrossCorr,2017PazSilvaMultiqubitGaussianSpectroscopy,2018NorrisOptBandControlNoise} and demonstrated experimentally for various systems and types of noise spectra \cite{2011AlvarezMeasureSpectrumColoredNoseDD,2011BylanderSpectroscopySuperconductingFluxQubit,2011AlmogSpectroscopyAtomicEnsembles1,2013KotlerSpectroscopyAtomicIon,2014MuhonenSpectroscopySemiconductorSpin,2015RomachSpectroscopyNVCenter,2016AlmogSpectroscopyAtomicEnsemble2,2017MalinowskiSpectroscopyGaAsSpinQubit,2020FreySimultSpectralEstQubitSlepian,2022BarrControlNoiseSpecDephasing}.
Once the frequency profile of the noise is understood, the control can be designed in a manner to enforce minimal frequency response of the system in the frequency regimes where the noise is strong; known as the filter design approach to noise mitigation.
This filter design approach to noise mitigation was traditionally developed to overcome dephasing noise \cite{1998BanPhotonEchoDD,1998ViolaDynamicalErrorSupp,*1999ViolaDDOpenSystems,1999ZanardiSymmetrizingEvol,1999ParityKicksDecoherence,2001KofmanDynamControlDecay,2007UhrigKeepQubitAlivePi,2011BiercukDDFilterDesign,2012SuFFMultiQubitDD,2014PazSilvaGenFFControl}, but has been extended to other sources of noise \cite{2013GreenArbControlUnivNoise} as well as tailored for logic gates \cite{2012GreenHighOrderFilterLogic}.  Filter design approaches to noise mitigation have been demonstrated in the lab on various experimental platforms \cite{2009BiercukDDonIons,*2009BiercukOptDDQuantumMem,*2014SoareExpNoiseFilter,2010deLangeUnivDDNVcenter,2011BluhmDDSemiconductorFF,2012MedfordDDSemiconductorSpinFF}.

An alternative (or complementary) approach to mitigating noise can be applied at a step higher in the quantum device software stack using the language of compiled quantum gates; that is to say gate primitives compiled from the hardware-level/pulse-level primitives from the previous approach in a robust manner.  The goal of this approach is to compile the hardware-level primitives into gate instructions in a manner that cancels out erroneous dynamics due to the noisy features of the device for construction of a gate.  This approach was traditionally developed for systematic control errors in the form of composite gate sequences in NMR \cite{1986LevittCompPulses,1994WimperisCompositePulse,2004BrownArbitraryAccurateCompPulses,2007AlwayArbitraryPrecisionCompositePulse} and has been extended to various noise scenarios  \cite{2012MerrillCPReview,2014KabytayevRobustnessCompPulsesTimeDepNoise} and implemented on a variety of quantum devices \cite{2003LeibfriedCPSingleIon,2008HaffnerCPIons,2008TimoneyCPsinglequbit,2009RakreungdetCPMicrowaveControl,2013ShappertCPMicrowaveGates,2014MerrillCPsAddressingErrors,2015MountErrorCompIonTrap}.
This approach was later generalized to control of open-system Hamiltonians through the development of Dynamically Corrected Gates (DCGs) \cite{2009KhodjastehDCGs,2009KhodjastehDCGSBoundedControls,2012KhodjastehAutoDCGs}.

In practice, both approaches to noise mitigation have advantages and disadvantages.
Filter design methods utilize intricate knowledge of the noise spectrum for developing control mitigation strategies, however the implementation of the designed controls require involved pulse shaping methods for constructing the control.
Gate-level mitigation methods adapt well into compiler-level optimizations utilizing pre-compiled gates within a gate set, but can often lack intricate consideration of the spectral features of the noise.
The recent development of Schr\"{o}dinger Wave Autoregressive Moving Average (SchWARMA) models \cite{2020SchultzSCHWARMA}, a technique for simulating time-correlated stationary processes in a quantum circuit language, has provided a formalism that allows for an approach to noise mitigation that can demonstrate the advantages of both flavors of noise mitigation.  In particular, the SchWARMA formalism allows for the representation of noise mitigation in a gate-level language adapting itself well to compiler-level optimizations while maintaining insight into the correlations of the noise present from the filter design approach.

To demonstrate this utility of the SchWARMA formalism for noise mitigation, we focus on the issue of mitigating time-correlated control noise for control noise dominant systems (with respect to the power of the various noise sources). This is accomplished via the minimization of the gate infidelity constructed from SchWARMA model representations of the noisy dynamics.
With our method, the computation of the optimal control that minimizes the gate error can be calculated at the compiler level as a simple convex optimization over the gate-level amplitudes of the applied control.
For this work, we will be optimizing over the integrated amplitudes of the control pulses for each gate at a coarse-grained scale that is agnostic to the  pulse profiles of the individual gates; as opposed to other direct minimization methods that require a basis for the control pulse such as GRAFS \cite{2018LucarelliGRAFS}.
While mitigation of control noise has previously been investigated in the language of DCGs \cite{2012KhodjastehAutoDCGs} and in the context of systematic control errors for ion traps systems \cite{2007AlwayArbitraryPrecisionCompositePulse, 2012MerrillProgressCompPulses, 2014KabytayevRobustnessCompPulsesTimeDepNoise}, our approach is unique in two aspects: (1) the focus on time-correlated stationary noise processes and (2) the compact representations of the noisy dynamics provided by the SchWARMA formalism.
With this approach, we achieve our main result of constructing optimal control for systems with only time-correlated stationary control noise that is provably optimal in the sense that it minimizes the first order term in the Magnus expansion of the noisy dynamics.
The SchWARMA model representations of the noise manifesting as terms in an analytical function of the infidelity of the dynamics allow for direct minimization of the error and avoid the need for the optimization in filter function space utilized by approaches~\cite{2012GreenHighOrderFilterLogic, Huang2017RobustQuantumGates}.
We also demonstrate the utility of extensions beyond pure control noise where we consider the addition of time-correlated dephasing noise.
In spite of loosing provable optimality of the approach when introducing dephasing, we show regimes where the errors from control noise plus dephasing dynamics are relatively close to the error observed from pure control optimal solutions when the dephasing noise is weak relative to the noise on the control. This provides confidence in the obtained solutions in spite of the nonconvex contributions from the introduction of dephasing.
Furthermore, we demonstrate the utility of the compact SchWARMA expressions of the infidelity by demonstrating the ability of the optimizer to make choices regarding the mitigation strategy when control noise and dephasing noise contributions compete.

In \cref{sec:DCGderive}, we will outline the background relating to our gate-based approach to quantum control using the language of dynamically corrected gates \cite{2009KhodjastehDCGs} followed by a representation of the effective error dynamics for the control noise problem in \cref{sec:ErrorsControlNoiseDom}.
With this development, we will show our main result of provably optimal control for time-correlated, stationary amplitude control noise in \cref{sec:controlham}: providing examples of optimal solutions, comparisons to other noise mitigation approaches, and a robustness analysis of the solutions.
We then discuss the extension of our approach to the case of control noise and time-correlated, stationary dephasing noise in \cref{sec:controldephasing}; utilizing the optimal control solutions to loosely define criteria for good solutions for noise models incorporating dephasing.
We conclude with a discussion about the results and future directions in \cref{sec:Discuss}.

%% file: Sections/2-GateBasedControl.tex
Our goal is to arrive at a circuit-level representation of the control dynamics under the influence of stationary noise.  
To that end, we will begin with a discussion of how to approximate the erroneous dynamics arising from the application of a gate sequence using the language from DCGs.
We can then use this representation of the dynamics of the noise to evaluate fidelity of a logic gate under the influence of these erroneous dynamics. 
Finally, we will translate the erroneous dynamics for a single qubit into the language of an ``error vector'' component that will allow us to rely on existing methods for evaluating the resulting expression for the fidelity with respect to the erroneous dynamics of the composite gate sequence.

\subsection{Effective Error Dynamics}
In accordance with our gate-based approach to quantum control, our goal is to evaluate the error dynamics of a sequence of quantum logic gates where the ideal gate is of the form $Q = Q_N Q_{N-1}...Q_{2} Q_1$ and each gate will have it's own associated contribution to the total error of the gate sequence.
Khodjasteh et al. \cite{2009KhodjastehDCGs} developed an approach to computing such dynamics through the development of DCGs.
They showed that the generator of the total error dynamics of a composite gate sequence, $\Phi_Q$, consisting of a sequence of erroneous gates can be approximated via
\begin{equation}\label{eqn:EPGcomposite}
    e^{-i\Phi_Q} = e^{-iP_{N-1}^\dagger \Phi_{Q_N} P_{N-1}}...e^{-iP_{1}^\dagger \Phi_{Q_2} P_{1}}e^{-iP_{0}^\dagger \Phi_{Q_1} P_{0}}
\end{equation}
where $\Phi_{Q_j} \equiv \Phi_{Q_j}(t_j, t_{j-1})$ is the generator of the error dynamics, in an interaction picture with respect to the ideal control $Q_j$, for an individual gate in the sequence (coined the ``error per gate", EPG) and $P_j$ is to total accumulated ideal control up through gate $j$, i.e. $P_j = Q_j Q_{j-1}...Q_2 Q_1 Q_0$ where $Q_0 = I$.
It was also shown that total error of the full gate sequence, $\Phi_Q$, can be expressed as an infinite series expansion (the Magnus expansion \cite{1954MagnusExpansion}) with respect to the noise contributions:
\begin{equation}\label{eqn:magnusexpans}
\Phi_Q = \sum_{i=1}^\infty \Phi_Q^{[i]}\,,
\end{equation}
where $\Phi_Q^{[i]}$ is the $i^{\rm th}$ order term in the Magnus expansion. The first order term in this expansion is given by
\begin{equation}
    \label{eqn:firstorderMagnus}
    \Phi_Q^{[1]}=\sum_{j=1}^N P_{j-1}^\dagger \Phi_{Q_j} P_{j-1} \,,
\end{equation}
and higher order terms in the expansion are determined by nested commutators~\cite{2009KhodjastehDCGs} (see \cref{sec:magnusexpansion} for examples).
To evaluate \cref{eqn:EPGcomposite}, all that remains is to formally define the EPG for a general gate and noise process.

We will now outline the definition of the EPG from \cite{2009KhodjastehDCGs}.  An individual target gate, $Q_j$, in the gate sequence will be generated from a system defined by the Hamiltonian
\begin{equation}
    H = H_{\rm gate} (t) + H_e\,,
\end{equation}
where $H_e = H_{S,e} + H_{SB} + H_{B}$ defines the error dynamics from the noisy control, unwanted system-bath interactions, and bath, respectively.  
$H_{\rm gate}(t) = H_{\rm ctrl}(t) + H_{S,g}$ defines the ideal control and is composed of the set of switchable control Hamiltonians, $H_{\rm ctrl}(t)$, and any ideal internal components used for gating, $H_{S,g}$, (procession for $\sigma_z$ gates, for example).
The ideal gate propagator can then be defined as
\begin{equation}
    U_{\rm gate}(t_j,t_{j-1}) = \mathcal{T}_+ \left[ e^{-i \int_{t_{j-1}}^{t_j} ds \, H_{\rm gate}(s)} \right]
\end{equation}
such that $U_{gate}(t_j,t_{j-1})=Q_j$.
The propagator for the erroneous evolution of gate $Q_j$ can then be defined as
\begin{equation}
    U(t_j,t_{j-1}) = \mathcal{T}_+ \left[e^{-i \int_{t_{j-1}}^{t_j} ds\,[H_{gate}(s) + H_{e}]} \right]\,.
\end{equation}
Now, we will transform into the interaction picture with respect to $U_{\rm gate}(t,0)$ to isolate the dynamics due to the error, that is
\begin{equation}
    \label{eqn:toggleframe}
    U(t_j,t_{j-1}) = U_{\rm gate}(t_j,t_{j-1})e^{-i \Phi(t_j,t_{j-1})}\,,
\end{equation}
and defining
\begin{equation}\label{eqn:EPG}
    e^{-i \, \Phi (t_j,t_{j-1})} = \mathcal{T}_+ \left[e^{-i \int_{t_{j-1}}^{t_j} ds \, \tilde{H}_{e}(s,0)} \right]
\end{equation}
where
\begin{equation}\label{eqn:Htilde}
\tilde{H}_{e}(t_j,t_{j-1}) = U_{\rm gate}(t_j,t_{j-1})^\dagger H_{e} U_{\rm gate}(t_j,t_{t_{j-1}})\,.
\end{equation}
where we arrive at the definition of the EPG, $\Phi(t_j,t_{j-1})$, from \cite{2009KhodjastehDCGs} for an individual gate in the sequence generating the total error dynamics in \cref{eqn:EPGcomposite}.

Now that we have an expression for the error dynamics for a composite gate sequence, our goal now is to use this description of the error dynamics to understand the contribution of the noise to the imprecision of a quantum logic gate
\cite{2013GreenArbControlUnivNoise}.  This can be done (at least in principle) straightforwardly by computing the Hilbert Schmidt inner product between the ideal and erroneous gate, $\mathcal{F}_{av} = \frac{1}{4} \langle |\mathrm{Tr}(U^\dagger_{\rm gate}(t) U(t) ) |^2\rangle$, which for our composite sequences:
\begin{equation} \label{eqn:infidelity}
    \mathcal{F}_{av} \approx \frac{1}{4} \left\langle \big|\mathrm{Tr}\left(e^{-i \, \Phi_Q} \right) \big|^2 \right\rangle \,.
\end{equation}
where the angle brackets denote an average over some semi-classical (wide-sense) stationary noise process defining the error Hamiltonian, $H_e$, and the expression is exact in the limit of infinitesimally small gates in the sequences (see \cref{eqn:toggleframe}).

\subsection{Error Vector}
For the case of a single qubit, $\mathcal{F}_{av}$ can be expressed in terms of the so-called ``error vector", $\vec{a}(t)$~\cite{2012GreenHighOrderFilterLogic}. The total error is related to the error vector via 
\begin{equation}
    \Phi_Q= \vec{a}(t)\cdot \vec{\sigma} = \sum_i \vec{a}_i(t)\cdot \vec{\sigma},
\end{equation}
where $\vec{a}_i(t)$ is related to the $i$th term in the Magnus expansion and $\vec{\sigma}=(\sigma_x, \sigma_y, \sigma_z)$. When the noise is sufficiently weak or the time scale of the control is small, $\vec{a}(t)\approx \vec{a}_1(t)$~\cite{2012GreenHighOrderFilterLogic,2013GreenArbControlUnivNoise,2014SoareExpNoiseFilter}. In this case, the dynamics are well-approximated by the first-order Magnus term, or equivalently, the first-order error vector with components
\begin{equation}
    a_{1,\mu}(t)
    =\frac{1}{2}\tr\left[\sigma_\mu \Phi^{[1]}_Q\right].
\end{equation}
The resulting fidelity can be shown to be related to $\vec{a}_1(t)$ through
\begin{equation}
\mathcal{F}_{av} \approx 1 - \braket{| \vec{a}_1 |^2}.
\label{eqn:FidErrorVectorExpansion}
\end{equation}
In the following sections, we will focus our attention on the first-order error vector and thus, the weak-noise limit. We compute the infidelity $1-\mathcal{F}_{av}$ for a quantum logic gate for pure control noise and control noise dominant systems. Through this analysis, we discuss the implications of these expressions as they pertain to an optimization of the control. Expressions for higher order terms in the Magnus expansion are discussed in \cref{sec:magnusexpansion} and an analysis of these high-order contributions to the Magnus expansion for the examples in this work are provided in \cref{sec:highorderterms}.

%% file: Sections/3-ControlNoiseExpressions.tex
Our goal now is to adapt the approach outlined in \cref{sec:DCGderive} to our control noise dominant systems with weak dephasing contribution.
There are additional details to the following derivation available in \cref{sec:firstorderdephasing}.
We will start with the full model, noting that the optimal pure control noise analysis will be a component of the full model analysis.
The first step is to calculate the EPG for each control pulse in the composite sequence.
We will start with single-axis control under the influence of dephasing
\begin{equation}
    \label{eqn:controlnoiseham}
    H = \frac{J(t)}{2} \sigma_z+\left(1+\epsilon(t) \right)\frac{\Omega(t)}{2} \sigma_x
\end{equation}
where the control is along $\sigma_x$, $\Omega(t)$ is the control drive, and $J(t)$ and $\epsilon(t)$ are semi-classical, wide-sense stationary noise processes defining the dephasing and control noise, respectively.
Recall that wide-sense stationary noise processes are stationary noise processes with a constant mean, autocorrelation that dependents strictly on lags, and with finite second moments.
Note that this is a case in \cref{sec:DCGderive} where $H_{\rm gate}(t) = H_{\rm ctrl}(t) = \Omega(t) \, \sigma_x/2$ and $H_e = H_{S,e} = J(t) \sigma_z/2 + \epsilon(t) \Omega(t) \sigma_x/2$.

\subsection{EPG and Error Vector Contributions}
We can now use this expression to compute the EPG.
To do this, we need to move to the interaction picture with respect to the ideal control, $H_{\rm gate}(t)$, giving
\begin{equation}
    \begin{gathered}
     \tilde{H}_{e}(t) =   \frac{J(t)}{2}\left(\cos(\Omega(t))\sigma_z + \sin(\Omega(t))\sigma_y \right)\\
     + \epsilon(t) \frac{\Omega(t)}{2} \sigma_x
    \end{gathered}
\end{equation}
For this work, we will restrict ourselves to piece-wise constant control for each gate, $Q_j$, where every gate has a uniform gate time, $\Delta t$, such that $\theta_j = \int_{t_j - \Delta t}^{t_j} ds \, \Omega(s)$ resulting in $Q_j = \exp(-i \, \theta_j \, \sigma_x/2)$.
With the new definition of the control, we can write the error per gate of gate $Q_j$ as
\begin{equation}
     \Phi_{Q_j} =   \frac{J_j}{2}\left(\cos(\theta_j)\sigma_z + \sin(\theta_j)\sigma_y \right)
     + \epsilon_j \frac{\theta_j}{2} \sigma_x
     \label{eqn:dephasingEPG0}
\end{equation}
where $\beta_j = \int_{t_j-\Delta t}^{t_j} ds \, \beta(s)$ for the semi-classical noise processes, $\beta \in \left[J,\, \epsilon \right]$, defining the dephasing and control noise.
Note, much like equation \cref{eqn:EPGcomposite}, the integration is implicit to the expressions for $J_j$, $\epsilon_j$, and $\theta_j$ and the limits of integration for each expression are defined by the gate index $j$.
We will further distinguish between the coherent and time-correlated components of the dephasing noise, $J_j$, by defining the mean $\mu_J = \sum_{j=1}^N J_j/(N\Delta t)$ and $\tilde{J}_j = J_j - \mu_J$ resulting in a zero-mean process for $\tilde{J}$.
This gives an equivalent form for the EPG of
\begin{equation}
    \label{eqn:dephasingEPG}
    \begin{split}
    \Phi_{Q_j} & =  \frac{\mu_J}{2} \left(\cos(\theta_j)\sigma_z + \sin(\theta_j)\sigma_y \right)\\
    & + \frac{\tilde{J}_j}{2}\left(\cos(\theta_j)\sigma_z + \sin(\theta_j)\sigma_y \right) \\
    & + \epsilon_j \frac{\theta_j}{2} \sigma_x
    \end{split}
\end{equation}
which will be used in further analysis.

Now that we have an expression for the EPG, our goal is to compute the first order term in the Magnus expansion of the noisy dynamics from \cref{eqn:firstorderMagnus} and convert that expression into an error vector. 
To do this, we need shift each error per gate at each time step $j$ to an interaction frame with respect to the ideal control leading up to time step $j$ (see \cref{eqn:firstorderMagnus}).
Specifically, we will evaluate the expression $P_{j-1}^\dagger \Phi_{Q_j} P_{j-1}$ where $P_j = Q_{j}Q_{j-1}...Q_2 Q_1 = \exp(-i \, \sum_{i=1}^j \theta_j \sigma_x/2)$ and $P_0=I$ for our control noise example.
Applying the transformation to results in the expression:
\begin{equation}\label{eq:magnustermsfull}
    \begin{split}
    P_{j-1}^\dagger \Phi_{Q_j} P_{j-1} = & \frac{\mu_J}{2} \left[
     \cos(\Theta_{j,1}) \, \sigma_z + \sin(\Theta_{j,1}) \, \sigma_y  \right]\\
     & + \frac{\tilde{J}_j}{2} \left[
     \cos(\Theta_{j,1}) \, \sigma_z + \sin(\Theta_{j,1}) \sigma_y  \right]\\
     & +\frac{\epsilon_j \Omega_j}{2} \, \sigma_x
    \end{split}
\end{equation}
where $\Theta_{b,a} = \sum_{k=a}^{b} \theta_k$ is the total accumulated ideal control from time step $a$ to $b$.

We can now represent the first-order term in the Magnus expansion as a sum of each error per gate in the sequence in its appropriate ideal control frame using \cref{eqn:firstorderMagnus}, $\Phi_{Q}^{[1]}= \sum_j^N P_{j-1}^\dagger \Phi_{Q_j} P_{j-1}$, and partition this expression into the $\sigma_x$, $\sigma_y$, and $\sigma_z$ components of the error to construct the error vector
\begin{equation}
    \label{eqn:errorvector}
    \vec{a}_1 = \left( \begin{array}{c} 
    \sum_j^N \epsilon_j \frac{\theta_j}{2}\\
    \sum_{j=1}^N\left(\frac{\mu_J}{2} + \frac{\tilde{J}_j}{2} \right) \sin(\Theta_{j,1})\\
    \sum_{j=1}^N\left(\frac{\mu_J}{2} + \frac{\tilde{J}_j}{2} \right) \cos(\Theta_{j,1})
    \end{array} \right).
\end{equation}
This expression can be used to evaluate the first-order term in the gate fidelity represented as a power series expansion in \cref{eqn:FidErrorVectorExpansion}, $\left\langle | \vec{a}_1 |^2 \right\rangle$.
Evaluating the first order term gives:
\begin{subequations}
    \label{eqn:DephasingErrorFull}
    \begin{equation}
    \label{eqn:DephasingErrorSum}
    \langle |\vec{a}_1|^2 \rangle = \frac{1}{4} \left[A + B + C\right]
    \end{equation}
    \begin{equation}
    \label{eqn:DephasingErrorCoherent}
    A = \mu_J^2 \left(N + \sum_{h=1}^N \sum_{i=h+1}^N \cos(\Theta_{i-h,i}) \right)
    \end{equation}
    \begin{equation}
    \label{eqn:DephasingErrorCorrelated}
    B = N \gamma_{\tilde{J}}(0)
    +2 \sum_{h=1}^N \gamma_{\tilde{J}}(h) \sum_{i = h+1}^N \cos (\Theta_{i-h,i})
    \end{equation}
    \begin{equation}
    \label{eqn:DephasingErrorControl}
    C = \gamma_{\epsilon}(0)\sum_{j=1}^N \theta_j^2 + 2 \sum_{h=1}^{N-1} \gamma_{\epsilon}(h) \sum_{i=h+1}^{N} \theta_j \theta_{j-h}
    \end{equation}
\end{subequations}
where $A$, $B$, and $C$ represent the coherent component of the dephasing noise, correlated dephasing noise component, and correlated control noise component, respectively. $\gamma_\beta(h) = \langle \beta_j \beta_{j-h} \rangle$ defines the autocovariance of the (wide-sense) stationary noise processes driving the dephasing, $\beta = \tilde{J}$, and control noise, $\beta = \epsilon$, error Hamiltonians.
All that is left to evaluate the above expression is to represent the statistical features of the noise, $\left\{\gamma(h)\right\}$, for weakly (wide-sense) stationary signals.  To do so, we will use a model-based approach developed from classical time-series analysis known as ARMA (autoregressive moving-average) modeling.

\subsection{\label{sec:ARMAmodels}ARMA Model Representation of Stationary Noise}

To compactly represent the statistical features of the noise, we will utilize Autoregressive Moving Average models from time series analysis \cite{1964FarlieARMA, 1976BoxARMA}.  ARMA(p,q) models are defined as:
\begin{equation}
    \beta_t = \sum_{i=1}^p a_i \beta_{t-i} + \sum_{j=0}^q b_j w_{t-j}
\label{eqn:ARMAgeneral}
\end{equation}
where $\beta_t$ is the signal at timestep $t$ (which will define the noise in our case), $\left\{a_i \right\}$ define the autoregressive portion of the model, $\left\{b_j \right\}$ define the moving average component of the model, and $w_i$ is a white-noise signal that defines the power of the noise.  ARMA models are general models for weakly (wide-sense) stationary Gaussian processes; i.e., given a wide-sense stationary signal, there exists a set of $\left\{a_i \right\}, \left\{b_j \right\}$, and $w_i$ that generates the same statistical features (up to arbitrary accuracy) of the true signal.  
In addition to the signal, the statistical features of the noise can be defined according to the ARMA formalism. We emphasize that the autocovariance $\gamma(h)$ of the noise is defined with respect to the ARMA model parameters and the statistical features of the white noise sequence.
Understanding this general applicability of an ARMA model representation, we will move forward in our discussion with respect to the noise strictly at the ARMA model level.

%% file: Sections/4-ControlNoiseAnalysis.tex
We will start will an analysis of the pure control noise Hamiltonian:
\begin{equation}
    \label{eqn:controlham}
    H(t) = \left(1+\epsilon(t) \right)\frac{\Omega(t)}{2} \sigma_x
\end{equation}
and show that the control solutions for this family of Hamiltonians are provably optimal in the sense that we can find solutions that globally minimize $\langle |\vec{a}_1|^2 \rangle$.
Furthermore due to the commutivity of the control with the noise, the minimization of $\langle |\vec{a}_1|^2 \rangle$ also minimizes all higher order terms in the power series expansion of the error vector in the expression of the infidelity (refer to \cref{eqn:controlhigherorder} for details).
Applying the treatment from the above sections to the pure control noise Hamiltonian of the form of \cref{eqn:DephasingErrorFull}, only the $C$ term (\cref{eqn:DephasingErrorControl}) is nonzero, i.e.
\begin{equation}
    \label{eqn:controlnoiseinfid}
    \langle |\vec{a}_1|^2 \rangle = \gamma_{\epsilon}(0)\sum_{j=1}^N \theta_j^2 + 2 \sum_{h=1}^{N-1} \gamma_{\epsilon}(h) \sum_{j=h+1}^{N} \theta_j \theta_{j-h} \, .
\end{equation}
This will be the objective function under analysis in this section.

\subsection{Optimality of the Control Sequences}\label{sec:controloptimality}

The main advantage of this approach is the ability to prove the optimality of a control sequence given an ARMA defined representation of stationary signals.  What we mean by optimal here is a control sequence that minimizes the first-order term in the expansion of the infidelity (\cref{eqn:controlnoiseinfid}) for the case of multiplicative control noise where the control noise has a stationary noise spectrum.  
To prove optimality, we need to show that equations of the form of \cref{eqn:controlnoiseinfid} have a global minimum.  Furthermore, we need an approach to finding such a minimum using a reasonable amount of time and resources for this result to be of practical interest.  
For the expression of the infidelity in \cref{eqn:controlnoiseinfid}, we can cover both requirements by showing that this function is convex.

To understand why this function is convex, we first recast the infidelity (\cref{eqn:controlnoiseinfid}) in a quadratic form:
\begin{equation}
    \label{eqn:controlnoisequad}
    \big\langle \left|a_1 \right|^2 \big\rangle = \vec{x}^T A \vec{x}
\end{equation}
where $\vec{x}^T = \left[\theta_1 \; \theta_2 \; ... \; \theta_N \right]$ is a vector of the control pulses and $A$ is the covariance matrix of the control noise $A_{i,j} = \sum_{i,j}^N \langle \epsilon_i, \epsilon_j \rangle = \sum_{i,j}^N \gamma(|i-j|)$.  Note that, for the case of pure control noise, the following objective function has a defined minimum when $\vec{x} = \vec{0}$; i.e., when there is no control given that this model is absent of other noise sources.  
To make this model of interest for the application to quantum logic gates, we require an additional constraint that the total accumulated integrated amplitude of the applied control is equivalent to the desired gate angle of the logic gate.  
Therefore given a desired logic gate $\exp(-i\,\theta_Q \sigma_x/2)$, the optimization requires a constraint on the form $\vec{E}^T\vec{x} = \theta_Q$ where $\vec{E}^T = [1\;1\;...\;1]$. 
This problem can therefore be cast as a quadratic program of the form
\begin{equation}
\begin{aligned}
\textrm{Minimize}\;\;\; \frac{1}{2} \vec{x}^T Q \vec{x}\\
\textrm{Subject to}\;\;\; \vec{E}\vec{x} = \theta_Q\,.
\end{aligned}
\end{equation}
Furthermore, the optimization is convex due to the fact that $Q$ is positive semi-definite by nature; being a covariance matrix.
This convex form provides the aforementioned provable optimality of the solutions.

\begin{figure}
    \centering
    \includegraphics[width=0.44\textwidth]{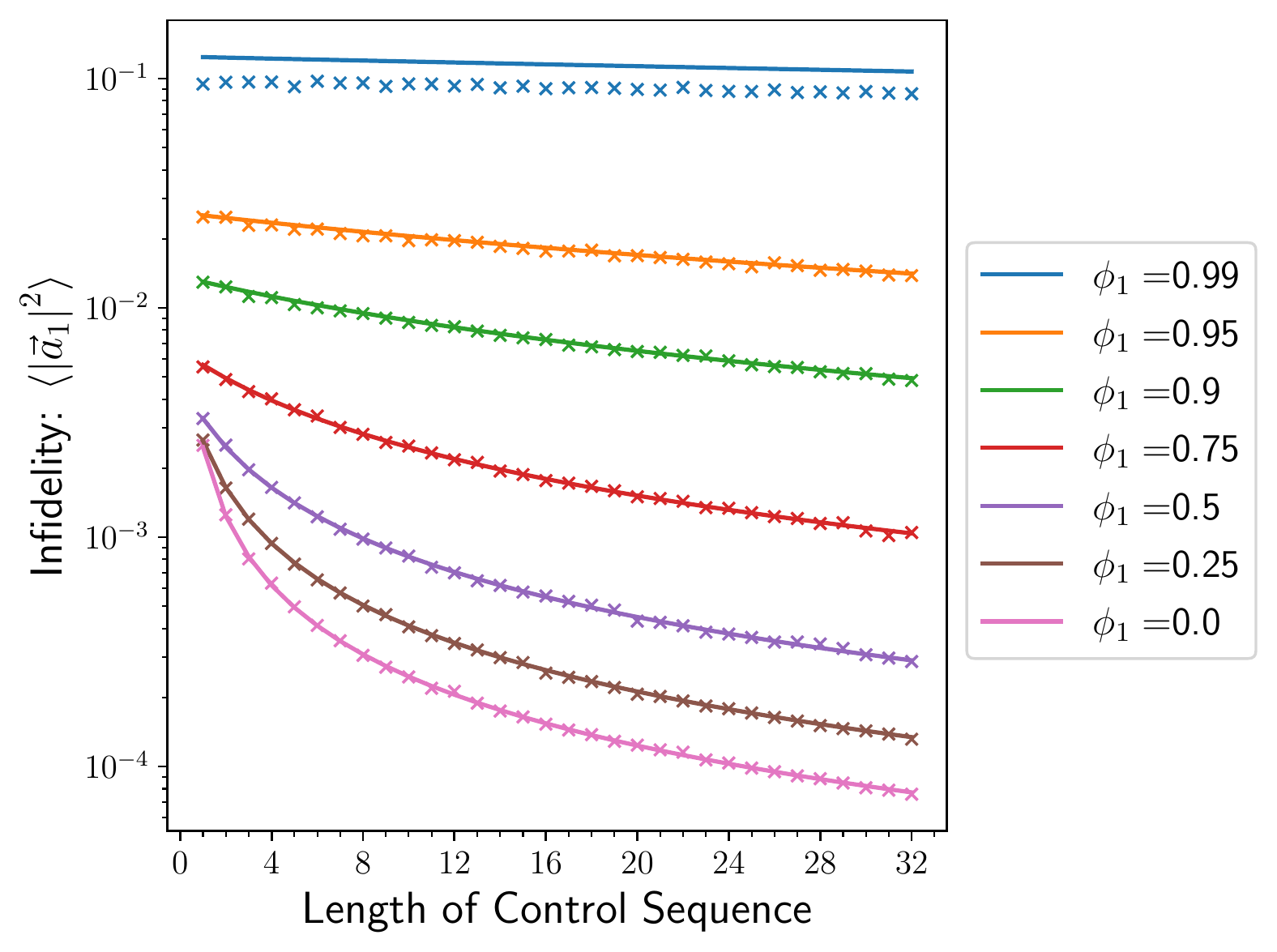}
    
    \caption{Infidelity of the optimal control for $1/f^2$ noise with low-frequency white noise cutoffs (AR(1) models from section \ref{sec:1fexample}) when applying an $R_X(\pi)$ logic gate of length $t$.  The solid lines are the theory (\cref{eqn:controlnoiseinfid}) and each marker is the average fidelity ($10^4$ ARMA trajectories) computed from Monte Carlo simulations of the noisy control (\cref{eqn:controlham}).  For all AR noise trajectories $\sigma_w^2 = 1.0 \times 10^{-3}$.  The Monte Carlo error and higher order contributions are discussed in \cref{sec:higherordercontrolnoise}}
    \label{fig:controlnoiseerror}
\end{figure}

\begin{figure*}
    \centering
    \subfloat[Power spectrum of $1/f^2$ noise with white noise cutoffs.]{\includegraphics[width=0.49\textwidth]{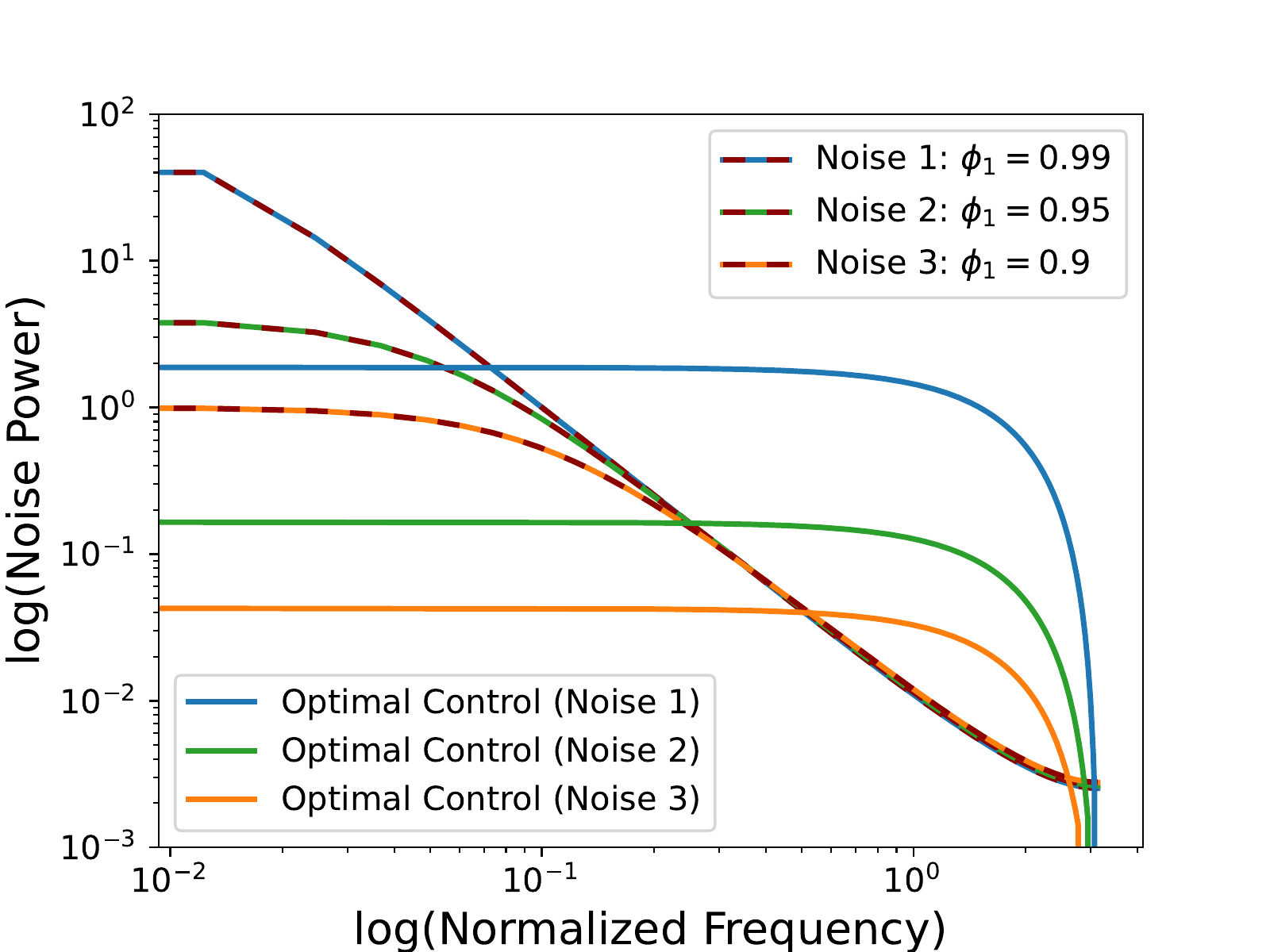}\label{fig:1fpowerspectrum}}
    \subfloat[Optimal control for $1/f^2$ noise with white noise cutoffs.]{\includegraphics[width=0.45\textwidth]{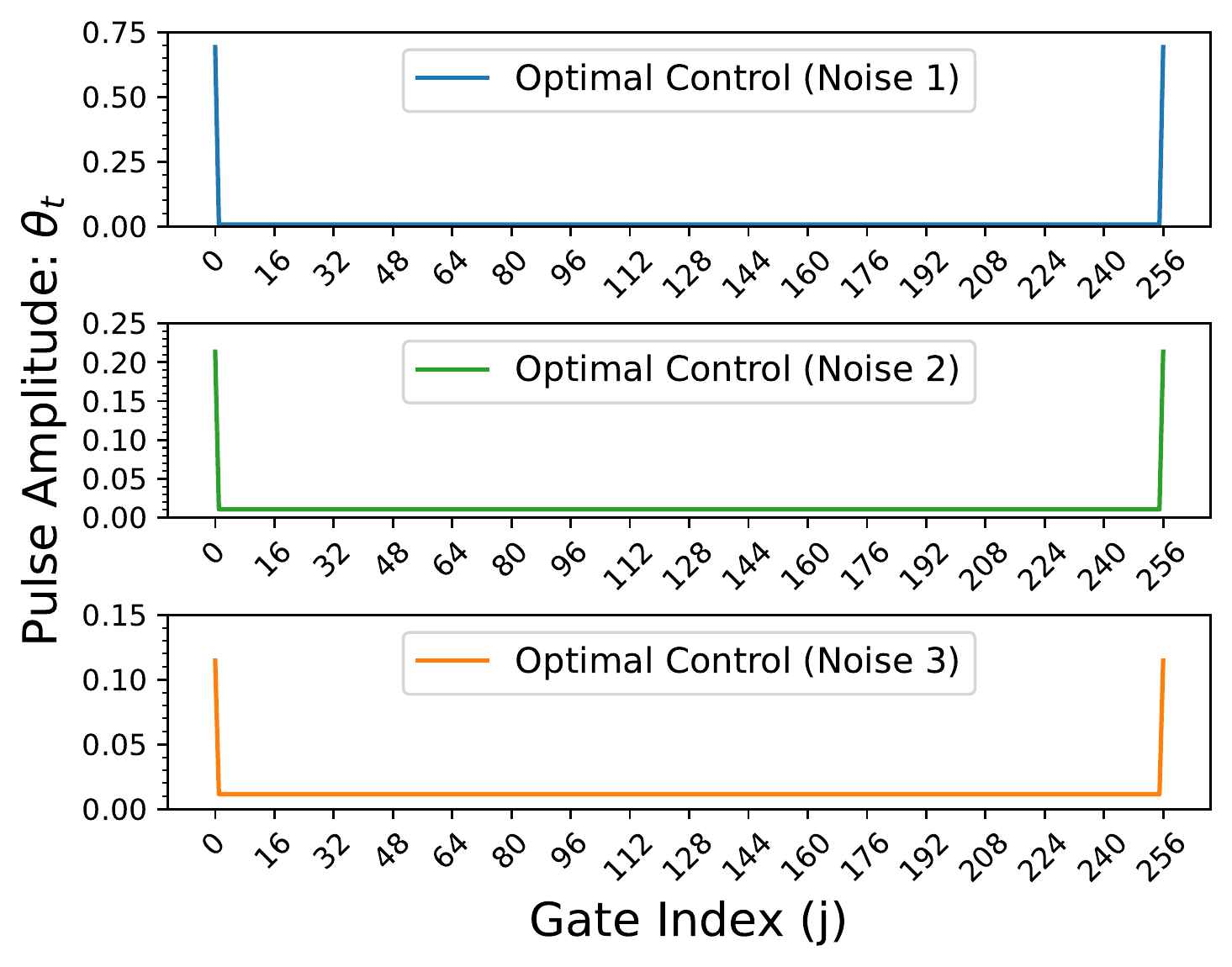}\label{fig:1fcontrolnoisesolutions}}

    \caption{Optimal control sequences for ARMA model defined stationary noise processes from \cref{sec:1fexample}.  For these examples, the power of the noise ($\sigma_w^2$) was scaled for visualization purposes.
    \cref{fig:1fpowerspectrum}: The power spectrum of the noise and optimal control for three AR(1) models representing $1/f^2$ noise spectra with various low-frequency white noise cutoffs.  The DC component of the control is not visible, but accounts for $8\%$, $48\%$, and $79\%$ of the total power of the control for noise 1, noise 2, and noise 3, respectively (as the spectrum has more white noise features).
    \cref{fig:1fcontrolnoisesolutions}: The time domain control noise solutions for optimally mitigating the various $1/f^2$ noise spectra.  As the white noise component of the noise increases, the control noise solutions have an increasingly more prevalent DC component.}
    \label{fig:controlnoisesolutionsmain}
\end{figure*}

The choice of remaining in the weak noise limit for our analysis has implications on the optimization as well.
Particularly, the expression for the infidelity for pure control noise in \cref{eqn:controlnoiseinfid} holds for the case where the total accumulated error in the gate angle is bounded; specifically:
\begin{equation}
    \left|\sum_{j =1}^N \epsilon_j \theta_j \right| < \pi.
\end{equation}
Without this restriction, the terms in the error vector in \cref{eqn:errorvector} would be modulo $\pi$; breaking the convexity condition.
In addition, we will also restrict our solutions to minimum gate angle solutions, i.e. $\sum_{j=1}^N \theta_j = \theta_Q$ in contrast to $(\sum_{j=1}^N \theta_j)\, \textrm{mod} \, 2\pi = \theta_Q$, for similar reasons.
However, we certainly have the ability to compare the performance of different solutions for different $2k\pi$ ($k \in \mathbb{Z}$) multiples of $\theta_Q$ to explore better potential solutions.
We can represent the bound on the total accumulated error angle with respect to the statistical features of the ARMA model representation of the noise:
\begin{equation}
    \max \left(\left|\langle \epsilon \rangle - \frac{Z \sqrt{\sigma_\epsilon^2}}{\sqrt{K}} \right|,\,\left|\langle \epsilon \rangle + \frac{Z \sqrt{\sigma_\epsilon^2}}{\sqrt{K}} \right|\right) \theta_g < \pi
\end{equation}
where $\sigma_\epsilon^2$ is the variance of the noise process, $K$ is the length of the control sequence, and $Z$ is the $z$ value defining the confidence interval of the noisy trajectory.
For instance, zero-mean white control noise requires $3.291 \sqrt{\sigma_w^2}\theta_Q/\sqrt{K}  < \pi$ for $Z = 3.291$ ($99.9\%$ confidence interval) and, for general ARMA model defined noise, $\sigma_\epsilon^2 = \mathcal{O}(\sigma_w^2)$ where $\sigma_w^2$ is the variance of the white noise process defining the power of the noise for the ARMA model.

\subsection{Single-Axis Optimal Control Sequences \label{sec:controlnoiseexamplesmain}}
In this section, we will demonstrate the utility of this approach for constructing provably optimal sequences for the control noise problem.
To do so, we will focus on the example of a family of white noise  and $1/f^2$ spectra with low-frequency white noise cutoffs.
While these examples are illustrative, we will continue to emphasize that this ARMA model formalism extends to all stationary noise processes and is therefore applicable to many more flavors of control noise than the example shown here.  
For instance, we have additional examples of bandlimited control noise in \cref{sec:controlnoiseexamplesappendix}.

For these examples we will also be analyzing the optimal control sequences from a filter design perspective.
Filter function approaches to quantum control aim to use control to tailor the systems frequency response relative to the power spectrum of the noise so that the two actions minimally overlap in the frequency domain.
To do this, we will be constructing the power spectrum of the ARMA model defined noise with power spectrum given by:
\begin{equation}
    \label{eqn:ARMApowerspectrum}
    S_\epsilon(\omega) = S_w(\omega) \frac{\left| \sum_{j=0}^q b_j \, \exp(-i j \omega) \right|^2}{\left| 1 + \sum_{k=1}^p a_k \, \exp(-i k \omega) \right|^2}
\end{equation}
where $S_w(\omega)$ is the power spectrum of the white noise process used to construct the ARMA model; i.e. $S_w(\omega)$ is the variance $\sigma_w^2$ of the white noise process.
In addition to the spectrum of the noise, we will also analyze the power spectrum of the control, the filter function $g(\omega,t)$, to gain intuition about the system's frequency response induced by the control with the understanding that the overlap of these two functions, up to some perturbative expansion of the dynamics, will inform the output state of the computation.

\subsubsection{White Noise} 
The first example chosen was white control noise due to the simplicity of the expression and the intuitive analysis of the optimal solution.  White noise is a Moving Average model of order 0, MA(0), (or ARMA(0,0)) with signal [See \cref{eqn:ARMAgeneral}]
\begin{equation}
    \epsilon_t = x_t
\end{equation}
where $x_t$ is a zero-mean white noise process for our control noise model.  Therefore, the noisy signal only is correlated at 0-lags which, from \cref{eqn:controlnoiseinfid}, results in the following expression for the infidelity:
\begin{equation}
    \label{eqn:whitenoiseinfid}
    \big\langle \left|\vec{a}_1 \right|^2 \big\rangle
    = \frac{\gamma(0)}{4} \sum_{i=1}^N \Omega_i^2 = \frac{\sigma_w^2}{4} \sum_{i=1}^N \Omega_i^2 \,.
\end{equation}
From observation of the equation above, the infidelity is reduced when the amplitude of the control at each time step is minimized under the constraint that the overall sum of the pulse angles equal a desired logic gate (a $\pi$ gate for instance).  Therefore, the optimal strategy is to apply a long, constant (DC) control pulse for $t$ time steps where the amplitude at each step is $\theta_g/t$.  This optimal control sequence has an intuitive frequency domain perspective as well in terms of the power spectrum, with the delta function feature of the DC control power spectrum sharply overlapping with the constant power spectrum of the white noise only at zero frequency.

\subsubsection{\texorpdfstring{$1/f^2$}{1f} with Low-Frequency White Noise Cutoffs \label{sec:1fexample}}
Another, perhaps more physically, relevant set of noisy signals are the family of $1/f^2$-like signals.
It was shown in Ref. \cite{2007PlaszczynskiLongStream1f} how ARMA models could be utilized to generate long streams of various flavors of $1/f^\alpha$ signals. Here, we focus on $1/f^2$ type noise due to it's simplicity in representation.  It was shown that $1/f^2$ signals with low frequency cutoffs can be represented as an autoregressive, AR(1), model (or ARMA(1,0) model):
\begin{equation}
    \epsilon_t = x_t + \phi_1 \, \epsilon_{t-1}
\end{equation}
where $\phi_1$ defines the location of the low-frequency cutoff of the signal with signals becoming $1/f^2$-like as $\phi_1 \rightarrow 1$ (pure $1/f^2$ signals at $\phi_1 = 1$ are non-stationary and therefore outside of the scope of our approach) and having more white noise features as $\phi_1 \rightarrow 0$.

Utilizing the covariance matrix from the various AR(1) processes, optimal control solutions of varying length were obtained. The first-order dynamics of the noisy control were simulated and compared to the theoretical estimates of the infidelity (\cref{eqn:controlnoiseinfid}) for an $R_X(\pi)$ logic gate shown in \cref{fig:controlnoiseerror}.
The largest deviation occurs for the $\phi_1 = 0.99$ AR(1) model which can be attributed to the influence of higher-order terms into the dynamics (see \cref{sec:higherordercontrolnoise}, \cref{fig:1fhigherordermontecarlo}).
Note that this deviation occurs between models because the total power of noise from each of these models in this example are different. More specifically, variance of the white noise does not set the total power of the noise of this sequence; it is filtered through the ARMA-model via \cref{eqn:ARMApowerspectrum}.
However, one can certainly scale the variance of the noise for the various ARMA models to noise of equivalent power as well which we utilize in some of the later sections.
Finally, note as the correlations of the signal are reduced ($\phi_1 \rightarrow 0$) the control noise solutions approach the white noise limit solutions.

\begin{figure*}
    \centering
    \subfloat[Error from the pulse-length 3 single-axis optimal control.]{\includegraphics[width=0.24\textwidth, valign=t]{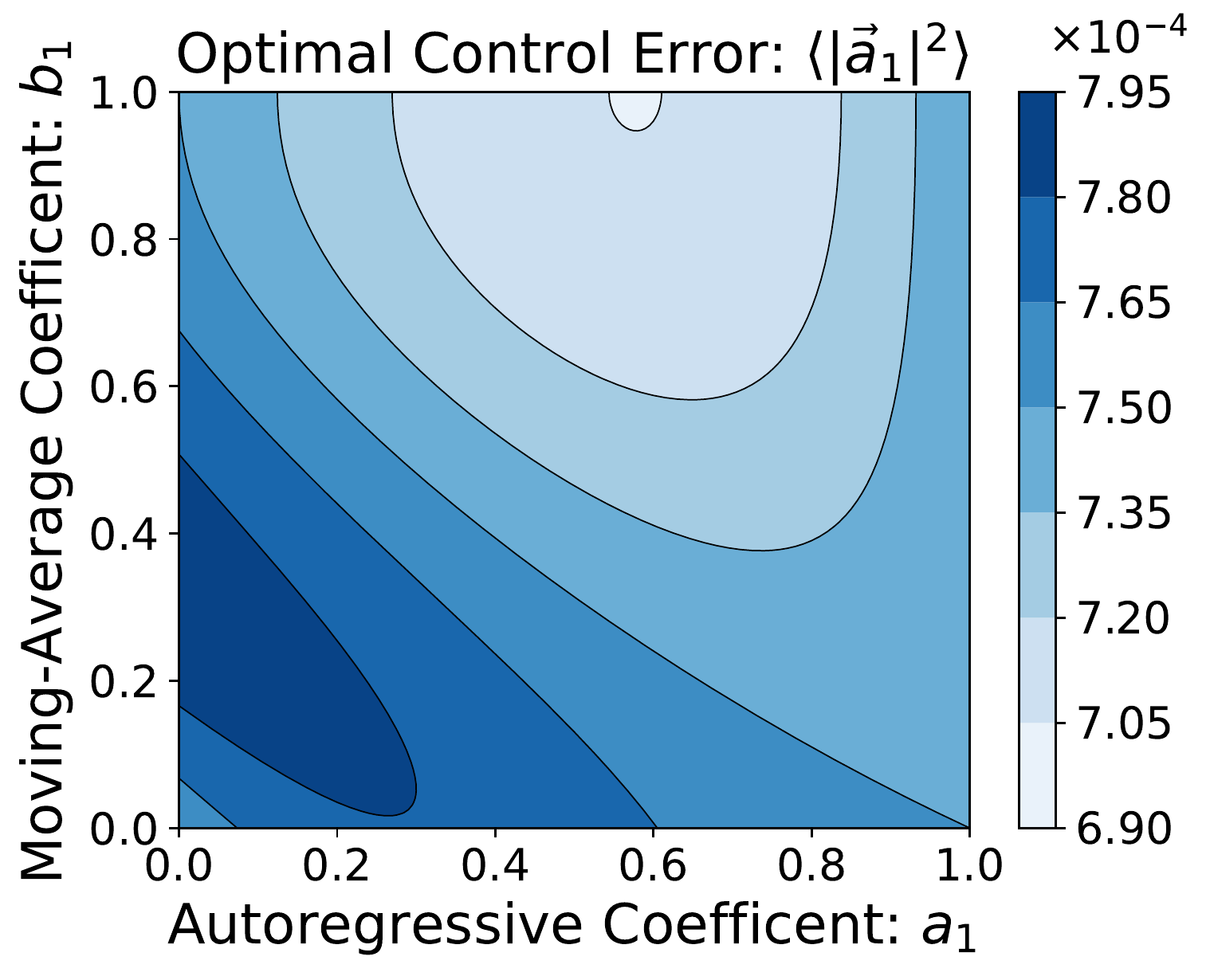}\label{fig:K3ARMA11Error}}
    \subfloat[Error from the SK1 CP sequence.]{\includegraphics[width=0.24\textwidth, valign = t]{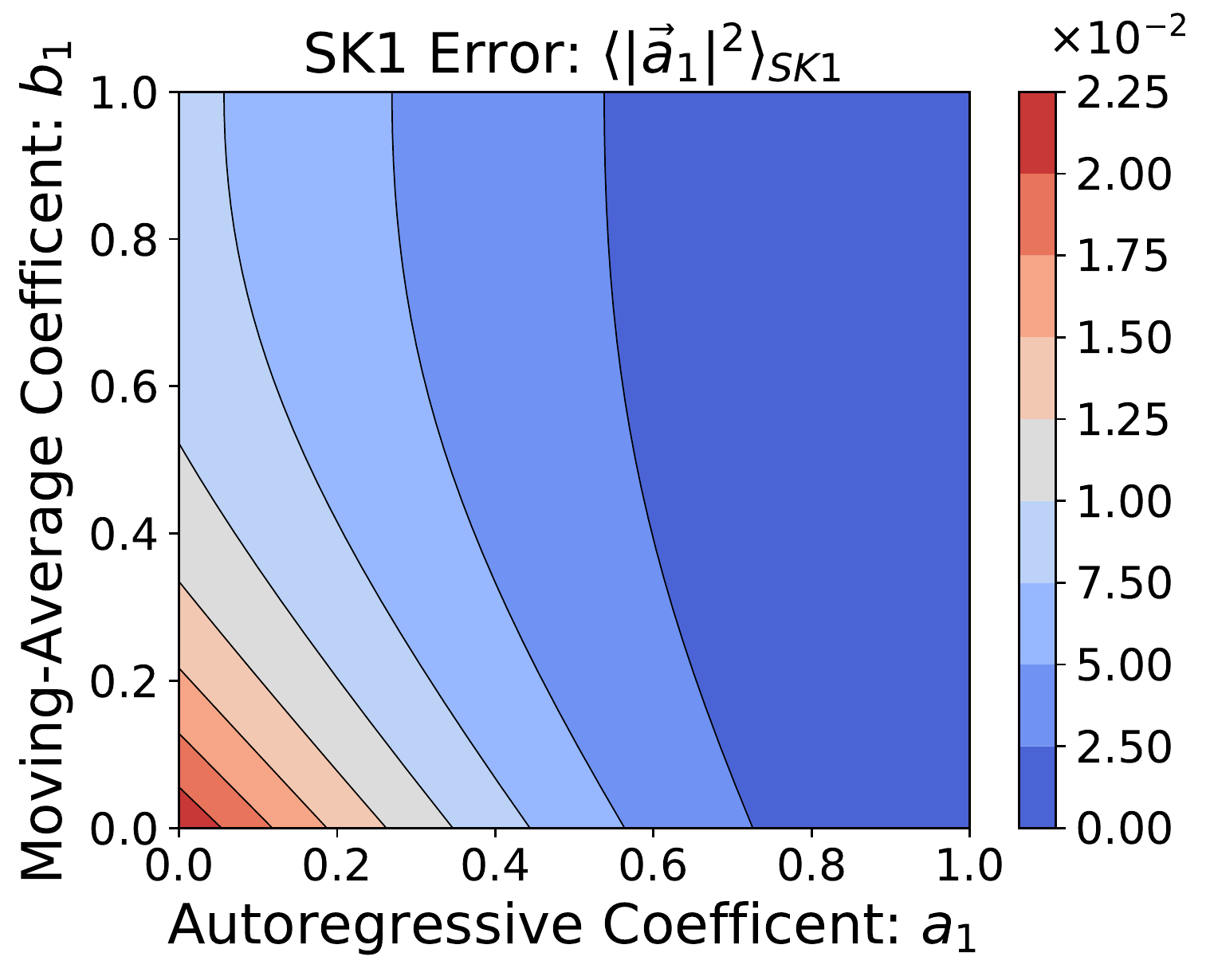}\label{fig:SK1ARMA11Error}}
    \subfloat[Difference between errors from \cref{fig:K3ARMA11Error} and \cref{fig:SK1ARMA11Error}.]{\includegraphics[width=0.24\textwidth, valign=t]{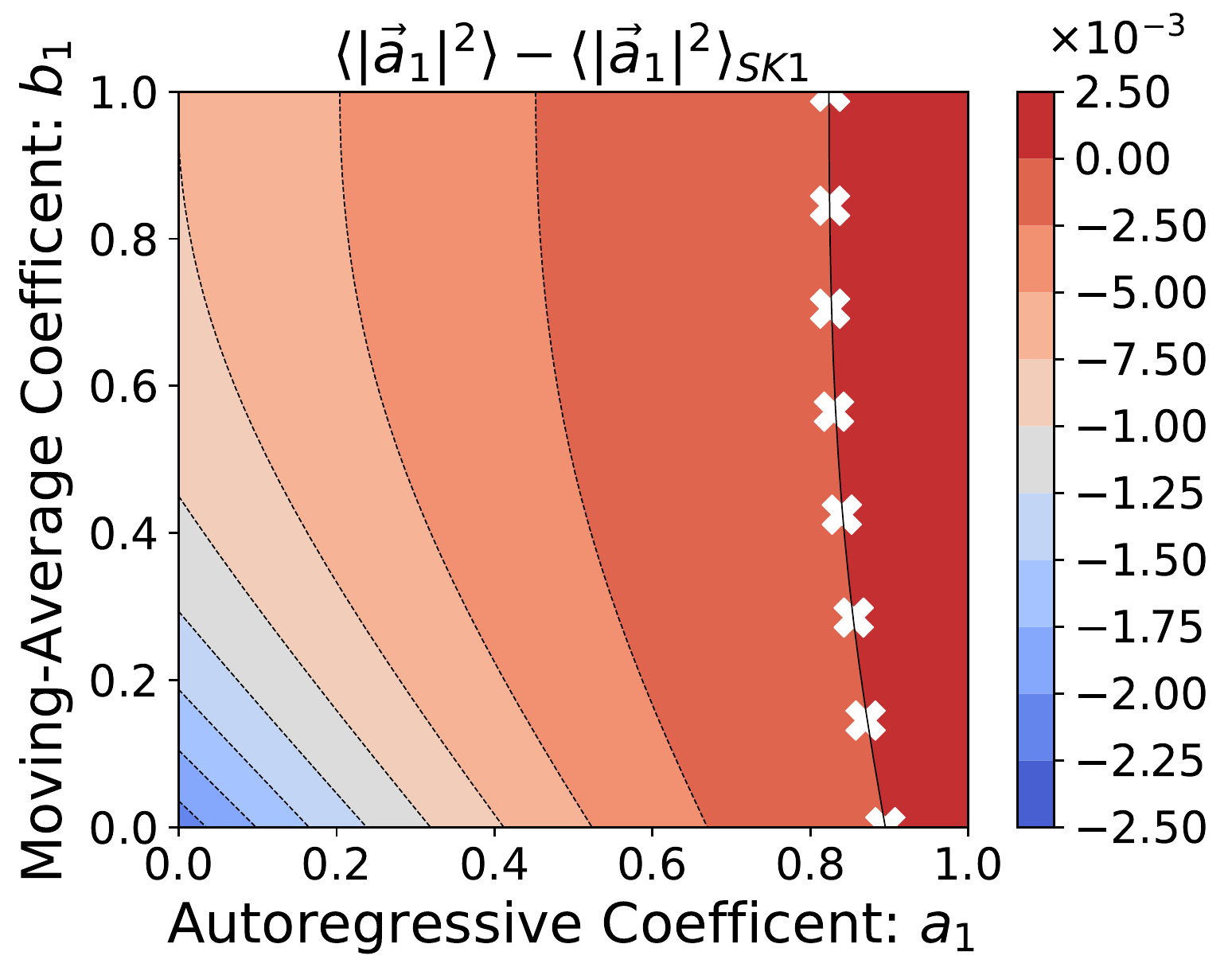}\label{fig:K3CompError}}
    \subfloat[Comparison between ARMA models and DC correlations.]{\includegraphics[width=0.24\textwidth, valign=t]{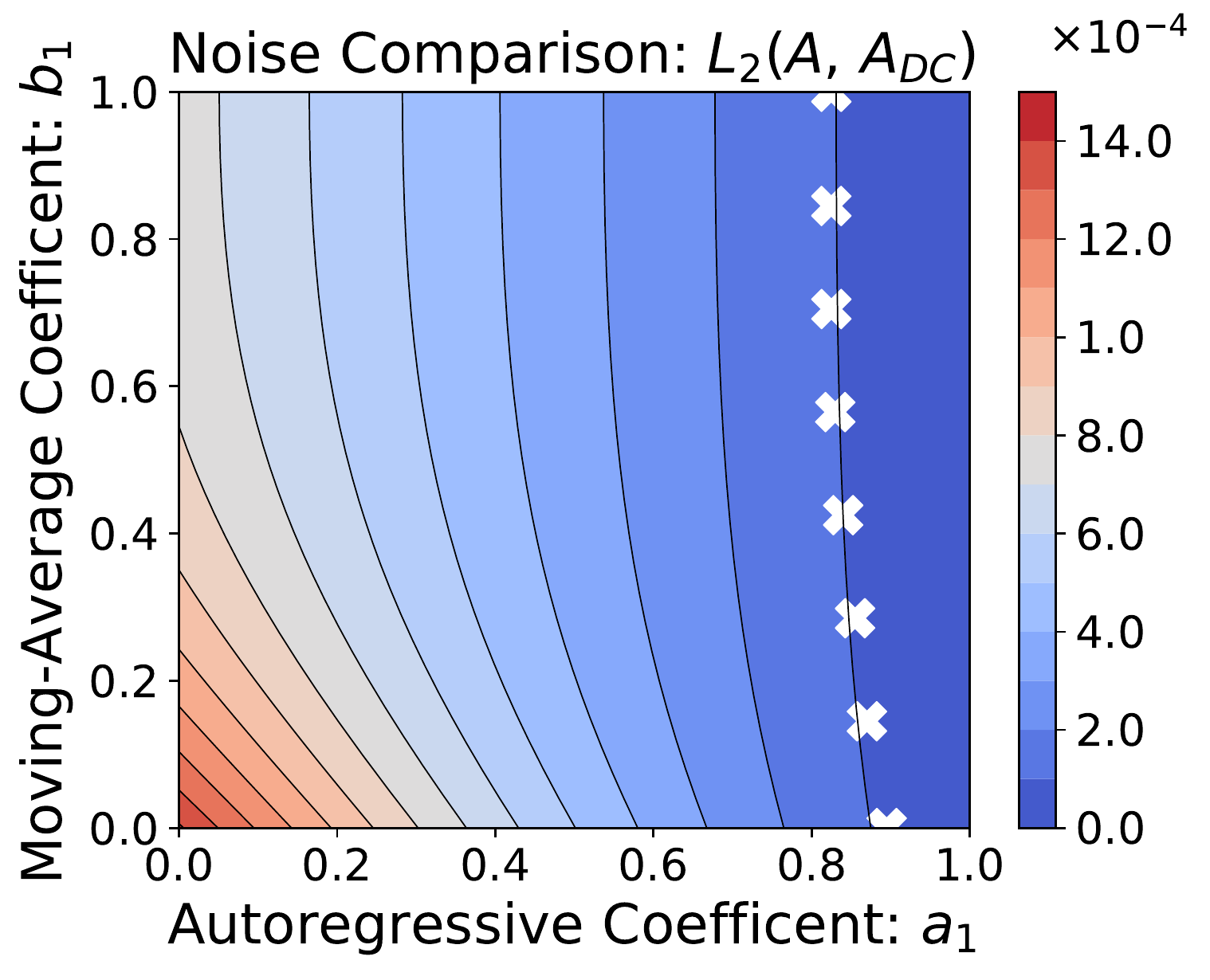}\label{fig:K3CompModels}}
    
    \subfloat[Error from the pulse-length 4 single-axis optimal control.]{\includegraphics[width=0.24\textwidth, valign=t]{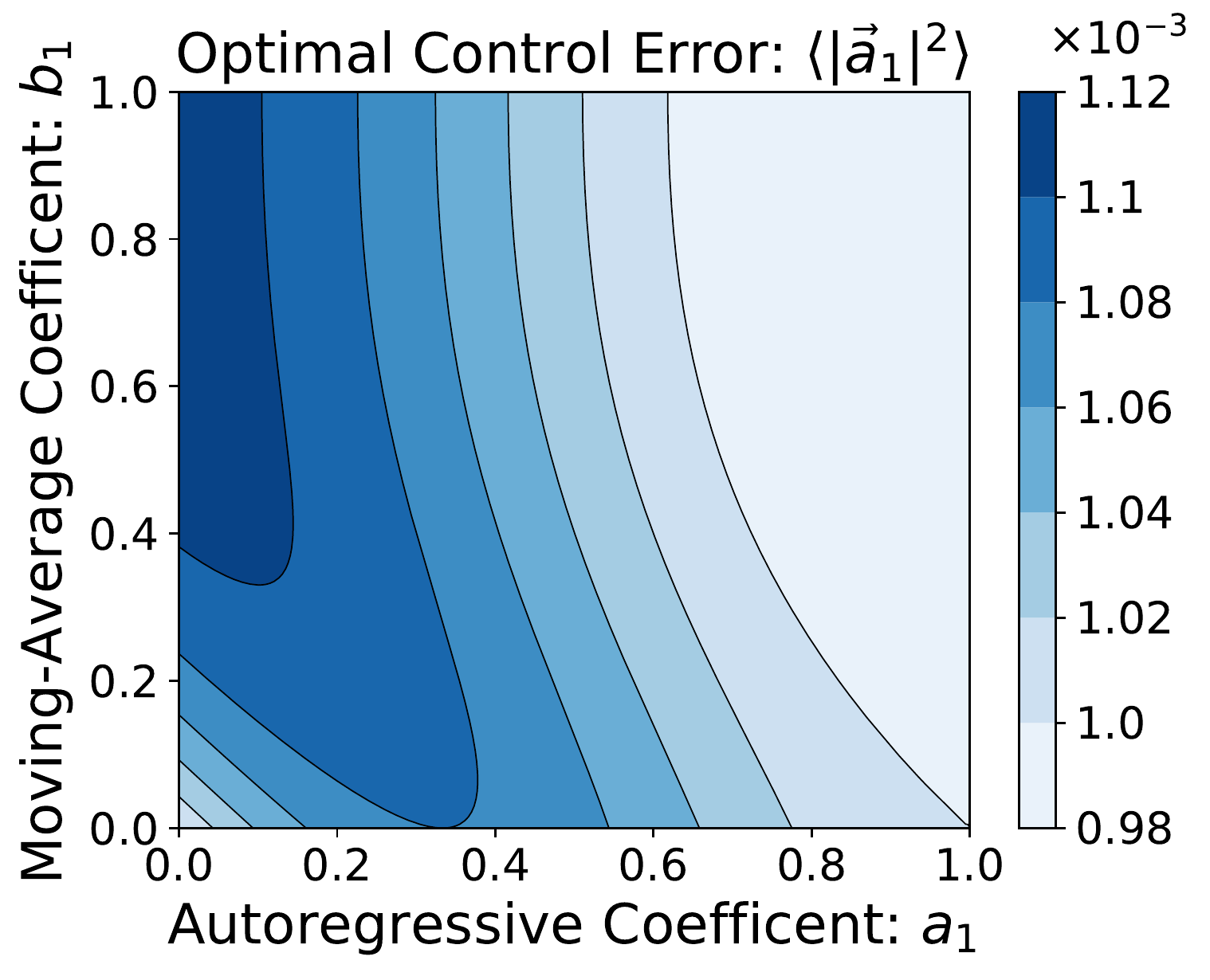}\label{fig:K4ARMA11Error}}
    \subfloat[Error from the BB1 CP sequence.]{\includegraphics[width=0.24\textwidth, valign=t]{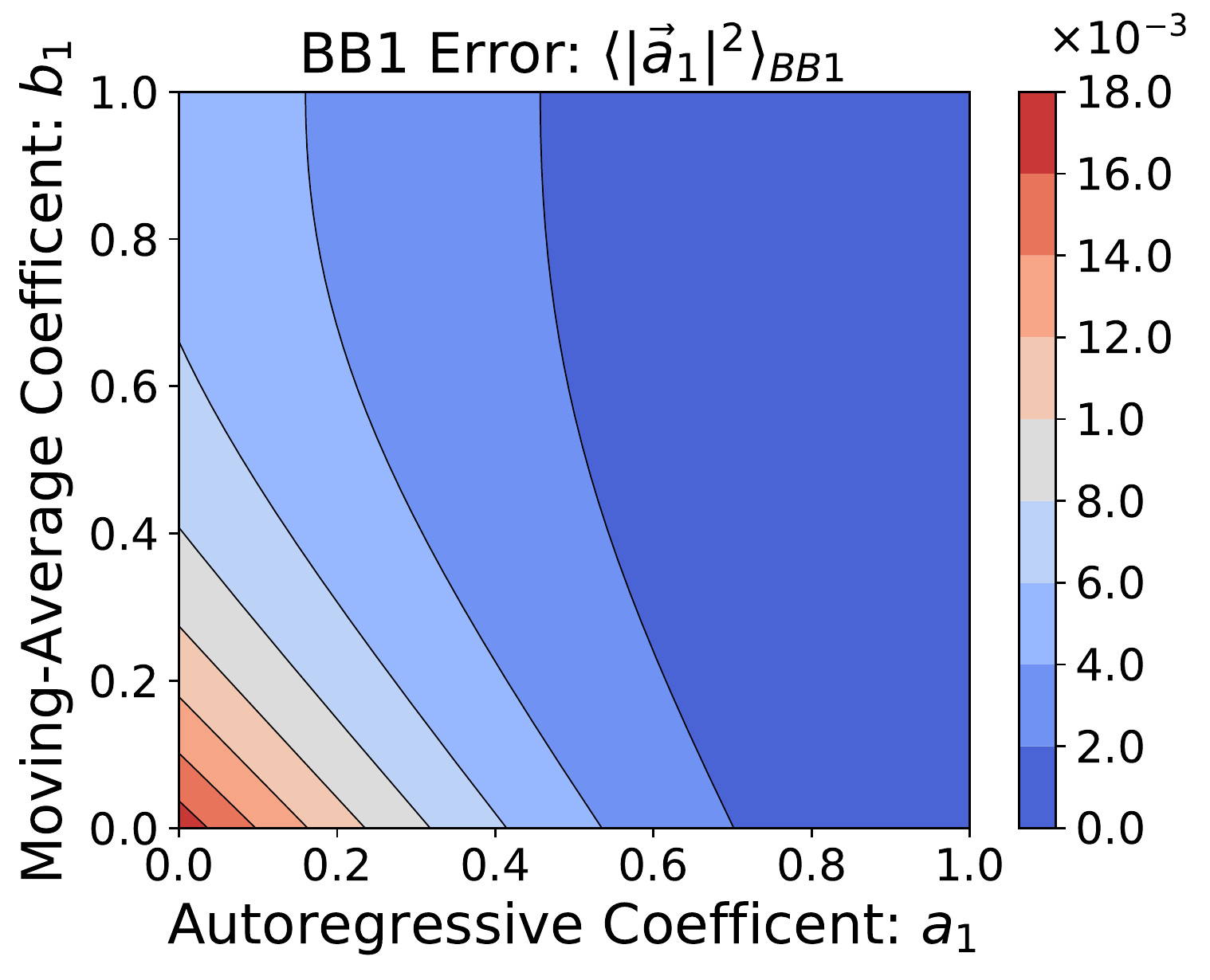}\label{fig:BB1ARMA11Error}}
    \subfloat[Difference between errors from \cref{fig:K4ARMA11Error} and \cref{fig:BB1ARMA11Error}.]{\includegraphics[width=0.24\textwidth, valign=t]{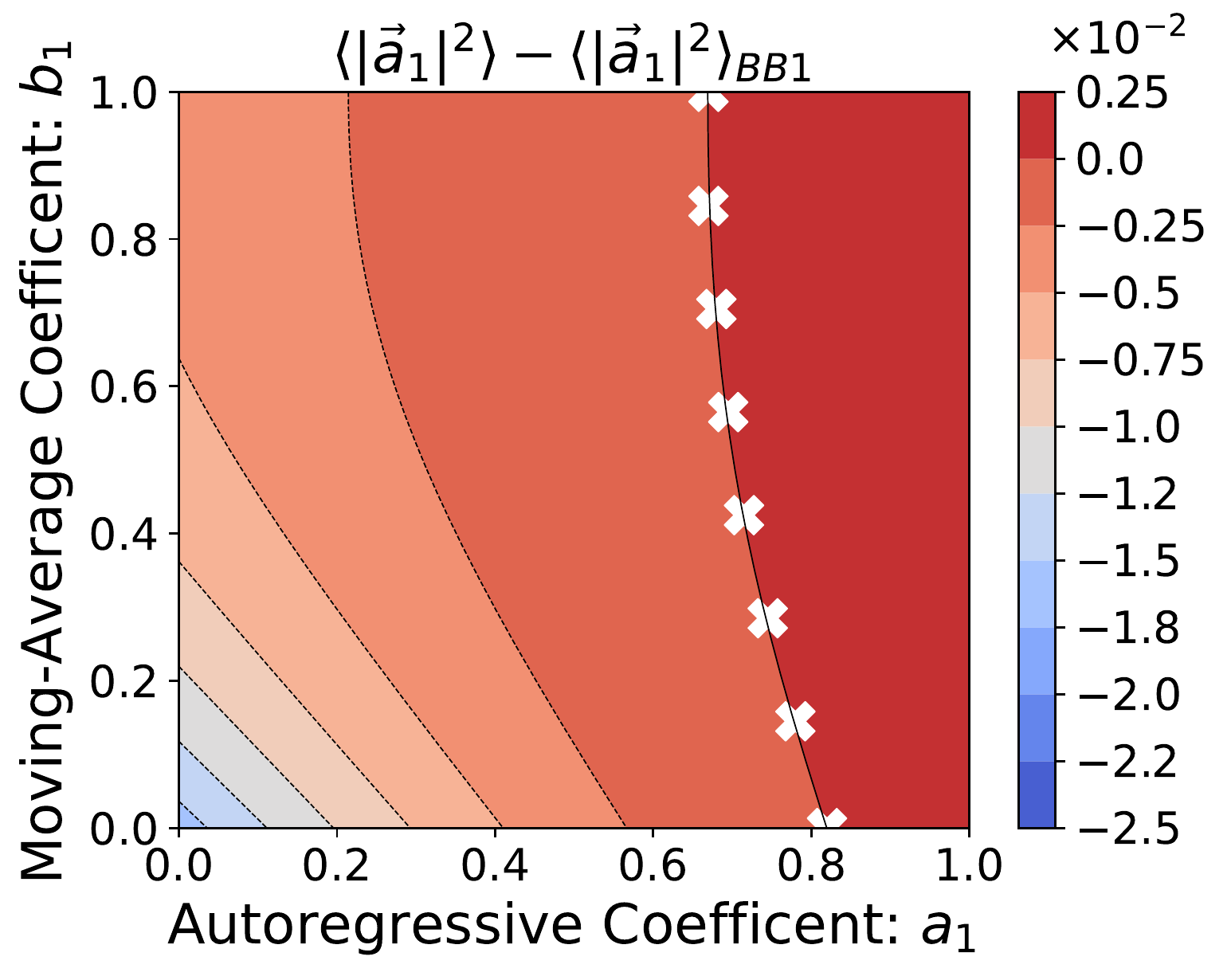}\label{fig:K4CompError}}
    \subfloat[Comparison between ARMA models and DC correlations.]{\includegraphics[width=0.24\textwidth, valign=t]{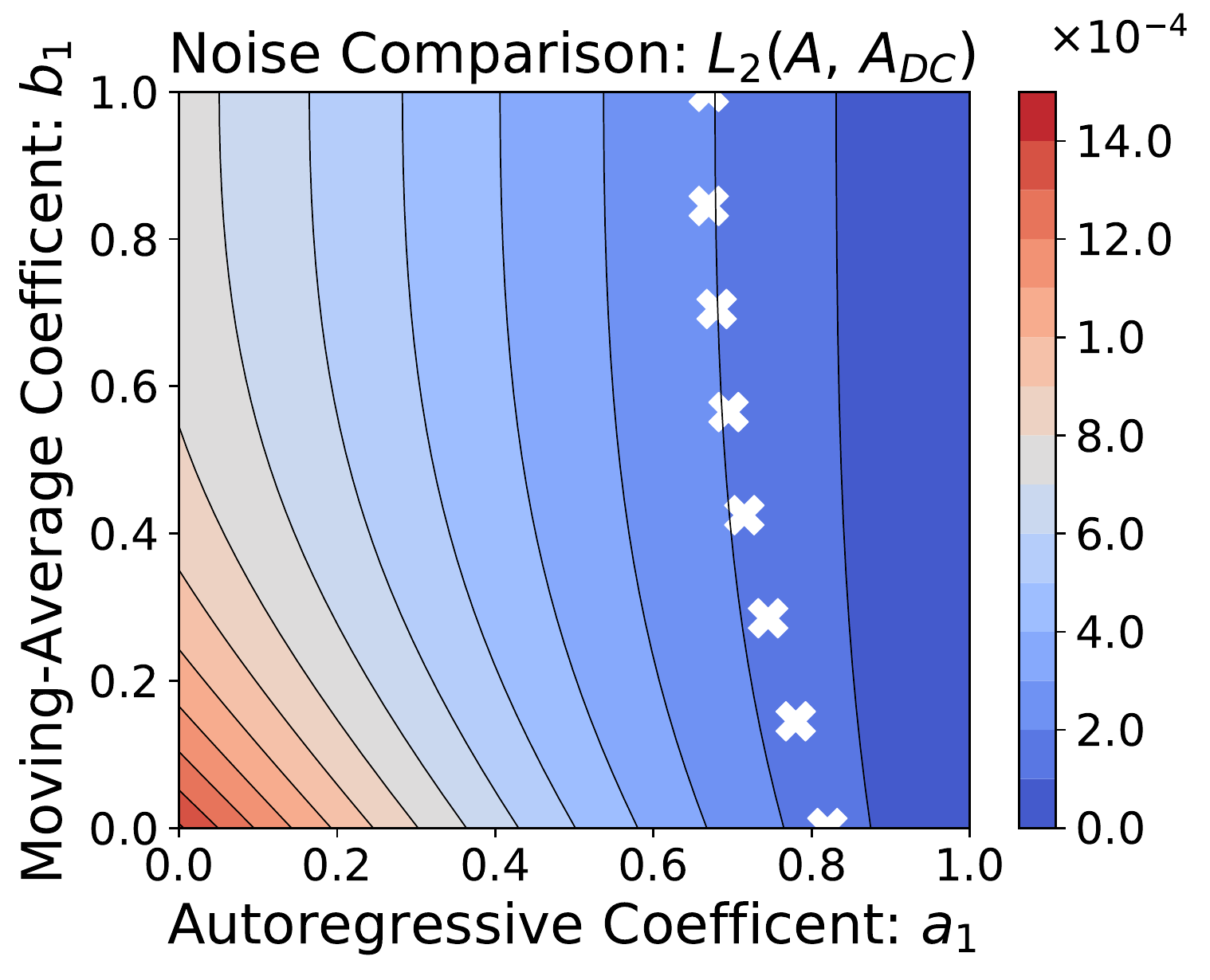}\label{fig:K4CompModels}}
    \caption{Comparison between the analytical infidelity of the composite pulse sequences (\cref{fig:SK1ARMA11Error}, \cref{fig:BB1ARMA11Error}) and equivalent-length single-axis optimal control sequences (\cref{fig:K3ARMA11Error}, \cref{fig:K4ARMA11Error}) over ARMA(1,1) model defined noise.  We can identify noise regimes sufficiently far away from DC where the single-axis optimal control is more advantageous; marked by the regimes to the left of the ``X" markers in \cref{fig:K3CompError}, \cref{fig:K4CompError}.  This crossover regime can be loosely related to the L2 norm between DC and the respective ARMA(1,1) autocovariance matrices; \cref{fig:K3CompModels}, \cref{fig:K4CompModels}}
    \label{fig:CPComparison}
\end{figure*}

In addition to the notion of smooth interpolation between control noise performance from the white noise limit, there is also an observed smooth interpolation between the control solutions from the white noise solution to the near $1/f^2$ spectrum.  
This is illustrated in \cref{fig:controlnoisesolutionsmain} (a) and (b) which contains a few examples of the power spectra of the control noise and the optimal control sequences for the AR(1) control noise power spectra in \cref{fig:controlnoiseerror}.  
What is not present in the plot of the power spectrum of the optimal control is the DC component of the control which constitutes $8\%$, $48\%$, and $78\%$ of the total power of the control for a $\phi_1$ of $0.99$ (Noise 1), $0.95$ (Noise 2), and $0.9$ (Noise 3), respectively.  
Therefore, the optimal control has more DC features as the low-frequency white noise bandwidth is increased; in other words as the noise has more white-noise features.  
This aspect is also reflected in the optimal control sequences in \cref{fig:controlnoisesolutionsmain} (b), where the control is showing greater DC features as the white noise contribution to the noise becomes more prevalent.

\begin{table}[b]
    \centering
    \begin{tabular}{|c|c|c|c|c|c|}
        \hline
        & $j$ & $1$ & $2$ & $3$ & $4$\\ \hline \hline
        \multirow{2}{*}{SK1} & $\theta_j$ & $\Phi_Q$ & $2\pi$ & $2\pi$ & - \\ \cline{2-6}
        & $\phi_j$ & $0$ & $-\phi_c$ & $\phi_c$ & - \\ \hline \hline
        \multirow{2}{*}{BB1} & $\theta_j$ & $\Phi_Q$ & $\pi$ & $2\pi$ & $\pi$ \\ \cline{2-6}
        & $\phi_j$ & $0$ & $\phi_c$ & $3\phi_c$ & $\phi_c$ \\ \hline
    \end{tabular}
    \caption{Prescription for composite pulse sequences tailored for a target gate of $R_X(\theta_{Q})$.  Each sequence utilizes the phase $\phi_c = \cos^{-1}\left(\theta_Q/(4\pi) \right)$.}
    \label{fig:CompositePulseParametes}
\end{table}

\subsection{Comparison to Composite Pulse Sequences}

We will now compare the optimal gate sequences with gate sequences traditionally developed for systematic (D.C.) control errors: composite gate sequences \cite{1986LevittCompPulses,1994WimperisCompositePulse, 2004BrownArbitraryAccurateCompPulses, 2007AlwayArbitraryPrecisionCompositePulse, 2012MerrillCPReview}.
Our goal will be to compare the performance of our single-axis ARMA-defined gates sequences to multi-axis composite gates designed for amplitude control errors. We show that despite the restricted control of the ARMA-defined gates, there are regimes where they outperform composite gate protocols.

For pure amplitude control noise, we will compare our sequences to two composite pulse sequences: BB1 and SK1 \cite{1994WimperisCompositePulse, 2004BrownArbitraryAccurateCompPulses}. The specifications for these sequences are shown in Table~\ref{fig:CompositePulseParametes}. Note that the total accumulated error specifies the total angle of rotation for the first pulse in both protocols.

For the composite gate sequences, we will be examining the following Hamiltonian
\begin{equation}
    H(t) = \Omega(t) \left[ (1 + \epsilon_a(t) ) \left(\vec{\rho}(t))\cdot \vec{\sigma} \right)\right]
\end{equation}
where $\Omega(t)$ is the amplitude of the control, $\vec{\rho}(t) = [\cos(\phi(t)),\,\sin(\phi(t)),\,0]$ defines the $x$/$y$ component of the control through $\phi(t)$, $\vec{\sigma}^T = [\sigma_x,\,\sigma_y,\,\sigma_z]$ is the Pauli vector, and $\epsilon_a$ is the amplitude error of the control.
Similar to the ARMA defined noise, the noisy trajectory, $\epsilon_a$, will be Gaussian (wide-sense) stationary stochastic noise processes.
Converting into the language of DCGs, we will have the $H_{\rm gate}(t) = \Omega(t)(\cos(\phi(t)) \,\sigma_x + \sin(\phi(t)) \,\sigma_y)$ and $H_e(t) = \Omega(t)\epsilon(t)(\cos(\phi(t)) \,\sigma_x + \sin(\phi(t)) \,\sigma_y)$.  Again, we will convert to the gate-based perspective where we will discuss in terms of integrated values, i.e. $\theta_j = \int_{t_j-\Delta t}^{t_j} ds \,\Omega(s)$, $\epsilon_j = \int_{t_j-\Delta t}^{t_j} ds \,\epsilon(s)$, and $\phi(t)\rightarrow \phi_j$ is fixed over the interval $(t_j-\Delta t, t_j]$.
Undergoing the same prescription from section \cref{sec:ErrorsControlNoiseDom}, we arrive at an EPG of
\begin{equation}
    \Phi_{Q_j} = \epsilon_j \frac{\theta_j}{2} \left(\cos(\phi_j) \sigma_x + \sin(\phi_j) \sigma_y \right)
    \label{eqn:EPGcompositepulse}
\end{equation}
for the composite pulse sequences.  Our goal in the following analysis is to compute the first-order term in the Magnus expansion (\cref{eqn:firstorderMagnus}) and consequently the infidelity (\cref{eqn:FidErrorVectorExpansion}) for the composite pulse sequences and compare them to our provably optimal single-axis control noise solutions.

\begin{figure*}
    \centering
    \subfloat[Sweep of ARMA(1,1) models.]{\includegraphics[width=0.3\textwidth, valign=t]{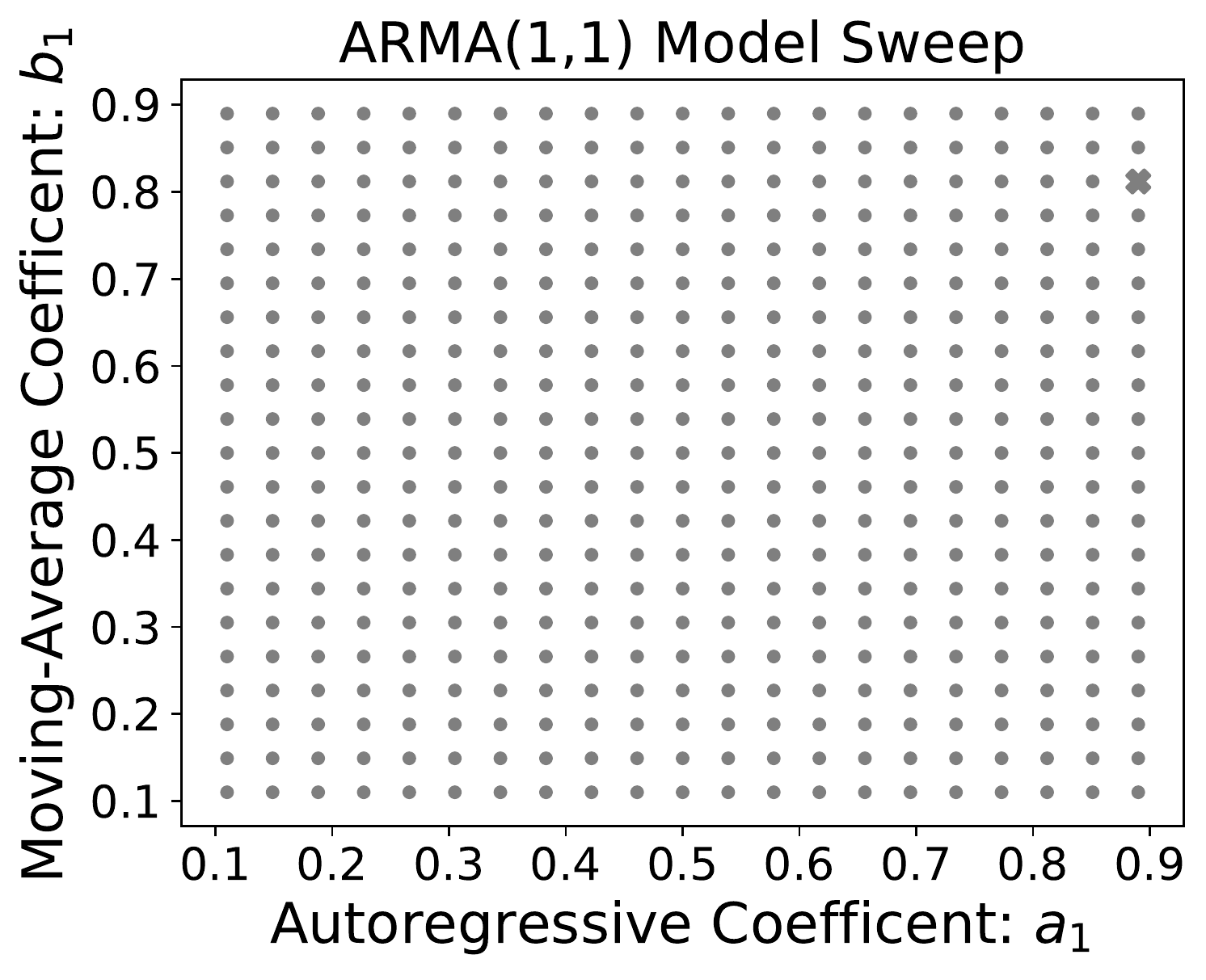}\label{fig:robustsweep}}
    \subfloat[Equivalent model error $\varepsilon$ curves.]{\includegraphics[width=0.3\textwidth, valign=t]{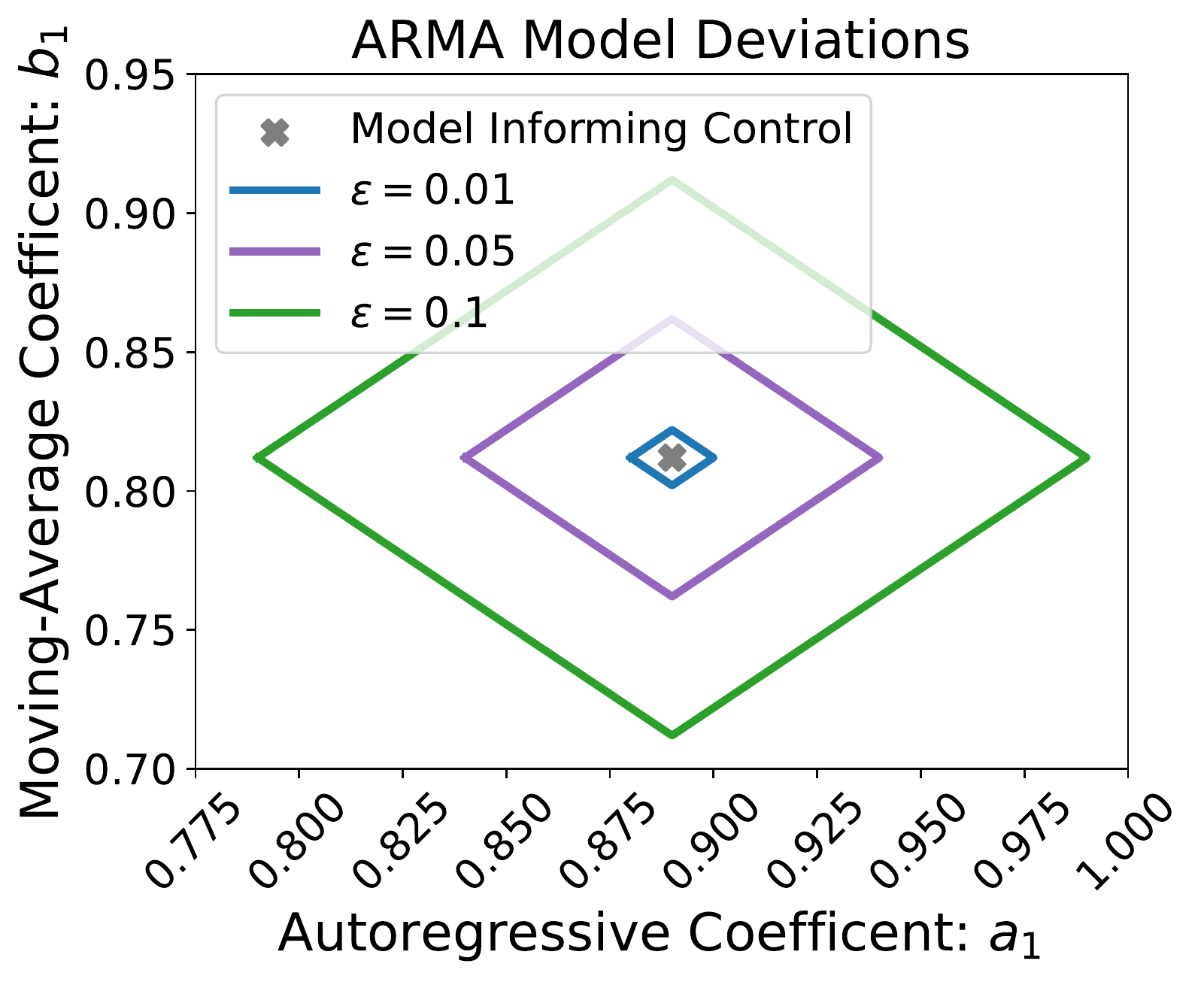}\label{fig:robustregion}}
    \subfloat[Worst-case increase in infidelity from model mismatch.]{\includegraphics[width=0.33\textwidth, valign=t]{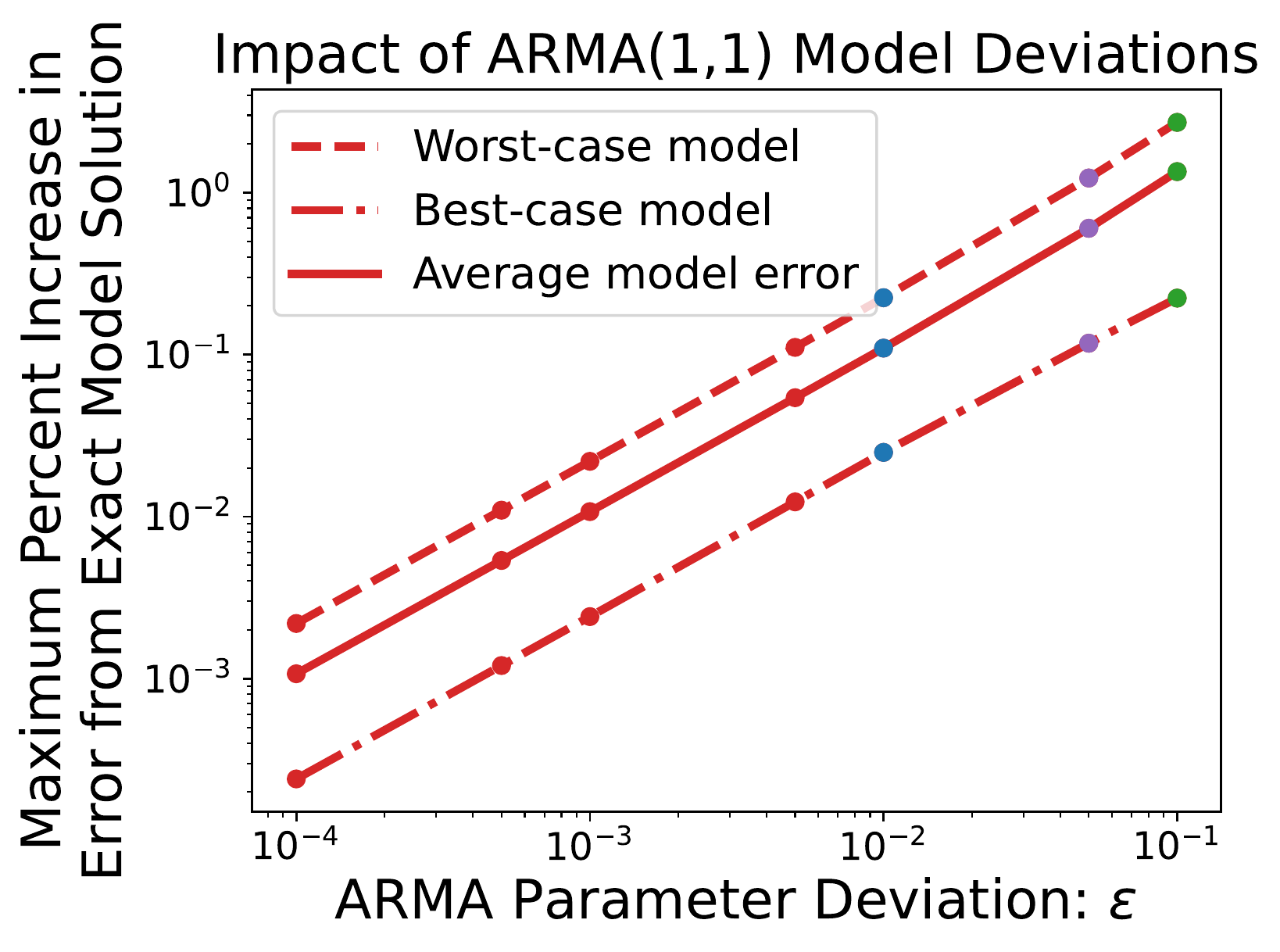}\label{fig:robustdeviation}}

\caption{Robustness of the optimal control sequences under model mismatches between the one informing the control and the true noise.  A variety of ARMA(1,1) models, red points in \cref{fig:robustsweep}, were utilized to generate optimal control solutions and applied to families of ARMA models that were a defined model distance, $\varepsilon$, from the model informing the control, \cref{fig:robustregion}.  The worst-case error (over all ARMA models \cref{fig:robustsweep} and deviations) results in less than a $5\%$ difference relative to the optimal result (when the models match) even up to a model deviation ($\varepsilon$) of $10 \%$.}
\label{fig:ARMAmodelfit}
\end{figure*}

The two sequences we will be examining are the SK1 and BB1 composite pulse sequences which are result in first- and second-order cancellation of systematic control errors, respectively.
Utilizing the single-axis $\sigma_x$ prescriptions for logic gates (shown in \cref{fig:CompositePulseParametes}) results in an error vector for SK1 of:
\begin{equation}
    \vec{a}_1 = \frac{1}{4}\left( 
    \begin{array}{c}
    \theta_{Q} \left(2\epsilon_1 - \epsilon_2 - \epsilon_3\right) \\
    (\epsilon_3-\epsilon_2) \cos(\theta_Q) \sqrt{16\pi^2-\theta_Q^2} \\
    -(\epsilon_3-\epsilon_2) \sin(\theta_Q) \sqrt{16\pi^2-\theta_Q^2}
    \end{array}
    \right).
    \label{eqn:SK1errvector}
\end{equation}
This leads to a first-order term in infidelity of
\begin{equation}
    \langle \left| \vec{a}_1 \right|^2 \rangle_{SK1} = 2\pi^2 \left(\gamma(0) - \gamma(1) \right) + \frac{\theta_Q^2}{4}\left(\gamma(0) - \gamma(2) \right).
    \label{eqn:infidSK1}
\end{equation}
Similarly, the BB1 sequence results in an error vector:
\begin{equation}
    \vec{a}_1 = \frac{1}{8}\left( 
    \begin{array}{c}
    \theta_{Q} \left(4\epsilon_1 - \epsilon_2 - 2\epsilon_3 - \epsilon_4 \right) \\
    (\epsilon_2-2\epsilon_3 + \epsilon_4) \cos(\theta_Q) \sqrt{16\pi^2-\theta_Q^2} \\
    - (\epsilon_2-2\epsilon_3 + \epsilon_4) \sin(\theta_Q) \sqrt{16\pi^2-\theta_Q^2}
    \end{array}
    \right)
    \label{eqn:BB1errvector}
\end{equation}
giving a first-order term in the infidelity:
\begin{equation}
    \begin{gathered}
    \langle \left| \vec{a}_1 \right|^2 \rangle_{BB1} = \frac{\pi^2}{2} \left(3\gamma(0) - 4\gamma(1) + \gamma(2) \right)\\
    + \frac{\theta_Q^2}{8} \left(2\gamma(0) + \gamma(1) - 2\gamma(2) - \gamma(3) \right).
    \end{gathered}
    \label{eqn:infidBB1}
\end{equation}
Upon inspection, the first-order terms above are $0$ in the DC limit; i.e. $\gamma(j)$ constant.

\cref{fig:CPComparison} shows the comparision between the infidelity, $\langle \left|\vec{a}_1 \right|^2 \rangle$, of composite pulse sequences and the optimal single-axis control from this work when applied to a wide variety of ARMA(1,1) defined noise.
For these calculations, the total power of the noise sequence was fixed to $10^{-3}$.
As expected, the advantage of the composite pulse seqeunces increase with the correlations of the noise (autoregressive coefficient $a_1 \rightarrow 1$), but with the single-axis optimal control sequences showing advantage when the noise is sufficiently far from DC.
The break-even regimes where the two approaches result in similar error are mapped out by the white X's in \cref{fig:K3CompError} and \cref{fig:K4CompError}.

While a simulation approach is useful for a comparison of the approaches, model-based methods for determining the appropriate control approach may be of the greater practical interest.
To that end, we analyzed the L2 norm between the autocovariance matrices of the ARMA(1,1) models and that of DC noise. This metric is strongly rooted in classical model-based approaches to characterizing systems. Moreover, it allows for a direct comparison between noise processes based on information (i.e., the autocovariance) used by our approach to construct optimal control sequences.

There is general agreement between the break-even regimes when utilizing the L2 norm to estimate the relative performance of the two approaches. Our optimal control sequences show advantage when the L2 norm is greater than $\sim10-20\%$ of the total power ($10^{-3}$ in our case) of the noise sequence: \cref{fig:K3CompModels} and \cref{fig:K4CompModels}.
This robustness of CP sequences up with $\sim10 \%$ mixing of non-DC time-correlations is consistent with other investigations into the robustness of CP sequences under time-dependent noise \cite{2014KabytayevRobustnessCompPulsesTimeDepNoise} and appears to hold for noise outside of the regimes previously studied.
Note that the L2 norm seems to be a respectable heuristic for approach selection, but these estimates are loose in some cases; see \cref{fig:K4CompModels} as $b_1\rightarrow 0$.
Tighter model-informed selection criteria and the extension of these approaches to more complex ARMA models is an open question for future work.

\subsection{ARMA Model Fit Robustness Analysis}
Our method critically relies on the characterization of the control noise of the system for generating provably optimal control.
While methods are established for estimating the noise of the qubit systems \cite{2003ShoelkopfQuantumNoise,2004FaoroDynamicalSupp1fNoise,2012YoungQubitsSpectrometers,2018NorrisOptBandControlNoise} and have been demonstrated \cite{2020FreySimultSpectralEstQubitSlepian}, inaccuracies in estimation can lead to imperfections in ARMA model fitting. In turn, such imperfections will influence the performance of our approach. Because of this, an analysis was performed on the robustness of this approach when there is a mismatch between the ARMA model used to generate the optimal control sequence and the true noise correlations of the system.

To do this, we again utilize a range of ARMA(1,1) models [see \cref{fig:robustsweep}] and calculate the optimal control for these models. We then apply that optimal control to a family of ARMA models with a defined model distance away, from the model informing the control; see \cref{fig:robustregion}.
For model deviations, we sampled a set of $\sim 400$ ARMA model parameter deviations, $a \rightarrow a+\varepsilon_{a}$ and $b \rightarrow b+\varepsilon_{b}$, such that $\left| \varepsilon_a \right| + \left| \varepsilon_b \right| = \varepsilon$ for a range of $\varepsilon$; examples in \cref{fig:robustregion}.
In the worst-case scenarios, i.e. when the error observed is greater than the model matching case, the observed infidelity had less than $5 \%$ difference from the exact model matching infidelity, \cref{fig:robustdeviation}.

Similar to the $1/f^2$ results from \cref{fig:1fcontrolnoisesolutions}, there appears to be a notion of relatively smooth interpolation between control noise solutions for equivalent ARMA models (equivalent $p$ and $q$) and similar model parameters ($a_1$, $b_1$).
This notion results in a natural robustness of this approach to uncertainties in the estimates of the ARMA model parameters.
Perhaps a more subtle point is that this analysis hints of a potential robustness relative to model reductions as well; i.e. when the ARMA model parameters ($p$, $q$) may be over-complete. 
For example, the $a \rightarrow 0$ and $b \rightarrow 0$ regimes from Fig.~\ref{fig:robustsweep} contain such scenarios; including examples where ARMA(1,1) defined control is applied to AR(1) or MA(1) noise.
This seems to imply that our approach is also robust to ARMA model mismatches due to anomalous introduction of additional artifacts into the ARMA model.
We conjecture that this model reduction robustness also extends to other flavors of model reduction mismatches, AR(2) control for AR(1) noise for instance, as well.
The requirement to find generally good but not exact representations of the noise may provide useful avenues for scalability of this approach. Furthermore, it suggests this approach potentially allows for restricted characterization of qubit control noise, but broadly applied control solutions that will deviate a small percentage from the strictly optimal solutions from properly calibrated control.

\begin{figure*}
    \centering
    \subfloat[Control for Various Noise Sources.]{\includegraphics[width=0.55\textwidth]{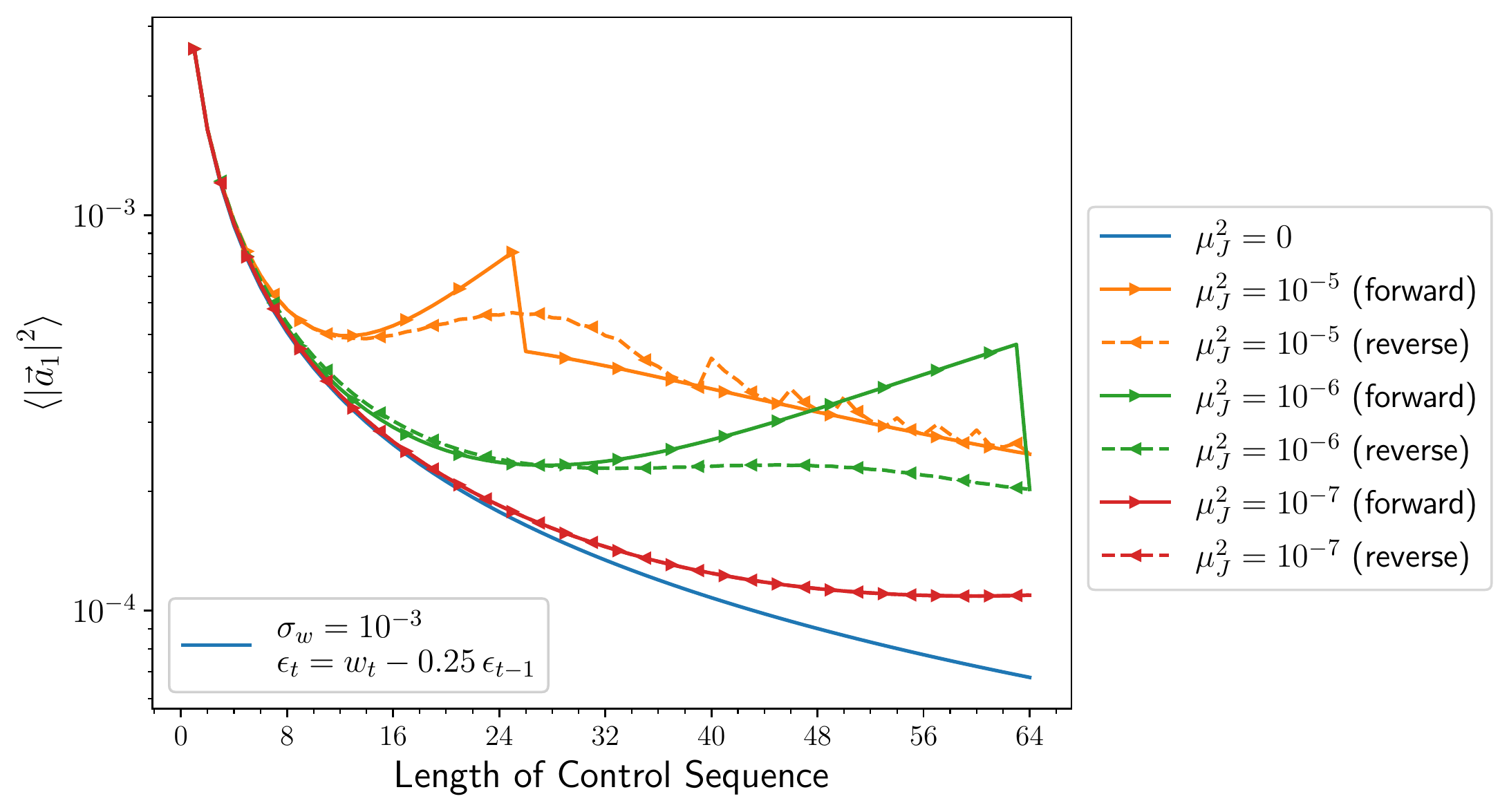}\label{fig:errorstrengthfull}}
    \subfloat[Deviations from optimal control.]{\includegraphics[width=0.39\textwidth]{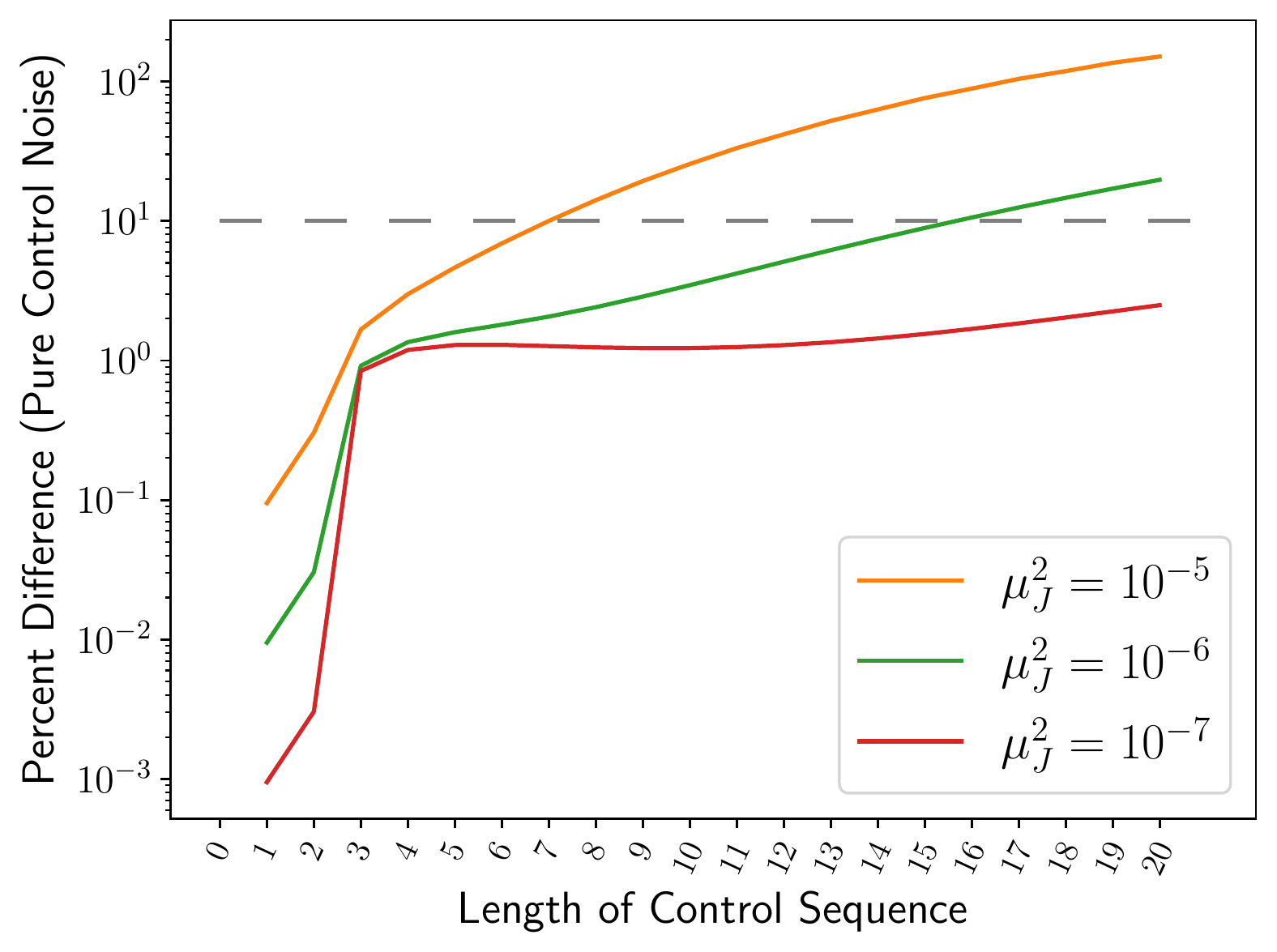}\label{fig:errorstrengthdeviate}}
    \caption{\cref{fig:errorstrengthfull} The error in the optimized control developed from a $1/f^2$ control noise spectra with the addition coherent dephasing at varying strengths.  The lowest blue curve is the optimal solutions without a dephasing contribution.  Due to the nonconvex nature of the optimization forward and reverse sweeps were performed (details in \cref{sec:controldephasing}). \cref{fig:errorstrengthdeviate} Percent change in the optimal control solution error when weak dephasing is introduced. There are regimes where the optimized solutions are within a $10\%$ difference (gray dashed line) from the optimal solutions.}
    \label{fig:errorstrength}
\end{figure*}

%% file: Sections/5-DephasingAddition.tex
We have shown in \cref{sec:controlham} how to construct
control sequences for pure control noise that optimally reduces the first-order term in the Magnus expansion of the error dynamics.  
This was possible given that this first-order term (under linear constraints) is a convex function.
With the added contribution of dephasing, the first-order term is no longer a convex function and we therefore lose the provable optimality of this approach in most cases.
However, in systems where the control noise is strong relative to the strength of the dephasing noise (or at the very least the dominant source of error during the application of the gate), we can rely on the convex features of the dominant control noise contributions to find good solutions to the non-convex problem \cref{eqn:DephasingErrorFull}.
The obtained solutions can be viewed as local perturbations to the optimal control solutions/performance due to the addition of some weaker non-convex landscape (terms A and B in \cref{eqn:DephasingErrorFull}) introduced at the minimum of the convex control noise term (term C in \cref{eqn:DephasingErrorFull}).

An example of this weak perturbation behavior relative to the optimal solution is shown in \cref{fig:errorstrength} which provides an example of $1/f^2$ control noise and pure coherent (DC) dephasing at various strengths.
Due to the nonconvex nature of the objective function, forward and reverse sweeps of the control solutions were performed to find good initial conditions for the optimizations due to the competition between the nonconvex dephasing contributions and the convex control noise components (see \cref{fig:errorstrength}).
For the forward sweeps, constant drive solutions were used for the initial conditions for the optimizations.
For the reverse sweep initial conditions, the lowest amplitude time step, $\theta_{min}$, was removed from the $K^{th}$ solution and all remaining $K-1$ amplitudes were re-scaled by $\theta_{min}/(K-1)$.
With this approach we were able to find that, for sufficiently weak noise or for sufficiently short control pulses, the errors observed from the optimized solutions with added dephasing are within a small percent change relative to the optimal solutions (\cref{fig:errorstrength}). This provides confidence in the achievement of a good solutions for this nonconvex case.
While finding good, minimal solutions in the weak noise regime are of interest, we will not restrict ourselves to that regime in the analysis below.
Furthermore, there are other interesting features of our approach outside of this regime that are discussed in \cref{sec:coherentdephasingappendix}.

\subsection{Control Noise Dominant Solutions}
Similar to \cref{sec:controlnoiseexamplesmain}, we will start with simple ARMA model representations of the control noise and build up complexity in the ARMA model representations of the dephasing noise.
For all examples below, we utilized the $1/f^2$ ARMA model for the control noise from \cref{fig:controlnoiseerror} and \cref{sec:1fexample} with $\phi_1 = 0.25$.
We will start with the white dephasing noise case and then introduce pure coherent dephasing (DC dephasing) followed by $1/f^2$ dephasing for the analysis.

\subsubsection{White Dephasing Noise}
For the first example, we will analyze the performance of our optimal control protocols under the influence of zero-mean white dephasing noise.
This example is a unique case of dephasing where we still retain the provable optimality of the control sequence.
For the white noise case, the expression for the infidelity is
\begin{equation}
    \label{eqn:DephasingErrorWhite}
    \langle \vec{a}_1^2 \rangle = \frac{1}{4} \left[\vec{x}^T A \vec{x} + N \gamma_J(0)\right]
\end{equation}
where we have used the quadratic form of the control noise component (see \cref{eqn:controlnoisequad}) for compactness.
In this case, the control has no effect on the error due to the dephasing and has a static contribution to the error proportional to the variance of the white \emph{dephasing} noise times the length of the control sequence ($N \gamma_J(0)$).

\subsubsection{Coherent Dephasing Noise} \label{sec:coherentdephasingmain}
The second model considered was pure DC dephasing noise with $1/f^2$ control noise.  The error vector component contains contributions from the $A$ and $C$ terms in \cref{eqn:DephasingErrorFull}.
\cref{fig:controlnoisedephasing} shows the added contribution from the dephasing noise to the infidelity with a pure dephasing rate of $\mu_J = \sqrt{10^{-5}}$.
In the coherent dephasing case, there was a noticeable discontinuity in the infidelity as a function of the length of the optimized control sequence.
This observation highlights one of the strengths of the approach: the compactness of the ARMA representations of the noise allow for optimizations over all contributions of the total infidelity of the dynamics. This is equivalent to a weighted (according to the strength of the noise) optimization over the various noise contributions (or equivalently, filter functions over the various noise axes).
The aforementioned discontinuity in the infidelity is an example of this, where the reduction in the infidelity is a result of the optimizer switching focus from mitigating control noise to dephasing (more details available in \cref{sec:coherentdephasingappendix}). 
Note that this behavior is apparent even after the reverse sweep to generate ``good'' initial conditions.

The ability to identify ``hard decision" regimes is an attractive feature of our approach. When the dephasing and control noise have nearly equivalent power, the optimizer does not easily converge to a ``good" solution. Essentially, the solutions correspond to local minima that cannot yield both control noise and dephasing suppression. However, given a characterization of the spectral features of the control and dephasing noise, we can begin to estimate relevant sequence lengths where such crossovers occur. This information enables use to develop an empirical bound on the maximum number of pulses permitted prior to reaching this regime. As an example, consider Fig.~\ref{fig:controlnoisedephasing}, where the optimal sequence length is approximately 10 gates. While this approach is not analytically rigorous, it does present a useful technique for optimal control design from an empirical perspective.

\subsubsection{\texorpdfstring{$1/f^2$}{1f2} Dephasing Spectra}
The final dephasing example implemented pure (i.e. zero mean) ARMA model dephasing and control noise.  For the dephasing noise, a $1/f^2$ ARMA model with $\phi_1 = 0.75$ (see \cref{sec:1fexample}) was implemented. 
For this example, we will be considering only the $B$ and $C$ contribution to the infidelity from \cref{eqn:DephasingErrorFull}.
The infidelity as a function of the length of the optimized control sequences displayed noticeably different behavior than the coherent dephasing example.
Where the control was able to mitigate the accumulation of the dephasing noise for the coherent case, the contributions from $1/f^2$ dephasing noise correlations dominate the error contribution for long length sequences. In fact, the longest pulse sequences performed similarly to naive single-pulse applications of the logic gate.
For this flavor of dephasing noise, a mitigation strategy was not found that could combat the dephasing contribution for long pulse lengths. However, identifiable improvement in infidelity are present for short sequence lengths, similar to the coherent dephasing case.
We conjecture that certain flavors of dephasing and control noise will allow for reductions in infidelity with increasing sequence length. It remains open to investigation as to whether such improvements can be achieved with noise spectra relevant to experimental systems.
Regardless, this approach can aid the user in finding high fidelity control solutions in the presence of control and dephasing noise where the control and dephasing noise exhibit temporal correlations.

%% file: Sections/6-Discussion.tex
\begin{figure}
    \centering
    \includegraphics[width=0.48\textwidth]{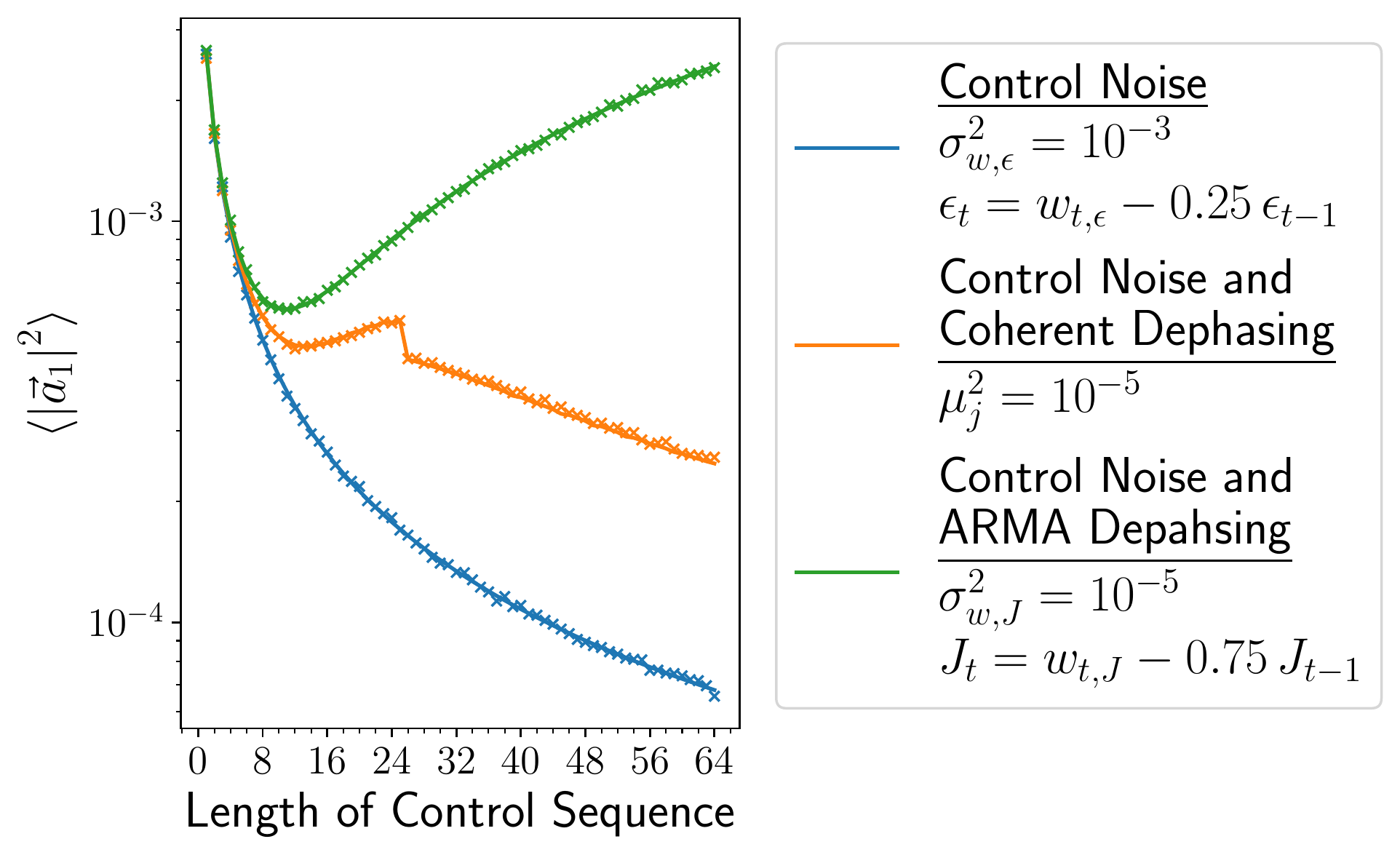}
    
    \caption{Infidelity of the optimized pulse sequences for the control noise and dephasing case.  The control noise shown is a $1/f^2$-like, AR(1), spectra and added pure coherent dephasing and $1/f^2$ dephasing spectra.  The lines are the theory from \cref{eqn:DephasingErrorFull} and markers are Monte Carlo simulations ($10^4$ Monte Carlo samples per point) of the first-order term in the Magnus expansion.  The jump in observed error for the coherent dephasing case is discussed in \cref{sec:coherentdephasingappendix}.}
    \label{fig:controlnoisedephasing}
\end{figure}

We have demonstrated the utility of the Schr\"{o}dinger Wave Autoregressive Moving Average model (SchWARMA) \cite{2020SchultzSCHWARMA} formalism in the context of mitigating noise in systems that are dominated by time-correlated semi-classical control noise.
This connection to SchWARMA models implies that the noisy control Hamiltonian can be cast as a quantum circuit with noisy gates defined by an ARMA model.  
The connection with noisy quantum circuits further infers an ability to fold the control sequence optimizations utilized in this work into a quantum compiler tool chain operating at the compiled quantum gate level.  
For more details about the SchWARMA model formalism for simulating time-correlated circuit or the use of SchWARMA models for time-correlation noise injection in experiments, refer to \cite{2020SchultzSCHWARMA} and \cite{2021MurphySchWARMANoiseInjection}, respectively.

For pure time-correlated control noise, we have shown that finding solutions that minimize the first-order term in the Magnus expansion of the dynamics is equivalent to a convex optimization; implying provable optimality of the designed control.
The utility of this approach was demonstrated for $1/f^2$ control noise and were compared to existing multi-axis control-noise-mitigating composite pulse protocols.
Furthermore, we have evidence that this approach is generally robust to errors in model estimation. This may allow for the broad use of solutions even in the case of model mismatch with tolerable deviations from optimal performance.
We have also discussed the extensions to systems with strong time-correlated control noise relative to a time-correlated dephasing contribution.
While the solutions lose provable optimality, we demonstrate regimes where the error from the designed control deviates by a small percentage from the optimal pure control solutions.
Finally, we show the power of the SchWARMA formalism to compactly represent the total infidelity of the noisy dynamics allowing for direct minimization of the infidelity of the system. This attribute gives the optimizer the ability to make informed decisions on mitigation strategies when dephasing and control noise contributions compete.

Although we have shown particular examples of control noise, the ARMA model formalism on which the SchWARMA models are built upon are representations (equivalent up to autocorrelation) of the general family of stationary noise processes \cite{1964FarlieARMA, 1976BoxARMA}; providing generality of this approach to many flavors of control noise. Our approach also naturally allows for: (1) the enforcement of amplitude constraints and (2) bandwidth constraints on the control. These features follow from the additive properties of the ARMA models, which allow the user to include frequency constraints via the mixing of high-power bandlimited ARMA signals into to the ARMA signals of the native noise.

The development of this approach provides a variety of follow-on directions to pursue.
A prime candidate for the experimental demonstration of these mitigation strategies are ion trap systems.
We envision a scenario where outputs of control noise spectroscopy protocols~\cite{2018NorrisOptBandControlNoise,2022BarrControlNoiseSpecDephasing} or simultaneous control and dephasing noise~\cite{2020FreySimultSpectralEstQubitSlepian} are used to inform our optimal control strategy. Estimated noise spectra could be to fit ARMA models and then leveraged thereafter for constructing optimized gate sequences.
While improvements on single-qubit control are advantageous, noise on two-qubit gates provide a strong contribution to the noise budget in many quantum computing approaches.
Extension of this noise mitigation approach to two qubit gates, particularly modeling two-qubit M{\o}lmer-S{\o}rensen entangling gates from ion traps, would provide an avenue for combating amplitude drifts for these slower and more error-prone gates. 
Finally, ARMA models 
are a subclass of a greater family of models from classical signal processing.  One particular family of models that may be of interest are Autoregressive Integrated Moving Average (ARIMA) models which represent non-stationary signals.  It remains an open question as to whether ARIMA models could be implemented in a closed-loop control approach for mitigating non-stationary noise.

%% file: Sections/7-Acknowledgements.tex
We thank Ken Brown for planting the initial seed for this work on utilizing the SchWARMA formalism in the context of mitigating time-correlated control noise.
CT, KS, PT, LN, GQ, and BC acknowledge funding from the U.S. Department of Energy (DOE), Office of Science, Office of Advanced Scientific Research (ASCR), Accelerated Research in Quantum Computing (ARQC) program, Grant No. DE-SC0020316. PT acknowledges funding from the DOE, Office of Science, ASCR, Quantum Computing Application Teams program, under fieldwork Proposal No. ERKJ347.
The views and conclusions contained herein are those of the authors and should not be interpreted as necessarily representing the official policies or endorsements, either expressed or implied, of the United States Department of Energy or the U.S. Government.

%% file: Sections/A1-MagnusExpansion.tex
In this section, we will outline the Magnus expansion as it pertains to the computation of the error dynamics of our composite control sequences.
In our case, the Magnus expansion is utilized to compute the effective error dynamics introduced from the noise Hamiltonians participating in the noisy control sequences.
To do so, we aim to solve the Schr\"{o}dinger equation:
\begin{equation} \label{eqn:a1schrodinereqn}
    \frac{\partial}{\partial t} U(t, t_0) = \bar{H}(t) U(t,t_0)
\end{equation}
where $\bar{H} = -i H(t)/\hbar$ with the initial condition $U(t_0, t_0)= I$.
For the examples in the main text, the Hamiltonians under analysis are the ``toggling frame" error Hamiltonians with respect to the ideal control, $\tilde{H}_e(t_j, t_{j-1})$ from \cref{eqn:Htilde}, with an additional toggling frame update with respect to the total accumulated ideal control due to the composition of multiple gates within a gate sequence, see \cref{eqn:EPGcomposite}.

A solution for \cref{eqn:a1schrodinereqn} is known and can be represented as an infinite series expansion with respect to the Hamiltonian, $\bar{H}(t)$; known as the Magnus expansion \cite{1954MagnusExpansion}.
For this work, we want to compute the time evolution of the noise Hamiltonian in the interaction picture with respect to the ideal gate frame (\cref{eqn:Htilde}), $\tilde{H}_{e}(t_j,t_{j-1}) = U_{\rm gate}(t_j,t_{j-1})^\dagger H_{e} U_{\rm gate}(t_j,t_{t_{j-1}})$.
In many instances, exact computation of the propagator for the error dynamics, $\tilde{U}(t_j, t_{j-1}) = \exp(-i\,\int_{t_{j-1}}^{t_j} ds\,\tilde{H}_e(s,0))=\exp(-i\,\Phi_{Q_j})$, is not possible and must be approximated.
To that end, the Magnus expansion \cite{1954MagnusExpansion} is typically employed which expands the exponent into an infinite sum of nested commutators; i.e.:
\begin{equation}
    \tilde{U}(t, 0) = \exp\left(\sum_{k=1}^\infty \Omega_k(t, 0) \right)
\end{equation}
where the terms in the Magnus expansion are of the form:
\begin{equation}
    \begin{split}
        \Omega_1(t, 0) = & \int_{0}^{t} d\tau \, \bar{\tilde{H}}(\tau, 0)\\
        \Omega_{n\ge 2}(t, 0) = & \sum_{j=1}^{n-1} \frac{B_j}{j!} \int_{0}^{t} d\tau \, S_n^{(j)}(\tau, 0)
    \end{split}
\end{equation}
where $B_j$ are Bernoulli numbers, $\bar{\tilde{H}}_e(t, 0) = -i \tilde{H}_e(t, 0)$, and $S_n^{(j)}$ is defined recursively where:
\begin{equation}
    S_n^{(k)}(t, 0) = \sum \left[\Omega_{i_1}, \left[...\left[\Omega_{i_k},\, \bar{\tilde{H}}_e(t, 0) \right]... \right] \right] \;\; \mathrm{s.t.} \;\; i_1 + ... + i_k = n-1
\end{equation}
Using the above equations, we can write the first couple of terms in the Magnus expansion:
\begin{equation}
    \begin{split}
        \Omega_1(t,0) = & \int_0^{t} dt_1 \, \bar{\tilde{H}}_e(t, 0)\\
        \Omega_2(t,0) = & \frac{1}{2}\int_0^{t} dt_1 \int_{0}^{t_1} dt_2 \, \left[ \bar{\tilde{H}}_e(t_1, 0), \bar{\tilde{H}}_e(t_2, 0) \right]\\
        \Omega_3(t,0) = & \frac{1}{6} \int_0^{t} dt_1 \int_{0}^{t_1} dt_2 \int_{0}^{t_2} dt_3 \, \left\{\left[\bar{\tilde{H}}_e(t_1, 0), \left[ \bar{\tilde{H}}_e(t_2, 0), \bar{\tilde{H}}_e(t_3, 0) \right] \right] + \left[ \left[ \bar{\tilde{H}}_e(t_1, 0), \bar{\tilde{H}}_e(t_2, 0) \right], \bar{\tilde{H}}_e(t_3, 0) \right] \right\}\,.
    \end{split}
\end{equation}

Green et al. \cite{2012GreenHighOrderFilterLogic} have shown the prescription for representing the terms in the Magnus expansion in terms of contributions to the error vector representation of the noise: 
\begin{equation}
    \tilde{U}(t, 0) = \exp\left(\sum_{k=1}^\infty \Omega_k(t, 0) \right) = \exp\left(-i\,\vec{a}(\tau)\cdot \vec{\sigma} \right)
\end{equation}
where $\vec{a}(\tau) \cdot \vec{\sigma} = \sum_{i=1}^\infty \vec{a}_i \cdot \vec{\sigma}$.
In the case where one can factor out the semi-classical variable defining the noise from the effective error dynamics, i.e. $\tilde{H}_e(t,0) = \beta(t,0)/2 \, \vec{s}_1(t,0) \cdot \vec{\sigma}$, one can equivalent represent the terms in the Magnus expansion with respect to the error vector as:
\begin{equation}
    \begin{split}
        \Omega_n(t,0) = -i \, \vec{a}_n(t,0) \cdot \vec{\sigma}
    \end{split}
\end{equation}
and the terms are recursively defined with respect to the control vector, $s_1(t, 0)$.
The first few terms are of the form:
\begin{equation}
    \begin{split}
        \vec{a}_1(t,0) = & \int_0^t dt_1 \,\left(\frac{\beta(t,0)}{2}\right) \vec{s}_1(t,0)\\
        \vec{a}_2(t,0) = & \int_0^t dt_1 \int_0^{t_1} dt_2 \,\left(\frac{\beta(t_1,0)}{2}\right)\left(\frac{\beta(t_2,0)}{2}\right) \vec{s}_2(t_1,t_2,0,0)\\
        \vec{a}_3(t,0) = & \int_0^t dt_1 \int_0^{t_1} dt_2 \int_0^{t_2} dt_3 \,\left(\frac{\beta(t_1,0)}{2}\right)\left(\frac{\beta(t_2,0)}{2}\right)\left(\frac{\beta(t_2,0)}{2}\right) \vec{s}_2(t_1,t_2,t_3,0,0,0)
    \end{split}
\end{equation}
where
\begin{equation}
    \begin{split}
        \vec{s}_2(t_1,t_2,0,0) \equiv & \; \vec{s}_1(t_2,0) \times \vec{s}_1(t_1,0)\\
        \vec{s}_3(t_1,t_2,t_3,0,0,0) \equiv & \; \vec{s}_1(t_3,0) \times \left[\vec{s}_1(t_2,0) \times \vec{s}_1(t_1,0)\right] + \left[\vec{s}_1(t_3,0) \times \vec{s}_1(t_2,0)\right] \times \vec{s}_1(t_1,0) \, .
    \end{split}
\end{equation}
For the case of only control noise (see \cref{sec:controlham}), one can factor out the control noise, $\epsilon(t)$, from the expression for $\tilde{H}_e$ and apply the construction for the higher-order terms with respect to the error vector as shown above; the analysis of higher order terms for the examples the main text (\cref{sec:controlnoiseexamplesmain}) are presented in \cref{sec:higherordercontrolnoise}.
For the case of control noise and dephasing, there are two variables defining the semi-classical noise, $\epsilon(t)$ and $\tilde{J}(t)$, that results in complications when constructing an error vector representation of the higher order terms in the Magnus expansion discussed in \cref{sec:higherorderdephasing}.

%% file: Sections/A2-FirstOrderDerivation.tex
In this section, we will show how we obtained the objective function in \cref{eqn:DephasingErrorFull} by applying the control frame and subsequent simplifications.
Starting with the definition of the error per gate from \cref{eqn:Htilde}, $\tilde{H}_{e}(t_j,t_{j-1}) = U_{\rm gate}(t_j,t_{j-1})^\dagger H_{e} U_{\rm gate}(t_j,t_{t_{j-1}})$,
and noting that our Hamiltonian from \cref{eqn:controlnoiseham} is of the form:
\begin{equation} 
    H(t) = \frac{J(t)}{2} \sigma_z+\left(1+\epsilon(t) \right)\frac{\Omega(t)}{2} \sigma_x\, ,
\end{equation}
we can obtain the Hamiltonian in the interaction picture with respect to the ideal control:
\begin{equation}
    \begin{gathered}
     \tilde{H}_{e}(t) = \frac{J(t)}{2}\left(\cos(\Omega(t))\sigma_z + \sin(\Omega(t))\sigma_y \right)\\
     + \epsilon(t) \frac{\Omega(t)}{2} \sigma_x \,.
    \end{gathered}
\end{equation}
We can then obtain an EPG of the form from \cref{eqn:dephasingEPG0}:
\begin{equation}
     \Phi_{Q_j} =   \frac{J_j}{2}\left(\cos(\theta_j)\sigma_z + \sin(\theta_j)\sigma_y \right)
     + \epsilon_j \frac{\theta_j}{2} \sigma_x
\end{equation}
where $\beta_j = \int_{t_j-\Delta t}^{t_j} ds \, \beta(s)$ for the semi-classical noise processes, $\beta \in \left[J,\, \epsilon \right]$, defining the dephasing and control noise.
Note, much like equation \cref{eqn:EPGcomposite}, the integration is implicit to the expressions for $J_j$, $\epsilon_j$, and $\theta_j$ and the limits of integration for each expression are defined by the gate index $j$.
We will further distinguish between the coherent and time correlated components of the dephasing noise, $J_j$, by defining the mean $\mu_J = \sum_{j=1}^N J_j/(N\Delta t)$ and $\tilde{J}_j = J_j - \mu_J$ resulting in a zero-mean process for $\tilde{J}$.
This gives an equivalent form for the EPG of
\begin{equation}
    \begin{split}
    \Phi_{Q_j} & =  \frac{\mu_J}{2} \left(\cos(\theta_j)\sigma_z + \sin(\theta_j)\sigma_y \right)\\
    & + \frac{\tilde{J}_j}{2}\left(\cos(\theta_j)\sigma_z + \sin(\theta_j)\sigma_y \right) \\
    & + \epsilon_j \frac{\theta_j}{2} \sigma_x
    \end{split}
\end{equation}
which was used in the analysis of the main text.

We can obtain each individual error per gate at times $j$ by making use of the Baker-Campbell-Hausdorff (BCH) \cite{2000BCHFormula} formula
\begin{equation}
    e^{i\lambda G}A e^{-i\lambda G} = A + i\lambda [G,A] + \frac{(i\lambda)^2}{2!} [G,[G,A]] + \frac{(i\lambda)^3}{3!}[G,[G,[G,A]]] + ... + \frac{(i\lambda)^n}{n!}[G,[G,...[G,A]...]]
\end{equation}
We can then compute each non-commuting term in $\tilde{H}_{e,j}$.  To start, we can compute the first few terms in the $\tilde{J}_j$ non-commuting terms (noting that the $\mu_J$ terms have a similar form): $[G,A]=[\sigma_x,\tilde{J}_j/2 \sigma_z] = -i \tilde{J}_j$, $[G,[G,A]] = [\sigma_x,-i \tilde{J}_j \sigma_y] = 2 \tilde{J}_j$, ...
;resulting in an error per gate at time $j$ (\cref{eqn:dephasingEPG}) of
\begin{equation}
 \begin{split}
 \tilde{H}_{e,j} & = A + B + C\\
 A & = \frac{\mu_J}{2}\left[\left(\sum_{k=0}^\infty (-1)^k \frac{\theta_j^{2k}}{(2k)!} \right)\sigma_z + \left(\sum_{k=0}^\infty (-1)^k \frac{\theta_j^{2k+1}}{(2k+1)!}\right)\sigma_y \right]
  = \frac{\mu_J}{2}\left[\cos(\theta_j)\sigma_z + \sin(\theta_j)\sigma_y \right]\\
 B & = \frac{\tilde{J}_j}{2}\left[\left(\sum_{k=0}^\infty (-1)^k \frac{\theta_j^{2k}}{(2k)!} \right)\sigma_z + \left(\sum_{k=0}^\infty (-1)^k \frac{\theta_j^{2k+1}}{(2k+1)!}\right)\sigma_y \right]
  = \frac{\tilde{J}_j}{2}\left[\cos(\theta_j)\sigma_z + \sin(\theta_j)\sigma_y \right]\\
 C & = \frac{\epsilon_j \theta_j}{2}\sigma_x\\
 \tilde{H}_{e,j} & = \frac{\mu_J}{2}\left[\cos(\theta_j)\sigma_z + \sin(\theta_j)\sigma_y \right] + \frac{\tilde{J}_j}{2}\left[\cos(\theta_j)\sigma_z + \sin(\theta_j)\sigma_y \right] + \frac{\epsilon_j \theta_j}{2}\sigma_x
 \end{split}
\end{equation}
Now, our goal is to compute the erroneous dynamics of multiple time steps of erroneous gates; truncated up to first order in the Magnus expansion of the noise.
The generator for the total error dynamics are represented as \cite{2009KhodjastehDCGs} $\exp(-i \Phi_A)$ where $\Phi_A = \sum_{i=1}^\infty \Phi^{[i]}_A$ is an infinite sum of the $i^{th}$ order terms in the Magnus expansion.
We will truncate this expansion up to first order, requiring the evaluation of the sum $\Phi_A^{[1]} = \sum_{i=1}^N P_{j-1}^\dagger \Phi_j P_{j-1}$ (\cref{eqn:firstorderMagnus}) where $P_j = \exp(-i \, \sigma_0 \sum_{l=1}^j \theta_l \sigma_x/2)$. 
Therefore, the final piece is to analyze the error per gate terms modulated according to the accumulated ideal control up to time $j$ (see \cref{eqn:EPG}).
We can again make use of the BCH formula to evaluate a single term for a given time step $j$:
\begin{equation}
    \begin{split}
    P_{j-1}^\dagger \Phi_{Q_j} P_{j-1} = & A' + B' + C' \\
    A' = & \,\frac{\mu_J}{2} 
    \left\{
    \cos(\theta_j)
    \left[
    \left(\sum_{k=0}^\infty (-1)^k \frac{\Theta_{j-1,1}^{2k}}{(2k)!} \right)\sigma_z + \left(\sum_{k=0}^\infty (-1)^k \frac{\Theta_{j-1,1}^{2k+1}}{(2k+1)!} \right)\sigma_y
    \right] \right.\\
    & \left.+ \sin(\theta_j) \left[\left(\sum_{k=0}^\infty (-1)^k \frac{\Theta_{j-1,1}^{2k}}{(2k)!} \right)\sigma_y + \left(\sum_{k=1}^\infty (-1)^k \frac{\Theta_{j-1,1}^{2k-1}}{(2k-1)!} \right)\sigma_z \right] \right\} \\
    = & \, \frac{\mu_J}{2} 
    \left\{
    \cos(\theta_j)
    \left[
    \cos(\Theta_{j-1,1})\,\sigma_z + \sin(\Theta_{j-1,1})\,\sigma_y
    \right]+ \sin(\theta_j) \left[
    \cos(\Theta_{j-1,1})\,\sigma_y - \sin(\Theta_{j-1,1})\,\sigma_z
    \right]\right\}\\
    = &  \frac{\mu_J}{2} \left[
     \cos(\theta_j+\Theta_{j-1,1}) \, \sigma_z + \sin(\theta_j+\Theta_{j-1,1}) \sigma_y  \right] \\
    = & \frac{\mu_J}{2} \left[
     \cos(\Theta_{j,1}) \, \sigma_z + \sin(\Theta_{j,1}) \sigma_y  \right]\\
    B' = &  \,\frac{\tilde{J}_j}{2} 
    \left\{
    \cos(\theta_j)
    \left[
    \left(\sum_{k=0}^\infty (-1)^k \frac{\Theta_{j-1,1}^{2k}}{(2k)!} \right)\sigma_z + \left(\sum_{k=0}^\infty (-1)^k \frac{\Theta_{j-1,1}^{2k+1}}{(2k+1)!} \right)\sigma_y
    \right] \right.\\
    & \left.+ \sin(\theta_j) \left[\left(\sum_{k=0}^\infty (-1)^k \frac{\Theta_{j-1,1}^{2k}}{(2k)!} \right)\sigma_y + \left(\sum_{k=1}^\infty (-1)^k \frac{\Theta_{j-1,1}^{2k-1}}{(2k-1)!} \right)\sigma_z \right] \right\}\\
    = &  \frac{\tilde{J}_j}{2} \left\{\cos(\theta_j)\left[\cos(\Theta_{j-1,1})\,\sigma_z + \sin(\Theta_{j-1,1})\,\sigma_y\right] + \sin(\theta_j)\left[\cos(\Theta_{j-1,1})\,\sigma_y - \sin(\Theta_{j-1,1})\,\sigma_z\right] \right\} \\
    = & \frac{\tilde{J}_j}{2} \left[
    \cos(\theta_j+\Theta_{j-1,1}) \, \sigma_z + \sin(\theta_j+\Theta_{j-1,1}) \sigma_y  \right]\\
    = & \frac{\tilde{J}_j}{2} \left[
    \cos(\Theta_{j,1}) \, \sigma_z + \sin(\Theta_{j,1}) \sigma_y  \right]\\
    C' = & \frac{\epsilon_j \theta_j}{2} \sigma_x\\
    P_{j-1}^\dagger \Phi_{Q_j} P_{j-1} = & \, \frac{\mu_J}{2} \left[
    \cos(\Theta_{j,1}) \, \sigma_z + \sin(\Theta_{j,1}) \sigma_y  \right] + \frac{\tilde{J}_j}{2} \left[
    \cos(\Theta_{j,1}) \, \sigma_z + \sin(\Theta_{j,1}) \sigma_y  \right] +\frac{\epsilon_j \theta_j}{2} \sigma_x\\
    \end{split}
\end{equation}
where $\Theta_{b,a} = \sum_{l=a}^{b} \theta_l$.
We can equivalently represent the first order term in the noisy dynamics $\exp(-i \, \Phi_A^{[1]})$ in an error vector representation $\exp(-i \, \vec{a}_1)$ where
\begin{equation}
    \label{eqn:dephasingerrorvector}
    \vec{a}_1 = \left(
    \begin{array}{c}
    \sum_{j=0}^N \frac{\epsilon_j \theta_j}{2}\\
    \sum_{j=0}^N\left(\frac{\mu_J}{2} + \frac{\tilde{J}_j}{2} \right) \sin(\Theta_{j,1})\\
    \sum_{j=0}^N\left(\frac{\mu_J}{2} + \frac{\tilde{J}_j}{2} \right) \cos(\Theta_{j,1})
    \end{array}
    \right)
\end{equation}
Our goal now is to use this error error vector to compute the necessary expression to understand the first order contribution (w.r.t the Magnus expansion) to the infidelity with respect to the dynamics of the noise.

To do this (up to first order in the Magnus expansion), we simply need to compute the expression $\langle \vec{a}_1^2 \rangle$.
We get the following expression:
\begin{equation} \label{eqn:Fcomponentsfirstorder}
    \begin{split}
        \langle \vec{a}_1^2 \rangle = F_{xx} + F_{yy} + F_{zz}\\
        F_{xx} = \frac{1}{4} \left[\gamma_{\epsilon}(0)\sum_{i=1}^N \theta_i^2 + 2 \sum_{h=1}^\infty \gamma_{\epsilon}(h) \sum_{i=h+1}^{N} \theta_i \theta_{i-h} \right]\\
        F_{yy} = \frac{1}{4} \left[ \mu_J^2 \sum_{i=1}^N \sin^2(\Theta_{i,1})+2 \mu_J^2 \sum_{h=1}^N \sum_{i=h+1}^N \sin(\Theta_{i,1})\sin(\Theta_{i-h,1}) +\gamma_{\tilde{J}}(0) \sum_{i=1}^N \sin^2(\Theta_{i,1})\right.\\
        \left. + 2 \sum_{h=1}^\infty \gamma_{\tilde{J}}(h) \sum_{i=h+1}^{N} \sin(\Theta_{i,1})\sin(\Theta_{i-h,1}) \right]\\
        F_{zz} = \frac{1}{4}\left[ \mu_J^2 \sum_{i=1}^N \cos^2(\Theta_{i,1})+2 \mu_J^2 \sum_{h=1}^N \sum_{i=h+1}^N \cos(\Theta_{i,1})\cos(\Theta_{i-h,1}) +\gamma_{\tilde{J}} (0) \sum_{i=1}^N \cos^2 (\Theta_{i,1})\right.\\
        \left.+2 \sum_{h=1}^\infty \gamma_{\tilde{J}} (h) \sum_{i=h+1}^{N} \cos(\Theta_{i,1}) \cos (\Theta_{i-h,1}) \right]
    \end{split}
\end{equation}
where we have dropped in the ARMA formalism for the representation of the noise to describe the noisy trajectory correlations.
Making use of trigonometric identities allows for the simplification into the form of \cref{eqn:DephasingErrorFull}:
\begin{equation}
    \begin{split}
    \langle \vec{a}_1^2 \rangle & = \frac{1}{4} \left[A + B + C\right]\\
    A & = \mu_J^2 \left(N + \sum_{h=1}^N \sum_{i=h+1}^N \cos(\Theta_{i,i-h}) \right)\\
    B & = N \gamma_{\tilde{J}}(0)
    +2 \sum_{h=1}^N \gamma_{\tilde{J}}(h) \sum_{i = h+1}^N \cos (\Theta_{i,i-h})\\
    C & = \gamma_{\epsilon}(0)\sum_{j=1}^N \theta_j^2 + 2 \sum_{h=1}^{N-1} \gamma_{\epsilon}(h) \sum_{i=h+1}^{N} \theta_j \theta_{j-h}
    \end{split}
\end{equation}
which is used for the dephasing analysis of this work.

%% file: Sections/A3-ExampleSolutions.tex
\subsection{Control Noise Only}\label{sec:controlnoiseexamplesappendix}

\subsubsection{Bandlimited Noise} \label{sec:bandlimitedexamples}

\begin{figure}[b]
    \centering
    \subfloat[Power spectrum of bandlimited noise.]{\includegraphics[width=0.49\textwidth]{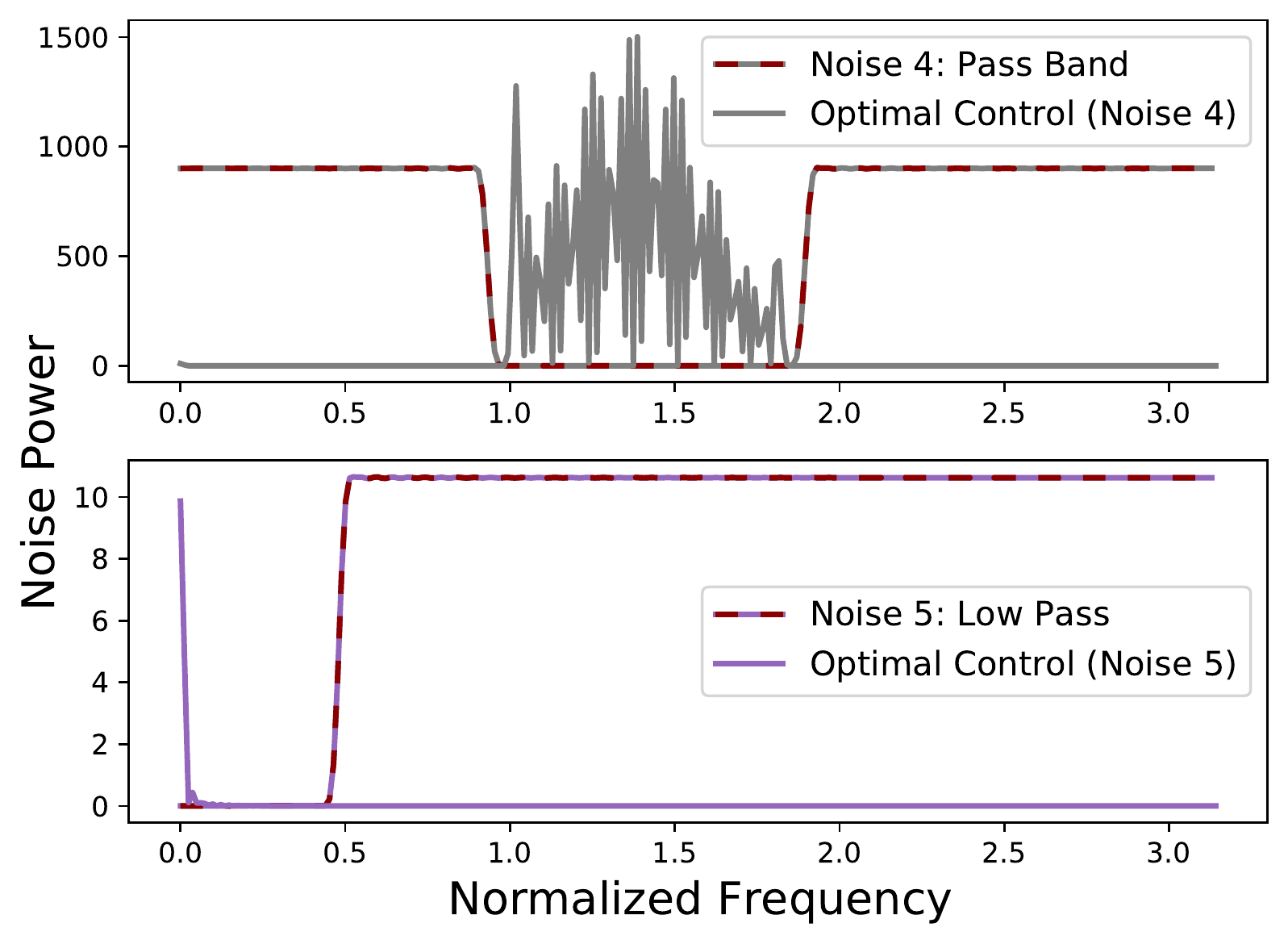}\label{fig:bandlimitedspectrum}}   
    \subfloat[Optimal control for the bandlimited noise examples.]{\includegraphics[width=0.49\textwidth]{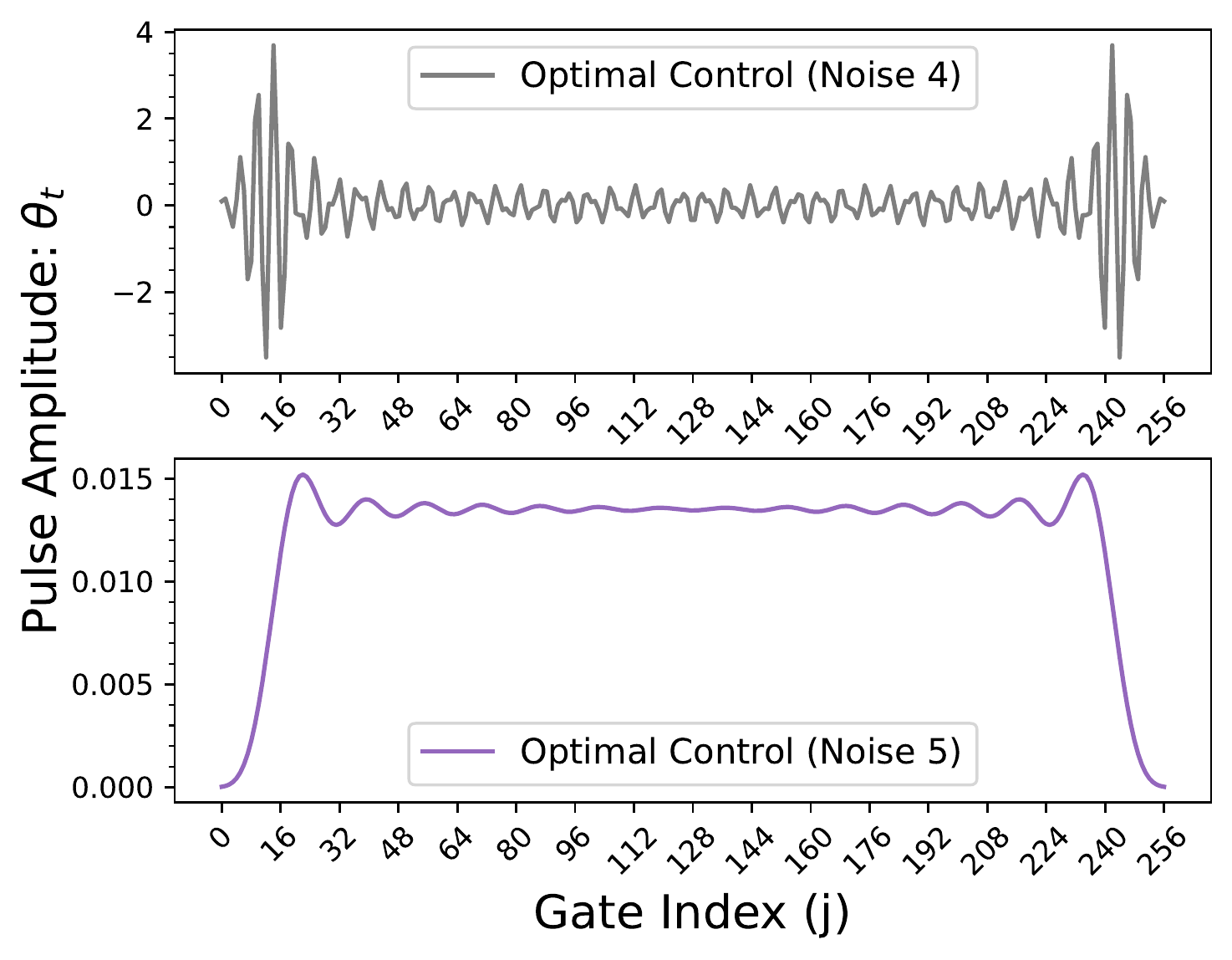}\label{fig:bandlimitedcontrol}}
    \caption{Optimal control sequences for ARMA model defined stationary noise processes from \cref{sec:bandlimitedexamples}.  For these examples, the power of the noise ($\sigma_w^2$) was scaled for visualization purposes.  
    (a) Power spectra for pass-band (top) and low-pass control noise and the power spectrum of the optimal control.
    Our approach successfully locates control noise solutions concentrated at the low power bands of the noise spectra.
    (b) The time domain optimal control sequences for pass-band (top) and low-pass (bottom) control noise.}
    \label{fig:controlnoisesolutions}
\end{figure}

For this additional example, we aim to demonstrate the ability of the approach to find control solutions that optimally mitigate the control noise.  To do so, we focused on two flavors of bandlimited control noise: low-pass and pass-band noise.  
The ARMA models for these processes are more complicated and therefore do not have an intuitive representation with respect to the model, but the ARMA model coefficients can be easily generated with standard finite-impulse response filters from signal processing.
\cref{fig:controlnoisesolutions}c. shows the power spectrum of the noise solutions, shown in \cref{fig:controlnoisesolutions}d., relative to power spectrum of the noise.
The resulting control noise solutions from the convex optimization generated filter functions that overlap with the appropriate band of the noise for both the pass-band and low-pass noise cases as shown in \cref{fig:controlnoisesolutions}c.

\subsection{Control Noise Dominant Solutions}

\begin{figure}
    \centering
    \subfloat[Relative noise contributions.]{\includegraphics[width=0.4\textwidth]{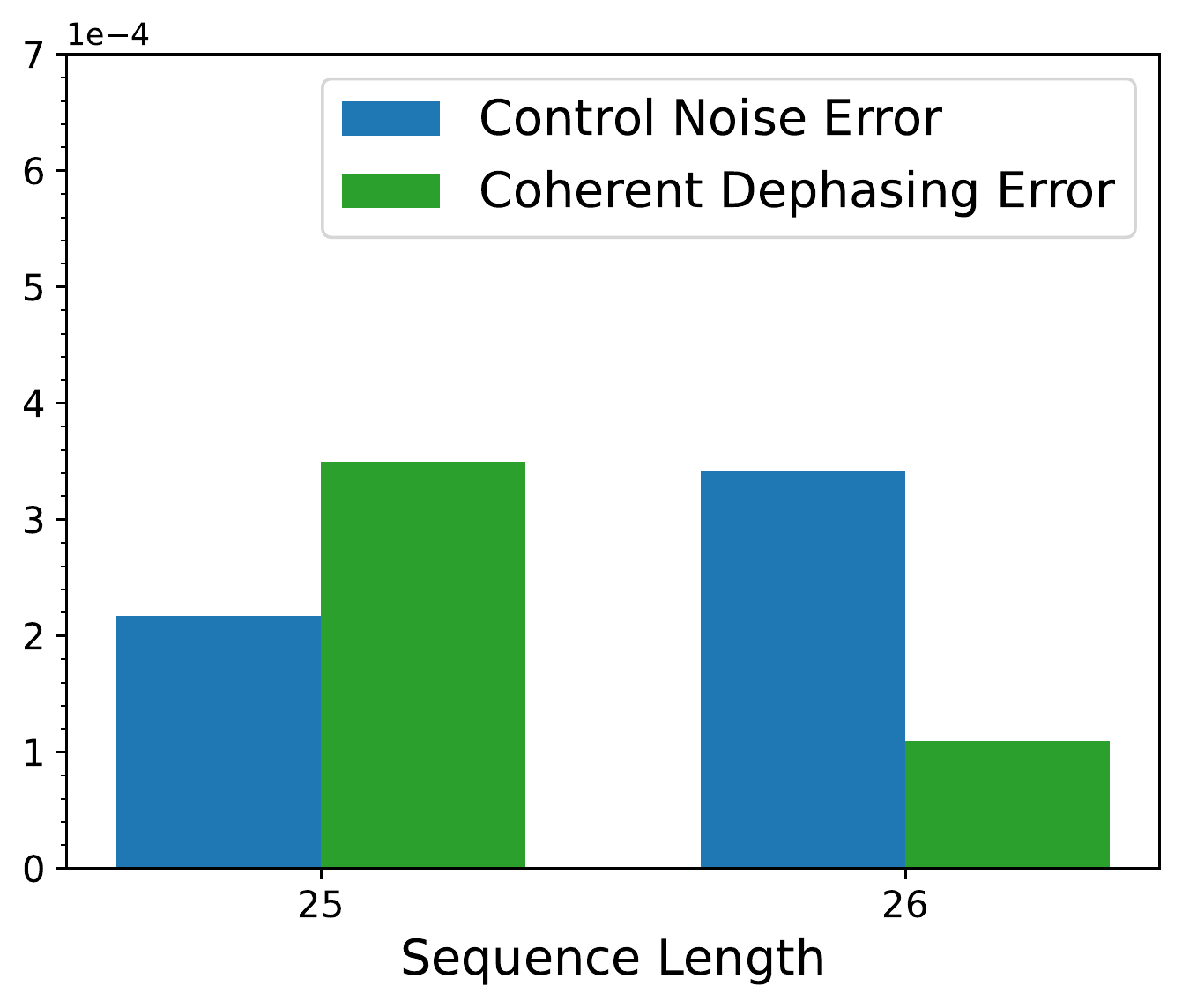}}
    \subfloat[Control noise solutions.]{\includegraphics[width=0.4\textwidth]{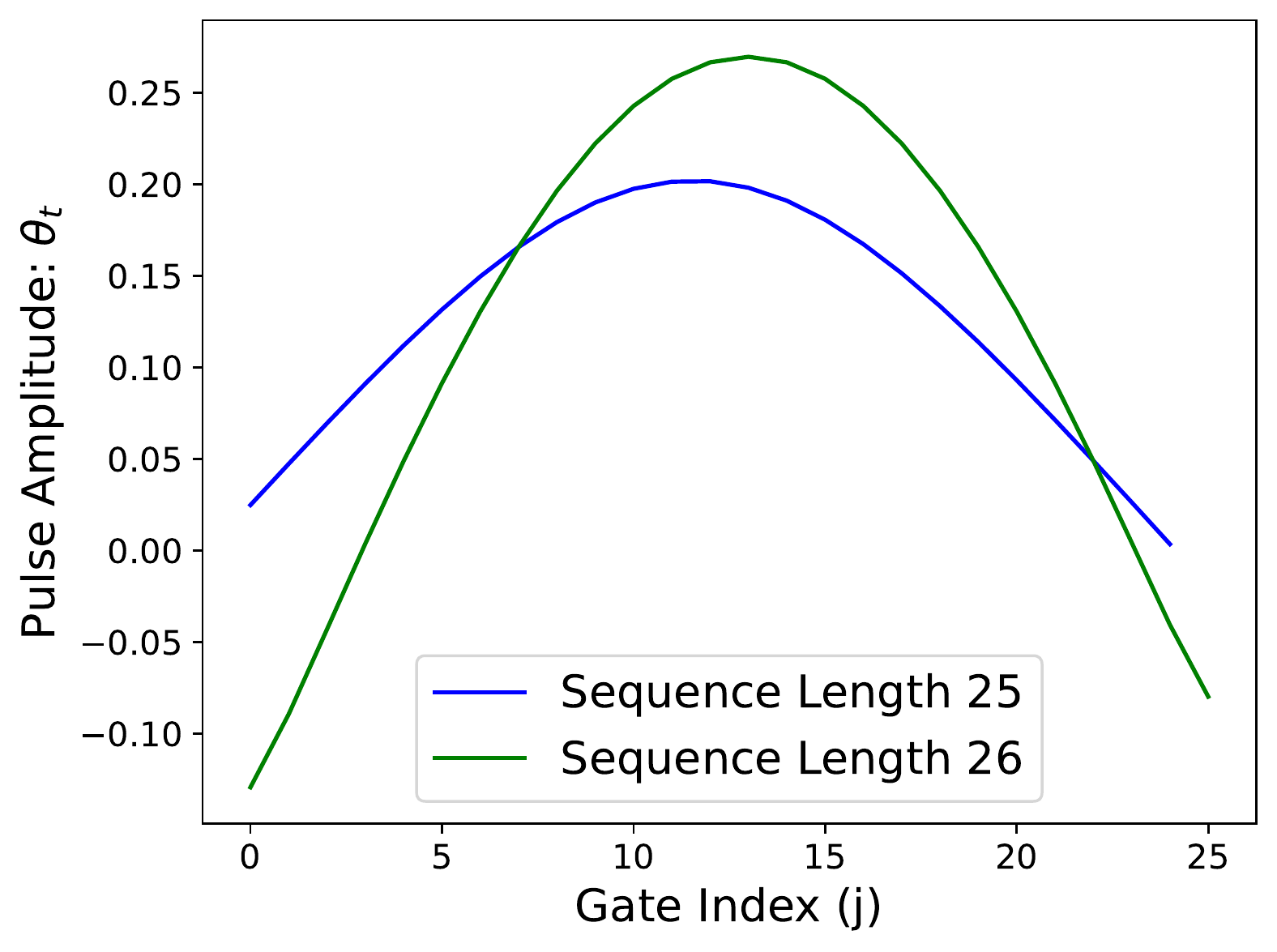}}
    
    \subfloat[Dephasing filter functions.]{\includegraphics[width=0.4\textwidth]{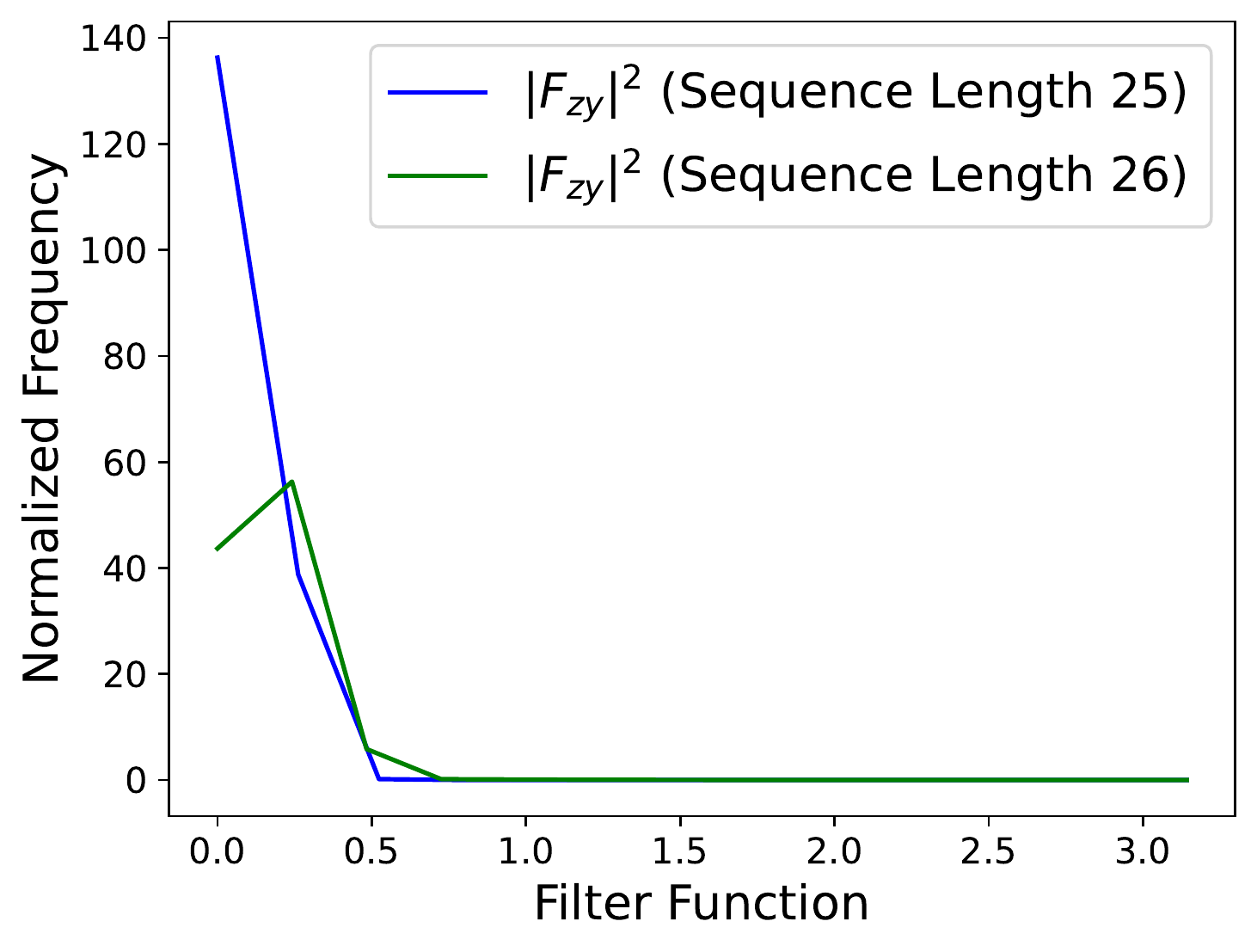}}
    \subfloat[Amplitude filter functions.]{\includegraphics[width=0.4\textwidth]{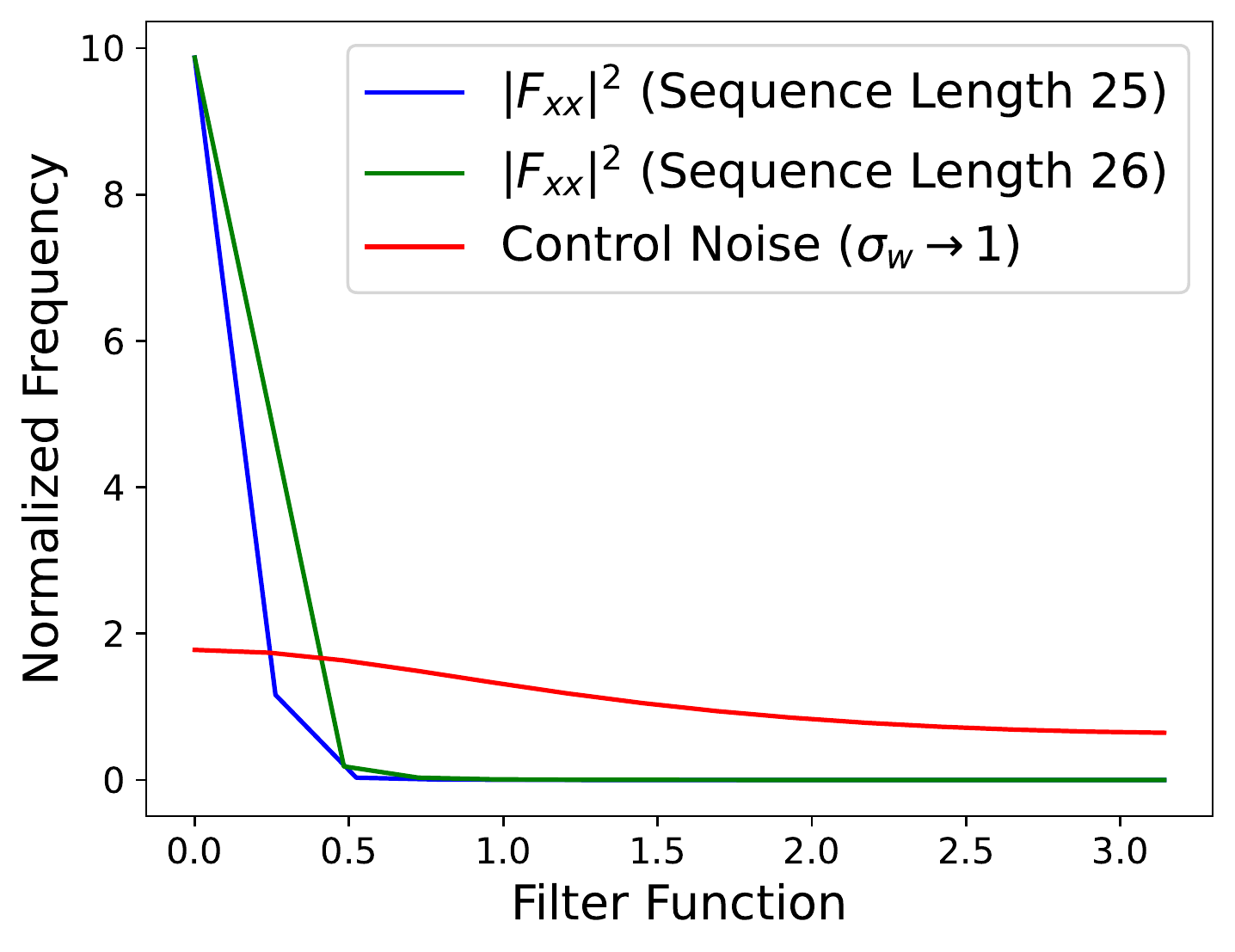}} 
    \caption{Filter function perspective on the discontinuity in the control solutions for the coherent dephasing case from figure \cref{fig:controlnoisedephasing} and \cref{sec:coherentdephasingmain}.  The discontinuity is a result of the optimizer switching between control pulses that focus on mitigating the control noise/dephasing contribution to the infidelity for pulse lengths $K=25$/$K=26$.  (a) The relative contribution to the total infidelity from the control noise and dephasing terms for the $K=25$ and $K=26$ solutions. (b)  The change in control solutions from the control noise focused to the dephasing noise focused mitigation schemes.  (c)  The corresponding dephasing filter functions for the aforementioned solutions.  The coherent dephasing noise (DC, i.e. zero frequency) has all of it's contribution at zero frequency, with the $K=26$ dephasing filter function having a reduced DC component relative to the $K=25$ control solution.  (d)  The corresponding control noise filter functions for the aforementioned solutions.  The power of the control noise has been rescaled for visual purposes.  The $K=25$ control noise solution filter function has a reduced low frequency overlap with the control noise spectrum relative to the $K=26$ solution.}
    \label{fig:FFplotscoherent}
\end{figure}

\subsubsection{Coherent Dephasing Noise} \label{sec:coherentdephasingappendix}

In this section, we will analyze the discontinuity in the control solutions for the coherent dephasing case in \cref{sec:coherentdephasingmain} (figure \cref{fig:controlnoisedephasing}).
This discontinuity is a consequence of our approach minimizing the total infidelity of the operation which contains both control noise and dephasing contributions to the total infidelity.
Therefore, the jump in observed error from the solutions is due to a switching of the optimizer from control noise focused to dephasing noise focused solutions at larger integration times of the noise (recall we have a static DC dephasing term in time here, term $B$ in \cref{eqn:DephasingErrorFull}).
These relative contributions to the total infidelity from the control noise and dephasing terms before ($K=25$) and after ($K=26$) the discontinuity are shown in figure \cref{fig:FFplotscoherent} as well as their associated control solutions.
To understand the change in the control solutions, we will transform into a filter function perspective in regards to the noise and control.
Specifically, we will plot the dephasing and control filter functions and compare them to the power spectrum of the control.
The control and dephasing filter function components are defined accordingly \cite{2020FreySimultSpectralEstQubitSlepian}:
\begin{equation}
\begin{gathered}
    F_{xx}(\omega, \tau) = \int_0^{\tau} ds \, \Omega(s)\, e^{i \omega s}\\
    F_{zy}(\omega, \tau) = \int_0^{\tau} ds \, \sin \,\Theta(s) \, e^{i \omega s}
\end{gathered}
\end{equation}
where $\Theta(t) = \int_0^\tau ds \, \Omega(s)$ and $\Omega(s)$ is the time-dependent amplitude of the control.
In the filter function picture, the choice of control from the optimization becomes clear with the relative contribution of the low frequency magnitudes of the control noise filter function, $|F_{xx}|^2$, and dephasing filter function, $|F_{zy}|^2$, swapping dominant contributions between $K=25$ (control noise mitigation focused) and $K=26$ (dephasing noise mitigation focused); recall both the DC dephasing noise and $1/f^2$ control noise spectra have significant low-frequency contributions.

%% file: Sections/A4-HigherOrderTerms.tex
\subsection{Higher Order Term Contribution for the Control Noise Solutions \label{sec:higherordercontrolnoise}}

In this section, we will discuss the weak noise limit assumption for our constructed optimal control for pure time-correlated control noise.
In the main text, we have provided insight into the minimization of the infidelity represented by the first order term in the Magnus expansion of the dynamics.
While optimal solutions can be found that minimize this approximation to the infidelity, there exists the question of whether the minimization of $\langle \vec{a}_1^2 \rangle$ is sufficient, or if higher order effects will significantly contribute to the dynamics.
In this section, we will run through such an analysis for the $1/f^2$ control noise solutions from \cref{sec:1fexample}.

\begin{figure}[b]
    \centering
    \includegraphics[width=0.45\textwidth]{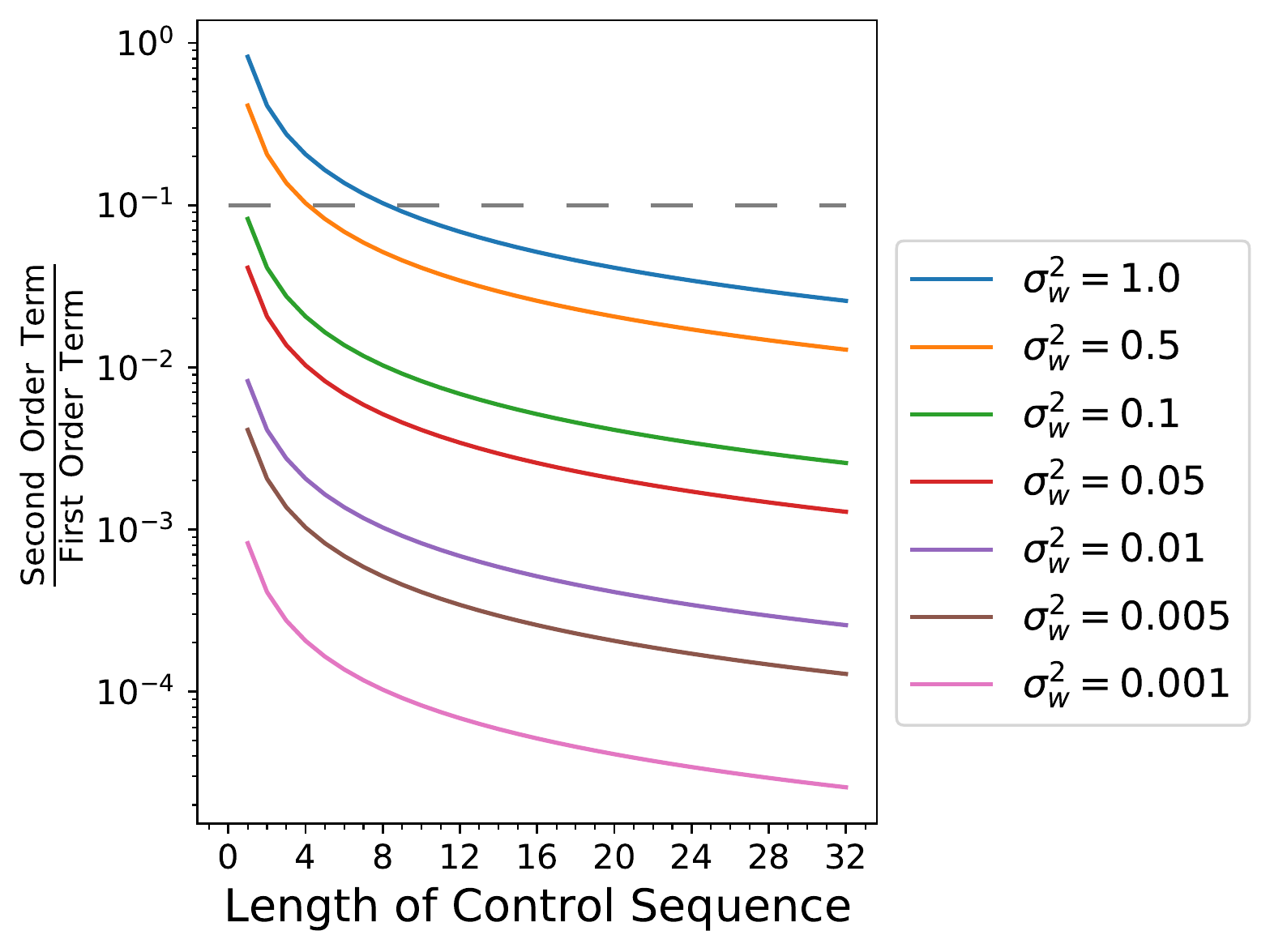}
    \caption{The ratio of the second order to first order term in the expansion of the error dynamics with respect to the infidelity for the optimal white control noise solutions for various powers of the noise, $\sigma_w^2$.  The target logic gate in this example is $\exp(-i \,\theta_g \sigma_x/2)$ with $\theta_g = \pi$.  The gray dashed line indicates a $10\%$ contribution of the second order term to the infidelity relative to the first order term.}
    \label{fig:controlWNlimit}
\end{figure}

To start, we will discuss a simple back of the envelope calculation for potential ARMA model representations of the noise that may lie within this weak noise limit regime recalling that these solutions will serve as a \emph{lower bound} on the erroneous dynamics of ARMA model representations of the noise with stronger temporal correlations.
For this approach, we will make use of the white noise limit solutions to understand bounds under which the noise power may be too strong to meet the ``weak noise limit" regime.
Recall, the white control noise optimal solutions are of the form (\cref{eqn:whitenoiseinfid}):
\begin{equation}
    \big\langle \left|a_1^{x} \right|^2 \big\rangle
    = \frac{\sigma_w^2}{4} \sum_{i=1}^N \Omega_i^2
    = \frac{\sigma_w^2}{4} \frac{\theta_g^2}{N}
\end{equation}
where the variance of the white noise ($S_w(\omega) = \sigma_w^2$) defines the power of the noise for the ARMA processes (\cref{eqn:ARMApowerspectrum}):
\begin{equation}
    S_\epsilon(\omega) = S_w(\omega) \frac{\left| \sum_{j=0}^q b_j \, \exp(-i j \omega) \right|^2}{\left| 1 + \sum_{k=0}^p a_k \, \exp(-i k \omega) \right|^2}
\end{equation}
Analyzing the higher order terms of the form \cite{2012GreenHighOrderFilterLogic,2013GreenArbControlUnivNoise,2014SoareExpNoiseFilter}: 
\begin{equation} \label{eqn:controlhigherorder}
\begin{gathered}
    \mathcal{F}_{av} = \frac{1}{2} \left[1 + \sum_{m=0}^{\infty} (-1)^m \frac{2^{2m}}{(2m)!} \left\langle \big| \vec{a}_1 \big|^2 \right\rangle ^m \right] = \\
    = 1 - \left\langle \big| \vec{a}_1 \big|^2 \right\rangle + \frac{\left\langle \big| \vec{a}_1 \big|^2 \right\rangle^2}{3} - \frac{2\left\langle \big| \vec{a}_1 \big|^2 \right\rangle^3}{45} + ...
\end{gathered}
\end{equation}
allow for us to compute the relative contributions of the second and first order terms in the expansion of the infidelity shown in figure \cref{fig:controlWNlimit}.
In figure \cref{fig:controlWNlimit}, we can identify regimes under which the first order term in the infidelity contributes to the bulk ($90\%$) of the contribution to the infidelity.

\begin{figure}[t]
    \centering
    \subfloat[Higher-order noise contributions as a function of noise power.]{\includegraphics[width=0.45\textwidth]{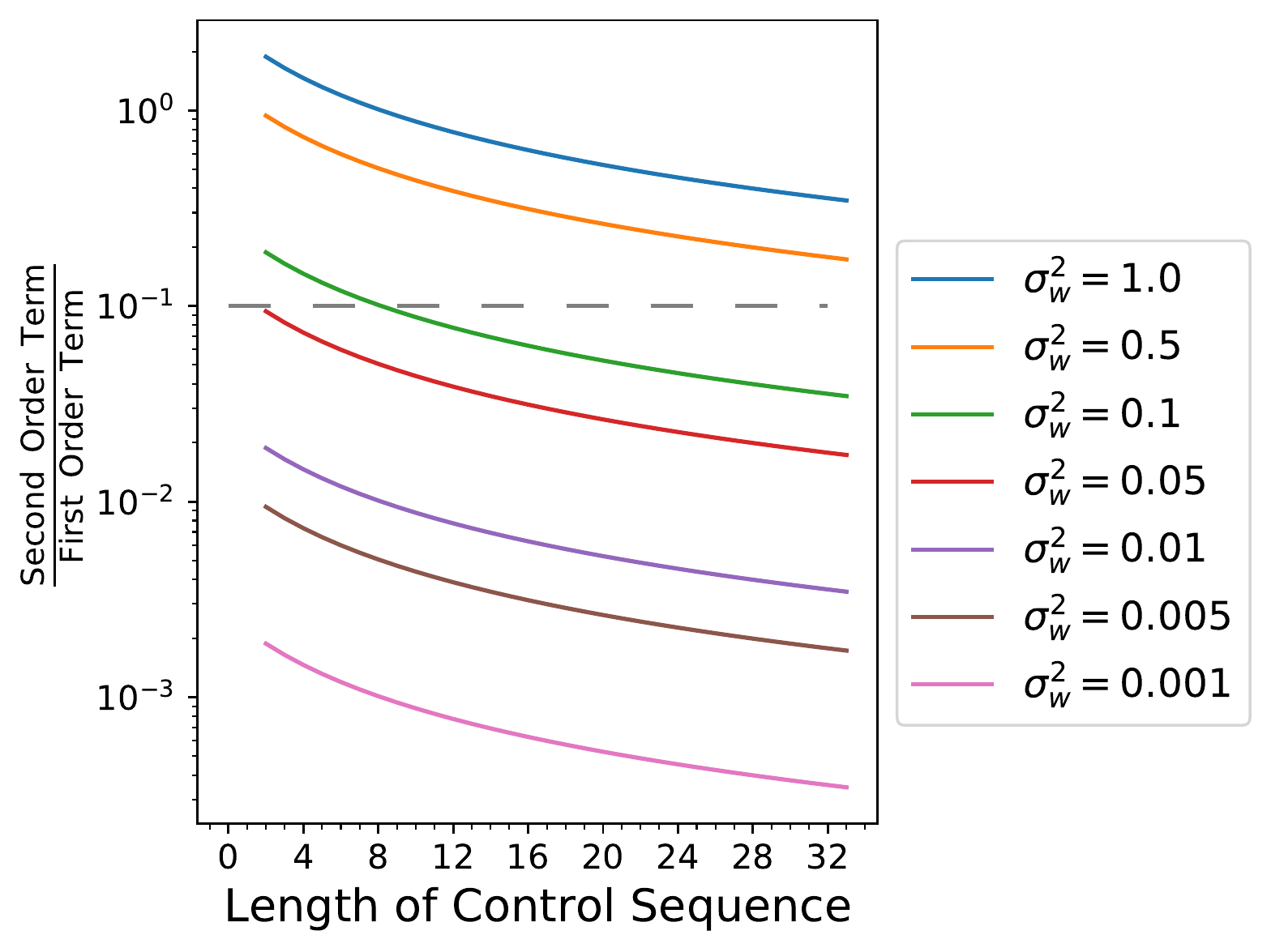} \label{fig:1fhigherorder}}
    \subfloat[Comparison of the higher order terms relative to the Monte Carlo error for the $1/f^2$ control noise solutions from \cref{sec:1fexample}, \cref{fig:controlnoiseerror}]{\includegraphics[width=0.45\textwidth]{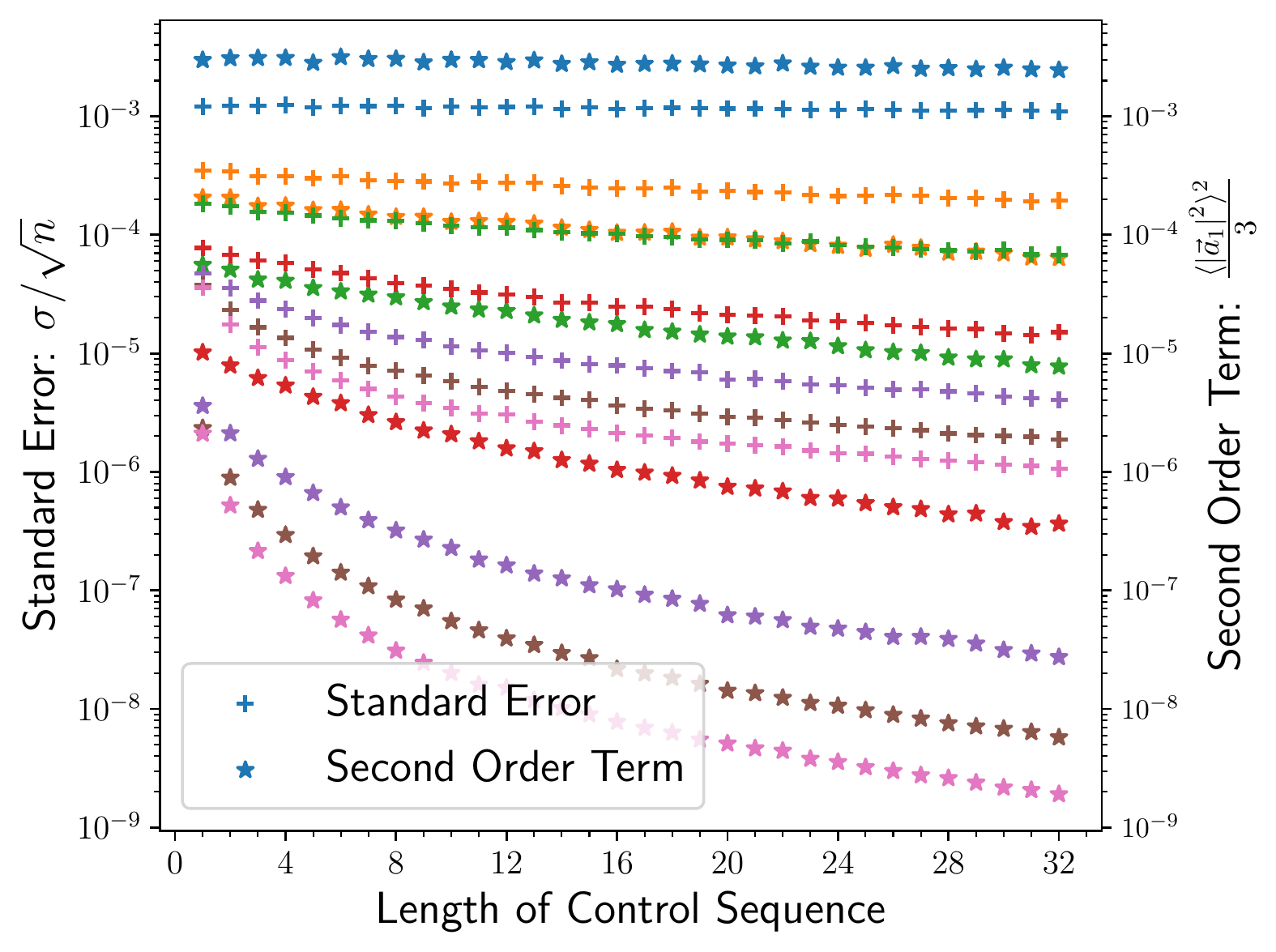} \label{fig:1fhigherordermontecarlo}}
    
    \caption{\cref{fig:1fhigherorder} The ratio of the second order to first order term in the expansion of the error dynamics with respect to the infidelity for the $1/f^2$ AR(1) model control noise noise solutions for various powers of the noise, $\sigma_w^2$.  The target logic gate in this example is $\exp(-i \,\theta_g \sigma_x/2)$ with $\theta_g = \pi$ and the AR(1) representation of the noise has an AR coefficient $a_1 = 0.75$.  The gray dashed line indicates a $10\%$ contribution of the second order term to the infidelity relative to the first order term. \cref{fig:1fhigherordermontecarlo}  The standard error from the Monte Carlo simulations (plus symbols) of the noisy trajectories and the second order term contribution (star symbols) to the infidelity (\cref{eqn:controlhigherorder}) from \cref{fig:1fcontrolnoisesolutions}.  In all but one case, the contribution from the second order term is below the standard error from the Monte Carlo simulations.}
    \label{fig:control1flimit}
\end{figure}

With noise powers that pass the ``white noise limit test", we can repeat the analysis of the higher order terms to gain a refined understanding of the weak noise limit for the noise of interest when additional correlations are at play.
We will repeat the above analysis, but with the $1/f^2$ control noise examples from \cref{sec:1fexample} to demonstrate this more refined analysis as well.  \cref{fig:control1flimit}, shows an example of this analysis for the $1/f^2$ control noise spectrum with white noise cutoff from \cref{sec:1fexample}.
For the example, we used an AR(1) model of the $1/f^2$ spectrum with $a_1 = 0.75$ and scaled the power, $\sigma_w^2$, of the noise to examine regimes that adhere to our ``weak noise limit" criterion.
Note that the correlated noise cases will scale less favorably than the white noise case, however good weak noise limit regimes are obtainable in spite of the increase in the error terms from the included correlations.

\subsection{Higher-Order Terms for Simultaneous Control and Dephasing Noise \label{sec:higherorderdephasing}} 

In this section we will provide the relevant details to understand the higher order terms in the Magnus expansion for the non-commuting case of time-correlated control noise and time-correlated dephasing.
Due to the existence of two semi-classical variables defining the control and dephasing noise, the analysis of the higher-order terms in the Magnus expansion (with respect to the error vector components) are subtlely different than the treatment in \cite{2012GreenHighOrderFilterLogic}.
We will run through the computation of such higher order terms as well as compute their contributions relative to the Monte Carlo error from the simulations in \cref{sec:coherentdephasingmain}.
Green at al. \cite{2012GreenHighOrderFilterLogic} have shown that the computation of the second order terms require the computation of the following terms:
\begin{equation}\label{eqn:secondorder}
    \mathcal{F}_{av} = 1 - \sum_p \langle \vec{a}_{1,p}^2 \rangle - \left[\langle \sum_p \left(\vec{a}_{2,p}^2 \rangle +2 \langle \vec{a}_{1,p} \vec{a}_{3,p}^T \rangle \right) - \frac{1}{3} \sum_{p,q} \langle \vec{a}_{1,p}^2 \vec{a}_{1,q}^2 \rangle  \right]
    + ...
\end{equation}
where the summations over $p,q$ are over the $x$, $y$, and $z$ components of the respective error vector and $\vec{a}_1$ is defined from \cref{eqn:dephasingerrorvector} in \cref{sec:firstorderdephasing} for our examples in the main text.
Recall from the pure control noise example that $\langle a_1^4 \rangle$ is simply $\langle a_1^2 \rangle^2$ (see \cref{sec:higherordercontrolnoise}); leaving only the evaluation of the terms $\langle \vec{a}_2^2 \rangle$ and $\langle \vec{a}_1 \vec{a}_3^T \rangle$ to be done.

For the analysis of these terms, we need to first revisit the first-order term, $\vec{a}_1$, which we defined as:
\begin{equation}
    \vec{a}_1(\tau) = \frac{1}{2} \int_{0}^{\tau} dt \, \bar{s}_1(t) \approx \frac{1}{2} \sum_{j=0}^N \bar{s}_1^j 
\end{equation}
where
\begin{equation}
    \tilde{s}_1(t) = \left(
    \begin{array}{c}
        \epsilon(t) \Omega(t) \\
        J(t) \sin\left(\Theta(t)\right)\\
        J(t) \cos\left(\Theta(t)\right)
    \end{array}
    \right) \approx
    \tilde{s}_1^j = \left(
    \begin{array}{c}
        \epsilon_j \Omega_j \\
        J_j \sin \left(\Theta_{1,j}\right)\\
        J_j \cos \left(\Theta_{1,j}\right)\\
    \end{array}
    \right)
\end{equation}
where we employed that same noise/control slicing and integration ideas from the main text; i.e. $\beta(t) \rightarrow \beta_j = \int_{t_{j-1}}^{t_j} ds \, \beta(s)$.
As mentioned before, this approach differs slightly from \cite{2012GreenHighOrderFilterLogic} due to the existence of multiple noise sources that do not allow factoring out of the noise to construct a true ``control vector", $s_1(t)$, but a vector that contains contributions from the noise and control, $\bar{s}_1(t)$.
Regardless, we can still implement the recursive definitions for the higher-order terms (see \cref{sec:magnusexpansion}) with respect to this ``dressed" control vector to compute the higher-order term contributions.

We will start with the computation of $\sum_{p,q} \langle a_{1,p}^2 a_{1,q}^2 \rangle$ due to the fact that a significant contribution of this expression has been pre-computed through the computation of $\langle a_{1,i}^2 \rangle$.
Particularly, we can recycle the results from the computation of $F_{xx}$, $F_{yy}$, and $F_{zz}$ to compute the $p=q$ contributions to the sum.
Expanding out the sum and using the Gaussian moment theorem to decompose any higher-order moments of the noise into expressions containing second-order moments:
\begin{equation}
    \langle \beta_j \beta_{j'} \beta_{k} \beta_{k'} \rangle = \langle \beta_j \beta_{j'}\rangle \langle \beta_{k} \beta_{k'} \rangle + \langle \beta_j \beta_{k} \rangle \langle \beta_{j'} \beta_{k'} \rangle + \langle \beta_j \beta_{k'} \rangle \langle \beta_{j'} \beta_{k} \rangle
\end{equation}
gives
\begin{equation}
    \sum_{p=q} \langle a_{1,p}^2 a_{1,p}^2 \rangle = F_{xx}^2 + F_{yy}^2 + F_{zz}^2
\end{equation}
where $F_{xx}$, $F_{yy}$, and $F_{zz}$ have been computed in \cref{eqn:Fcomponentsfirstorder}.
We can upper bound theses expressions as well:
\begin{equation}
    \begin{split}
        F_{xx}^2 = & \; \frac{1}{16}\sum_{j,j',k,k'} \langle \epsilon_j \epsilon_{j'} \epsilon_k \epsilon_{k'} \rangle \Omega_j \Omega_{j'} \Omega_k \Omega_{k'} \\
        = & \; \frac{3}{16} \left(\sum_{j,j'} \langle \epsilon_j \epsilon_{j'} \rangle \Omega_j \Omega_{j'}\right)^2 \le \; \frac{3}{16} \, \Omega^2_{\rm max} \left(\sum_{j,j'} \langle \epsilon_j \epsilon_{j'} \rangle\right)^2 = 3 \,\Omega_{\rm max}^2 \left(\sum_{h=0}^{N-1} c(h) (N-h) \gamma_{\epsilon \epsilon}(h) \right)^2 \\
        F_{yy}^2 + F_{zz}^2 = & \; \frac{1}{16} \sum_{j,j',k,k'} \langle J_j J_{j'} J_k J_{k'} \rangle \sin\left(\Theta_{1,j} \right) \sin\left(\Theta_{1,j'} \right) \sin\left(\Theta_{1,k} \right) \sin\left(\Theta_{1,k'}  \right)\\
        & \;+ \frac{1}{16}\sum_{j,j',k,k'} \langle J_j J_{j'} J_k J_{k'} \rangle \cos\left(\Theta_{1,j} \right) \cos\left(\Theta_{1,j'} \right) \cos\left(\Theta_{1,k} \right) \cos\left(\Theta_{1,k'} \right)\\
        = & \; \frac{3}{16} \left[\sum_{j,j'} \langle J_j J_{j'} \rangle \sin\left(\Theta_{1,j} \right) \sin\left(\Theta_{1,j'} \right)\right]^2 + \frac{3}{16} \left[\sum_{j,j'} \langle J_j J_{j'} \rangle \cos\left(\Theta_{1,j} \right) \cos\left(\Theta_{1,j'} \right)\right]^2\\
        = & \; \frac{3}{16} \left[\sum_{j,j'} \langle J_j J_{j'} \rangle \cos\left(\Theta_{1,j}-\Theta_{1,j'} \right) \right]^2 \le \; \; \frac{3}{16} \left(\sum_{j,j'} \langle J_j J_{j'} \rangle \right)^2 = \frac{3}{16} \left(\sum_{h=0}^{N-1} c(h) (N-h) \gamma_{JJ}(h) \right)^2
    \end{split}
\end{equation}
with
\begin{equation}
    c(h) = \left\{
    \begin{array}{cc}
         1 & {\rm if}\;\; h = 0 \\
         2 & {\rm if}\;\; h \neq 0
    \end{array}
    \right.
\end{equation}
in a very similar manner to \cite{2012GreenHighOrderFilterLogic}, but with the noise defined with respect to the ARMA models characterization of the noise sources instead of the root mean square error of the noise utilized in \cite{2012GreenHighOrderFilterLogic}.
Moving forward, we will defined the ARMA model defined representation of the noise correlations as
\begin{equation}
    \Gamma_{\beta \beta'} = \sum_{h=0}^{N-1} c(h) (N-h) \gamma_{\beta \beta'} (h) \,.
\end{equation}
where $\gamma_{\beta \beta'} = \langle \beta(t) \beta'(t) \rangle$.  Note that this expression, $\Gamma_{\beta\beta'}$, is simply the sum of all of the terms of the autocovariance matrix of the semi-classical noise, $\beta$.
Now, we are left with the computation of the $p \neq q$ cases in the above sum.
These are encompassed with three cases:
\begin{equation}
    \begin{split}
    \langle a^2_{1,x} a^2_{1,y} \rangle = \langle a_{1,y}^2 a_{1,x}^2 \rangle = & \; \frac{1}{16} \sum_{j,j',k,k'} \langle \epsilon_j \epsilon_{j'} J_k J_{k'} \rangle \Omega_{j} \Omega_{j'} \sin\left(\Theta_{1,k} \right) \sin\left(\Theta_{1,k'} \right) \\
     = & \frac{1}{16} \left(\sum_{j,j'} \langle \epsilon_j \epsilon_{j'} \rangle \Omega_j \Omega_{j'} \right)\left(\sum_{k,k'} \langle J_k J_{k'}  \rangle \sin\left(\Theta_{1,k} \right) \sin\left(\Theta_{1,k'} \right) \right)\\
    & + \frac{1}{8} \left(\sum_{j,k} \langle \epsilon_j J_k \rangle \Omega_j \sin\left(\Theta_{1,k} \right) \right)^2 \le \frac{1}{16}\Omega_{\rm max}^2 \Gamma_{\epsilon \epsilon} \Gamma_{JJ} + \frac{1}{8} \left(\Omega_{\rm max} \Gamma_{\epsilon J} \right)^2\\
    \langle a^2_{1,x} a^2_{1,z} \rangle = \langle a_{1,z}^2 a_{1,x}^2 \rangle \le & \; \frac{1}{16}\Omega_{\rm max}^2 \Gamma_{\epsilon \epsilon} \Gamma_{JJ} + \frac{1}{8} \left(\Omega_{\rm max} \Gamma_{\epsilon J} \right)^2\\
    \langle a^2_{1,y} a^2_{1,z} \rangle = \langle a_{1,z}^2 a_{1,y}^2 \rangle = & \frac{1}{16} \sum_{j,j',k,k'} \langle J_j J_{j'} J_k J_k' \rangle \sin\left(\Theta_{1,j} \right) \sin\left(\Theta_{1,j'} \right) \cos\left(\Theta_{1,k} \right) \cos\left(\Theta_{1,k'} \right) \le \frac{3}{16} \Gamma_{JJ}^2 \, .
    \end{split}
\end{equation}
Putting it all together, we get an expression for the upper bound of:
\begin{equation}
    \sum_{p,q} \langle \vec{a}_{1,p}^2 \vec{a}_{1,q}^2 \rangle \le \; \frac{3}{16} \, \Omega_{\rm max}^2 \Gamma_{\epsilon \epsilon}^2 + \frac{9}{16} \, \Gamma_{J J}^2 + \frac{1}{4} \left(\Omega_{\rm max}^2 \Gamma_{\epsilon \epsilon} \Gamma_{JJ} + 2 \left(\Omega_{\rm max} \Gamma_{\epsilon J} \right)^2 \right)
\end{equation}
which can be compactly represented due to the ARMA model representations of the noise correlations encompassed in the $\Gamma_{\beta \beta'}$ terms.

Now, we will compute $\langle \vec{a}^2_2(\tau) \rangle$ where
\begin{equation}
    \vec{a}_2(\tau) = \frac{1}{4}\int_0^\tau dt_2 \int_0^{t_2} dt_1 \, \vec{s_2}(t_1,t_2) \approx \frac{1}{4}\sum_{j=0}^N \sum_{j'=0}^j \vec{s_2}^{j,j'}
\end{equation}
and
\begin{equation}
    s_2^{j,j'} = s_1^j \times s_1^{j'} = \left(
    \begin{array}{c}
        J_j J_{j'} \sin \left(\Theta_{1,j} - \Theta_{1,j'} \right)  \\
        J_j \Omega_{j'}\epsilon_{j'} \cos\left(\Theta_{1,j} \right) - J_{j'} \Omega_{j}\epsilon_{j} \cos\left(\Theta_{1,j'} \right)\\
        J_{j'} \Omega_{j}\epsilon_{j} \sin\left(\Theta_{1,j'} \right) - J_{j} \Omega_{j'}\epsilon_{j'} \sin\left(\Theta_{1,j} \right)
    \end{array}
    \right)\, .
\end{equation}
With the above expression, we can compute the components, $\langle \vec{a}_{2,p}^2 \rangle$:
\begin{equation}
    \begin{split}
    \langle \vec{a}_{2,x}^2 \rangle = & \frac{1}{16} \sum_{j=0}^{N} \sum_{j'=0}^j \sum_{k=0}^{N} \sum_{k'=0}^k \langle J_j J_{j'} J_k J_{k'} \rangle \sin \left(\Theta_{1,j} - \Theta_{1,j'} \right) \sin \left(\Theta_{1,k} - \Theta_{1,k'} \right) \le \frac{1}{16} \sum_{j=0}^{N} \sum_{j'=0}^j \sum_{k=0}^{N} \sum_{k'=0}^k \langle J_j J_{j'} J_k J_{k'} \rangle\\
    \le & \; \frac{1}{16} \sum_{j=0}^{N} \sum_{j'=0}^N \sum_{k=0}^{N} \sum_{k'=0}^N \langle J_j J_{j'} J_k J_{k'} \rangle = \frac{3}{16} \Gamma_{JJ}^2\\
    \langle \vec{a}_{2,y}^2 \rangle \le & \frac{1}{16} \sum_{j=0}^{N} \sum_{j'=0}^j \sum_{k=0}^{N} \sum_{k'=0}^k \left( \langle J_j J_k \epsilon_{j'} \epsilon_{k'} \rangle \Omega_{j'} \Omega_{k'} - \langle J_j J_k' \epsilon_{j'} \epsilon_{k} \rangle \Omega_{j'} \Omega_{k} - \langle J_{j'} J_k \epsilon_{j} \epsilon_{k'} \rangle \Omega_{j} \Omega_{k'} + \langle J_{j'} J_{k'} \epsilon_{j} \epsilon_{k} \rangle \Omega_{j} \Omega_{k} \right)\\
    \le & \frac{1}{16} \Omega_{\rm max}^2 \sum_{j=0}^{N} \sum_{j'=0}^j \sum_{k=0}^{N} \sum_{k'=0}^k \left( \langle J_j J_k \epsilon_{j'} \epsilon_{k'} \rangle + \langle J_{j'} J_{k'} \epsilon_{j} \epsilon_{k} \rangle \right) \le \; \frac{1}{8} \Omega_{\rm max}^2 \sum_{j=0}^{N} \sum_{j'=0}^N \sum_{k=0}^{N} \sum_{k'=0}^N  \langle J_j J_k \epsilon_{j'} \epsilon_{k'} \rangle\\
    \le & \; \frac{1}{16}\Omega_{\rm max}^2 \Gamma_{\epsilon \epsilon} \Gamma_{JJ} + \frac{1}{8} \left(\Omega_{\rm max} \Gamma_{\epsilon J} \right)^2\\
    \langle \vec{a}_{2,z}^2 \rangle \le & \; \frac{1}{16}\Omega_{\rm max}^2 \Gamma_{\epsilon \epsilon} \Gamma_{JJ} + \frac{1}{8} \left(\Omega_{\rm max} \Gamma_{\epsilon J} \right)^2
    \end{split}
\end{equation}
Putting it all together, we are able to represent the contribution from the second order terms as:
\begin{equation}
    \sum_p \langle a_{2,p}^2 \rangle \le \; \frac{3}{16} \Gamma_{JJ}^2 + \frac{1}{8}\Omega_{\rm max}^2 \Gamma_{\epsilon \epsilon} \Gamma_{JJ} + \frac{1}{4} \left(\Omega_{\rm max} \Gamma_{\epsilon J} \right)^2
\end{equation}
which is the second piece of the computation of the second-order terms.  There is one term to go.

For the final term in the Magnus expansion, we first need to compute the term, $\bar{s}_3^{j,j',j''}$:
\begin{equation}
    \begin{split}
        \bar{s}_{3,x}^{j,j',j''} = & \; 2 J_j \epsilon_{j'} J_{j''} \Omega_{j'} \cos\left(\Theta_{1,j} - \Theta_{1,j''} \right) - J_j J_{j'} \epsilon_{j''}  \Omega_{j''} \cos\left(\Theta_{1,j} - \Theta_{1,j'} \right) - \epsilon_{j} J_{j'}  J_{j''} \Omega_{j} \cos\left(\Theta_{1,j'} - \Theta_{1,j''} \right)\\
        \bar{s}_{3,y}^{j,j',j''} = & J_j J_{j'} J_{j''} \sin\left(\Theta_{1,j'} - \Theta_{1,j''} \right) \cos\left(\Theta_{1,j} \right) - J_j J_{j'} J_{j''} \sin\left(\Theta_{1,j} - \Theta_{1,j'} \right) \cos\left(\Theta_{1,j''} \right)\\
        & - J_j \epsilon_{j'} \epsilon_{j''} \Omega_{j'} \Omega_{j''} \sin\left(\Theta_{1,j} \right) + 2 \epsilon_{j}  J_{j'} \epsilon_{j''} \Omega_{j} \Omega_{j''} \sin\left(\Theta_{1,j'} \right) - \epsilon_{j}  \epsilon_{j'} J_{j''} \Omega_{j} \Omega_{j'} \sin\left(\Theta_{1,j''} \right)\\
        \bar{s}_{3,z}^{j,j',j''} = & J_j J_{j'} J_{j''} \sin\left(\Theta_{1,j} - \Theta_{1,j'} \right) \sin\left(\Theta_{1,j''} \right) - J_j J_{j'} J_{j''} \sin\left(\Theta_{1,j'} - \Theta_{1,j''} \right) \sin\left(\Theta_{1,j} \right)\\
        & - J_j \epsilon_{j'} \epsilon_{j''} \Omega_{j'} \Omega_{j''} \cos\left(\Theta_{1,j} \right) + 2 \epsilon_{j}  J_{j'} \epsilon_{j''} \Omega_{j} \Omega_{j''} \cos\left(\Theta_{1,j'} \right) - \epsilon_{j}  \epsilon_{j'} J_{j''} \Omega_{j} \Omega_{j'} \cos\left(\Theta_{1,j''} \right)
    \end{split}
\end{equation}
Utilizing the expressions above, we can again bound the contributions:
\begin{equation}
    \begin{split}
        \langle \vec{a}_{1,x} \vec{a}_{3,x} \rangle \le & \; \frac{1}{16} \sum_{j=0}^N \sum_{k=0}^N \sum_{k'=0}^k \sum_{k''=0}^{k'} \vec{s}_{1,x} \vec{s}_{3,x} \le \; \frac{1}{8} \sum_{j=0}^N \sum_{k=0}^N \sum_{k'=0}^N \sum_{k''=0}^{N} \langle \epsilon_{j} J_k \epsilon_{k'} J_{k''} \rangle \Omega_{j} \Omega_{k'}\\
        \le & \; \frac{1}{8}\Omega_{\rm max}^2 \Gamma_{\epsilon \epsilon} \Gamma_{JJ} + \frac{1}{4} \left(\Omega_{\rm max} \Gamma_{\epsilon J} \right)^2\\
        \langle \vec{a}_{1,y} \vec{a}_{3,y} \rangle \le & \; \frac{1}{16} \sum_{j=0}^N \sum_{k=0}^N \sum_{k'=0}^k \sum_{k''=0}^{k'} \vec{s}_{1,y} \vec{s}_{3,y} \le \; \frac{1}{16}\sum_{j=0}^N \sum_{k=0}^N \sum_{k'=0}^N \sum_{k''=0}^{N} \left( \langle J_j J_k J_{k'} J_{k''} \rangle + 2 \langle J_j \epsilon_{k}  J_{k'} \epsilon_{k''}\rangle \Omega_{k} \Omega_{k''} \right)\\
        \le & \; \frac{1}{16}\Gamma_{JJ}^2 + \frac{1}{8}\Omega_{\rm max}^2 \Gamma_{\epsilon \epsilon} \Gamma_{JJ} + \frac{1}{4} \left(\Omega_{\rm max} \Gamma_{\epsilon J} \right)^2\\
        \langle \vec{a}_{1,z} \vec{a}_{3,z} \rangle \le & \; \frac{1}{16}\Gamma_{JJ}^2 + \frac{1}{8}\Omega_{\rm max}^2 \Gamma_{\epsilon \epsilon} \Gamma_{JJ} + \frac{1}{4} \left(\Omega_{\rm max} \Gamma_{\epsilon J} \right)^2
    \end{split}
\end{equation}
results in the contribution from these terms of:
\begin{equation}
    \sum_{p}  \langle \vec{a}_{1,p} \vec{a}_{3,p} \rangle = \frac{3}{8}\Omega_{\rm max}^2 \Gamma_{\epsilon \epsilon} \Gamma_{JJ} + \frac{3}{4} \left(\Omega_{\rm max} \Gamma_{\epsilon J} \right)^2 + \frac{1}{8}\Gamma_{JJ}^2
\end{equation}
which is the final peice to estimate the second-order terms from \cref{eqn:secondorder}.

\begin{figure}[t]
    \centering
    \subfloat[Monte Carlo error of simulations.]{\includegraphics[width=0.7\textwidth]{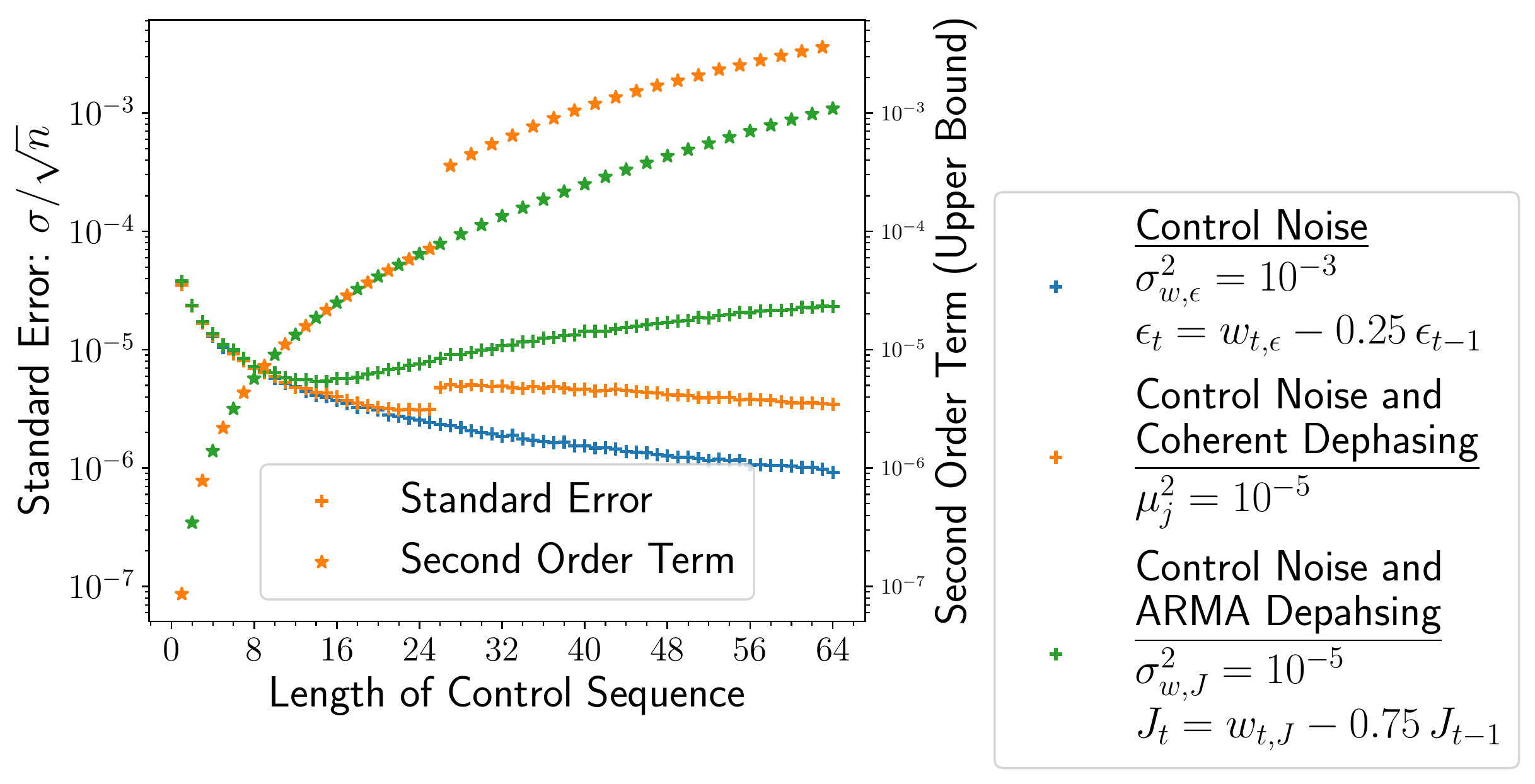}}
    \caption{Monte Carlo error for the simulated noisy control and contribution of the higher order terms in the Magnus expansion for the coherent noise and ARMA dephasing noise examples from \cref{fig:controlnoisedephasing}.}
    \label{fig:higherorderdephasing}
\end{figure}

Putting it all together, we arrive at an upper bound on the second-order terms (with respect to the noise) of:
\begin{equation}
    \sum_p \left(\langle \vec{a}_{2,p}^2 \rangle +2 \langle \vec{a}_{1,p} \vec{a}_{3,p}^T \rangle \right) - \frac{1}{3} \sum_{p,q} \langle \vec{a}_{1,p}^2 \vec{a}_{1,q}^2 \rangle \le \; \frac{7}{16} \Gamma_{JJ}^2 + \frac{7}{8}\Omega_{\rm max}^2 \Gamma_{\epsilon \epsilon} \Gamma_{JJ} + \frac{7}{4} \left(\Omega_{\rm max} \Gamma_{\epsilon J} \right)^2
\end{equation}
We can use the above expression to bound the contribution of the second order terms in the simulations in the examples from \cref{sec:controldephasing}.
To do so, we simply need estimates for the values of $\Omega_{\rm max}$, $\Gamma_{\epsilon \epsilon}$, $\Gamma_{J J}$, and $\Gamma_{\epsilon J}$.
Due to implications on the weak noise limit, we have already discussed bounds on the maximum amplitude of the control (see \cref{sec:controloptimality}) which can loosely set as $\Omega_{\rm max} \le \pi$.
The other expressions are related to the total power of the noise during the application of the gate:
\begin{equation}
    \begin{split}
    \Gamma_{\epsilon \epsilon} \le & \; \mathcal{O} (N^2 \sigma^2_{w, \epsilon})\\
    \Gamma_{J J} \le & \; \mathcal{O} (N^2 (\mu_J^2 + \sigma^2_{w,J}) )\\
    \Gamma_{\epsilon J} \approx & \; \mathcal{O} (N^2 \sigma^2_{w, \epsilon} (\mu_J^2 + \sigma^2_{w,J}) )
    \end{split}
\end{equation}
These estimates were utilized to obtain upper bounds on the second-order contributions to the effective error dynamcis and plotted relative to the Monte Carlo error of the simulation in \cref{fig:higherorderdephasing}.

